\documentclass[aps,prd,preprint,superscriptaddress,tightenlines,nofootinbib]{revtex4}

\usepackage{rotating}

\usepackage{graphicx}
\usepackage{epsfig}
\usepackage{dcolumn}
\usepackage{bm}

\newcommand{\dedx}{dE/dx}
\newcommand{\enu}{e^+ \nu_e}

\newcommand{\dpke}{D^+ \to \bar{K}^0 e^+\nu_e}

\newcommand{\bdzke}{{\cal B}( D^0 \to K^- e^+\nu_e)}
\newcommand{\dppie}{ D^+ \to \pi^0e^+\nu_e}

\newcommand{\dzke}{D^0\to K^-e^+\nu_e}

\newcommand{\dzpie}{D^0\to \pi^-e^+\nu_e}

\newcommand{\vcs}{|V_{cs}|}
\newcommand{\vcd}{|V_{cd}|}

\newcommand{\ksp}{K^0_S\pi^0}

\newcommand{\piz}{\pi^0}
\newcommand{\beqn}{\begin{equation}}
\newcommand{\eeqn}{\end{equation}}

\newcommand{\ra}{\rightarrow}
\def\kenu{D^0\to K^- e^+ \nu_e}
\def\penu{D^0\to \pi^- e^+ \nu_e}
\def\ksenu{D^+\to \bar{K}^0 e^+ \nu_e}
\def\p0enu{D^+\to \pi^0 e^+ \nu_e}

\def\ksp0enu{D^+\to K^0_S (\pi^0) e^+ \nu_e}

\def\cleoc{\hbox{CLEO-c~}}

\begin{document}

\preprint{CLNS 08/2039}
\preprint{CLEO 08-21}

\title{\boldmath Study of $D^0 \to \pi^- e^+ \nu_e$,
      $D^+ \to \pi^0 e^+ \nu_e$,
      $D^0 \to K^- e^+ \nu_e$,
      and
      $D^+ \to \bar{K}^0 e^+ \nu_e$
in Tagged Decays of the $\psi(3770)$ Resonance}

\author{J.~Y.~Ge}
\author{D.~H.~Miller}
\author{V.~Pavlunin}
\altaffiliation{Now at University of California at Santa Barbara, Santa Barbara, California 93106, USA}
\author{B.~Sanghi}
\altaffiliation{Now at Fermi National Accelerator Laboratory, Batavia, Illinois 60510, USA}
\author{I.~P.~J.~Shipsey}
\author{B.~Xin}
\affiliation{Purdue University, West Lafayette, Indiana 47907, USA}
\author{G.~S.~Adams}
\author{D.~Hu}
\author{B.~Moziak}
\author{J.~Napolitano}
\affiliation{Rensselaer Polytechnic Institute, Troy, New York 12180, USA}
% ...
\collaboration{CLEO Collaboration}
\noaffiliation

\date{February 2, 2009}

\begin{abstract}

Using $\psi(3770)\ra D \bar D $ events collected with the
\cleoc detector at the Cornell $e^+ e^-$ storage ring, tagged by
fully reconstructing one $D$ meson in a hadronic decay mode, we
measure absolute branching
fractions and differential decay rates
for $D^0 \rightarrow \pi^- e^+ \nu_e$,  $D^+ \rightarrow \pi^0 e^+
\nu_e$, $D^0 \rightarrow K^- e^+ \nu_e$, and $D^+ \rightarrow
\bar{K}^0 e^+ \nu_e$. The measured decay rates are used to study
semileptonic form factors governing these transitions and to test
unquenched Lattice QCD (LQCD) calculations. We average our results
with previously published \cleoc measurements of the same quantities using a neutrino
reconstruction technique.
Combining LQCD calculations of form factor absolute
normalizations~$f_+(0)$ and measurements of $f^{\pi}_+(0)
\vcd$ and $f^K_+(0)\vcs$, we find $\vcd =
0.222(8)(3)(23)$ and $\vcs = 1.018(10)(8)(106)$, where
the uncertainties are statistical, experimental systematic,
and from LQCD, respectively.

\end{abstract}

\maketitle

\section{Introduction}

The quark mixing parameters are fundamental constants of the
Standard Model of particle physics. They determine the nine
weak-current quark coupling elements of the
Cabibbo-Kobayashi-Maskawa~(CKM) matrix~\cite{ckm}.
In the Standard Model the CKM matrix is unitary.
Measuring the quark couplings tests the unitarity of the matrix.

The extraction
of the quark couplings is difficult because quarks are bound
inside hadrons by the strong interaction. Semileptonic decays are
the preferred way to determine the CKM matrix elements as the
strong interaction binding effects are confined to the hadronic
current. They are parameterized by form factors that are
calculable, for example, by lattice quantum chromodynamics~(LQCD)
and QCD sum rules. Nevertheless, form factor uncertainties
dominate the precision with which the CKM matrix elements can be
determined~\cite{VubVcb}.

Studies of the semileptonic decays of $D$ mesons play an important
role in understanding the CKM matrix. First these decays allow
the robust determination  of the
couplings $\vcs$ and $\vcd$ by combining measured branching fractions
with form factor calculations. Second
 $\vcs$ and $\vcd$ are tightly constrained when
the CKM matrix is assumed to be unitary. Therefore measurements of charm
semileptonic decay rates, when combined with the
values of $\vcs$ and $\vcd$ constrained by the unitarity of the CKM matrix, rigorously test
theoretical predictions of $D$ meson semileptonic form factors.

Recently using $\psi(3770)\ra D \bar D $ events and a
neutrino reconstruction technique combined with an independent
measurement of the number of $D$ mesons, CLEO reported the most
precise determinations of the absolute branching fractions and
differential decay rates $d \Gamma /d q^2$ for the
decays $\dzpie$, $\dppie$, $\dzke$, and $\dpke$~\cite{Nadia}.
(Throughout this
paper charge-conjugate modes are implied.)
The differential decay rates were used
to determine the absolute magnitude and shape of the
semileptonic form factors and to determine $\vcs$ and $\vcd$.
In this paper we present a complementary analysis which measures the
same quantities with similar precision in a common data set but
with a different technique that is independent of the number of
$D$ mesons in the data sample. The two analyses obtain consistent
results,
providing increased confidence in
their correctness, and each represents a marked
improvement in our understanding of charm semileptonic decays.

As the two analyses use a common data set, the results are
correlated. We calculate average values of the branching
fractions, form factors and $\vcs$ and $\vcd$ measured in the two
analyses, taking into account correlations between them. The
average values represent the best determinations of these
quantities with the \cleoc 281~pb$^{-1}$ data set.

The paper is organized as follows. We review the semileptonic
decay formalism in Sec.~\ref{formalism}.  The data sample and
\cleoc detector are described in
Sec.~\ref{dataAndDetector}. The analysis technique to identify
semileptonic decays is introduced in Sec.~\ref{sect_recon}.
In Secs.~\ref{BFs} and~\ref{decayRates}
we describe the use of
this technique to measure the absolute branching fractions, differential
decay rates and form factor parameters
for $D^0$~($D^+$) decays to $\pi^- e^+ \nu_e$ ($\pi^0 e^+ \nu_e$)
and $K^- e^+ \nu_e$ ($\bar{K}^0 e^+ \nu_e$).
The extraction of CKM parameters is described in Sec.~\ref{overview}.
In Sec.~\ref{average} we average the results presented here with
the results obtained in~\cite{Nadia}. Finally, in Sec.~\ref{summary} a
summary is provided.

\section{Semileptonic Decay Formalism}

\label{formalism}

The matrix element for a semileptonic decay $M_i (q_i
\bar{q}^\prime) \rightarrow M_f (q_f \bar{q}^\prime) \ell^+
\nu_\ell$ where $M_i$ and $M_f$ are the initial and final state mesons, $q_i$ and $q_f$ are the initial and final state quarks, and
$\bar q^\prime$ is a spectator anti-quark, can be written as
\begin{equation}
\mathcal{M}(M_i \rightarrow M_f \ell^+ \nu_\ell) = -i \frac{G_F}{\sqrt{2}}
\nonumber V_{q_i q_f}^* L^\mu H_\mu,
\end{equation}
\noindent where $G_F$ is the Fermi constant, $V_{q_i q_f}$ is the
appropriate CKM matrix element, and $L^\mu$  and $H_\mu$ are the
leptonic and hadronic currents, respectively. The leptonic current
is known and can be written in terms of the lepton and neutrino
Dirac spinors, $u_\ell$ and $v_\nu$,
\begin{equation}
L^\mu = \bar{u}_\ell \gamma^\mu (1 - \gamma_5) v_\nu. \nonumber
\end{equation}
\noindent The underlying simplicity of the weak transition $q_i
\rightarrow q_f W^+$ is obscured by the strong interaction as the
initial and final state quarks are bound within hadrons. The
hadronic current can be written as
\begin{equation}
H_\mu = \langle M_f | \bar{q}_f \gamma_\mu (1 - \gamma_5)
                                q_i | M_i \rangle. \nonumber
\end{equation}
\noindent The hadronic current describes the non-perturbative
strong interaction physics of hadron formation. Usually, one
exploits the fact that the hadronic current transforms as a four
vector under Lorentz transformations by parameterizing it with a
set of invariant form factors. This is achieved by constructing
all possible quantities with transformation properties of four
vectors from the momenta of particles involved in the decay, their
spin - polarization vectors and invariant tensors,  and expanding
the hadronic current in terms of these with
an invariant form factor multiplying each of them. The form
factors can only be functions of Lorentz scalars.  In  $M_i
\rightarrow M_f \ell^+ \nu_\ell$, there is one such invariant, which is
usually chosen to be $q^2$, the square of the invariant mass of the virtual
$W$.

In pseudoscalar-to-pseudoscalar semileptonic decays
($P_i(q_i\bar{q}) \rightarrow P_f(q_f \bar{q}) \ell^+
\nu_\ell$),
the hadronic current has a simple structure:
\begin{eqnarray}
& & \langle P_f(p_f)| V^\mu | P_i(p_i)  \rangle  \nonumber \\
& & \quad= f_+(q^2)\left((p_i + p_f)^\mu -\frac{m_i^2 -m_f^2}{q^2}(p_i-p_f)^\mu \right) \nonumber \\
& & \qquad+ f_0(q^2) \frac{m_i^2 -m_f^2}{q^2}(p_i-p_f)^\mu,
\end{eqnarray}
\noindent where $p_i$ ($m_i$) and $p_f$ ($m_f$) are the four-momenta (masses) of the initial
$P_i$ and final $P_f$ mesons, and $f_+(q^2)$
and $f_0(q^2)$ are the form factors
governing the transition.
Kinematic constraints require $f_+(0) = f_0(0)$.
In the limit of negligible lepton mass, which is applicable for $\ell = e$,
only one form factor remains,
\begin{equation}
\langle P_f(p_f)| V^\mu | P_i(p_i)  \rangle = f_+(q^2)(p_i +
p_f)^\mu. \nonumber
\end{equation}
\noindent The form factor $f_+(q^2)$ measures the probability
to form the final state hadron; it is largest when the daughter
meson is stationary in the parent meson rest frame $q^2=q^2_{\rm
max}$, and smallest when the daughter meson is moving with maximum
velocity in the parent meson rest frame $q^2=0$.

The differential decay rate is given by
\begin{equation}
\frac{d \Gamma}{d q^2} = \frac{G^2_F |V_{q_i q_f}|^2 p^3_{P_f} } {
24 \pi^3} |f_+ (q^2)|^2,
\end{equation}
\noindent where $p_{P_f}$
is the magnitude of the three-momentum of the $P_f$ meson in the
rest frame of $P_i$.
The shape of the
$q^2$ distribution is dominated by the dependence on $p^3_{P_f}$,
which arises because the decay proceeds via a $P$-wave.  This
dependence significantly enhances the rate at low $q^2$. We
perform fits to the differential decay rate to measure the four
semileptonic modes  $D \ra K e^+ \nu$ and $D \ra \pi e^+ \nu$. In
this paper we denote the form factor governing $D \ra K e^+ \nu$
and $D \ra \pi e^+ \nu$ by $f_+^K(q^2)$ and $f_+^{\pi}(q^2)$,
respectively.

\subsection{\boldmath Parametrization of the Form Factor $q^2$ Dependence}

The dependence of the form factors on $q^2$ is unknown, as it is
determined by non-perturbative QCD.
One may express the form factors in
terms of a dispersion relation, an approach that has been
well established in the literature (see, for example, Ref.~\cite{BSW}
and references therein):

\begin{equation}
f_+\left(q^2\right)  = \frac{f_+\left(0\right)}{1-\lambda}\frac{1}{1-\frac{\textstyle q^2}{\textstyle M^2_{\rm pole}}}
  +  \frac{1}{\pi}\int_{(m_D+m_P)^2}^\infty {\frac{{\rm Im}(f_+\left(t\right))}{t-q^2-i\varepsilon}dt}.
\label{eq:dispersion-int}
\end{equation}
where $M_{\rm pole}$ is the mass of the lowest lying $(q_i
\bar{q}_f)$ meson with the appropriate quantum numbers: for
$D \ra K e^+ \nu_e$ it is $D_s^{\ast +}(1^-)$ and for $D \ra \pi e^+
\nu_e$ it is $D^{*+}(1^-)$,
the parameter $\lambda$ gives the
contribution from the vector pole at $q^2=0$,
$m_D$ is the mass of the $D$ meson,
and $m_P$ is the mass of the final state pseudoscalar meson.
It is common to write the dispersive
representation in terms of an explicit pole and a sum of
effective poles,
\begin{equation}
 f_+(q^2) = \frac {\textstyle f_+(0)} {\textstyle 1- \lambda}
\frac {\textstyle 1} {\textstyle 1 -  \frac {\textstyle q^2} {\textstyle M^2_{\rm pole}}}
+ \sum_{k = 1}^{N} \frac{\textstyle \rho_k} {\textstyle 1 - \frac {\textstyle q^2}{\textstyle \gamma_k M^2_{\rm pole}}},
\label{eq:dispersion}
\end{equation}
where $\rho_k$ and $\gamma_k$ are
expansion parameters that are not predicted.

A series expansion around $q^2 = t_0$
~\cite{Boyd,Boyd-2,Arnesen,rhill}, where $t_0$ is defined below,
is commensurate with the dispersion relations. As expansions in
$q^2$ suffer from convergence problems due to the presence of
nearby poles, the expansion is formulated as an analytic
continuation into the $t=q^2$ complex plane. There is a branch cut
on the real axis for $t >   M_{K,\pi}^2$ corresponding to $D
(K,\pi)$ production, that is mapped onto the unit circle by the
variable $z$ defined as
\begin{eqnarray}
z(q^2, t_0) &=& \frac{ \sqrt{t_+ - q^2 } - \sqrt{t_+ - t_0 } } {
\sqrt{t_+ - q^2 } + \sqrt{t_+ - t_0 }}, \nonumber\\
t_{\pm} &=& \left( m_D \pm m_P \right)^2,
\end{eqnarray}
where $t_0$ is the arbitrary $q^2$ value that maps to $z=0$. We choose
$t_0 = t_+ (1-\sqrt{ 1 -t_-/t_+})$ because this choice minimizes
the maximum value of $z$ in the decay ($|z_{{\rm max}}| = 0.051
{\rm ~for~} D \ra K e^+ \nu_e {\rm~and~} |z_{{\rm max}}| = 0.17 {\rm
~for~} D \ra \pi e^+ \nu_e$).

The form factor is given by
\begin{equation}
\label{eq:series}
f_{+} (q^2)= \frac{ a_0} { P(q^2) \phi(q^2,t_0)} \left( 1 +\sum_{k
= 1}^{\infty}a_k(t_0)z(q^2, t^0)^k \right),
\end{equation}
where $P(q^2)=1$ for $ D \ra \pi$ and $P(q^2) = z(q^2,
m_{D_s^{*+}}^2)$ for $(D \ra K)$,  and $\phi$ is arbitrary.
Physically $P$ accounts for the presence of the pole, and $\phi$
is chosen to enable a simple expression for the series in terms of
the $a_k$. We follow Ref.~\cite{rhill}:
\begin{eqnarray}
 \phi(q^2, t_0) &=& c \left( \frac{ z(q^2, 0)} { - q^2}\right)^{5/2}
\left( \frac{ z(q^2, t_0)} { t_0 - q^2}\right)^{-1/2} \times\nonumber\\
& & \left( \frac{ z(q^2, t_-)} { t_- - q^2}\right)^{-3/4}
\frac{ t_+ - q^2 } { (t_+ - t_0)^{1/4}}.
\end{eqnarray}
This choice leads to the constraint
\begin{equation}
\sum_{k = 1}^{n_{c}}a_k^2 \leq 1
\end{equation}
for any choice of $n_c$. To leading order the coefficient
$c$ is given by
\begin{equation}
c = \sqrt { \pi m_c^2 /3},
\end{equation}
where $m_c$ is the charm quark mass, which we take to be 1.2
GeV/c$^2$. An advantage of the $z$ expansion is that it is model
independent and satisfies analyticity and unitarity. In addition,
measuring the $a_i$ in $D \ra \pi \ell^+ \nu_{\ell}$ constrains the
class of form factors needed to fit $B \ra \pi \ell^+ \nu_{\ell}$
and hence may improve the determination of $|V_{ub}|$. Finally, in
Heavy Quark Effective Theory (HQET)~\cite{HQET} there exist relations between the $a_i$ in $D$ and $B$
semileptonic decays.

The expansion parameters are not predicted. As $z$ is small, the
series is expected to converge quickly. Recently
BABAR~\cite{BABAR-06}, using a data sample of 75,000 $\dzke$ events,
found the differential rate to be well described with only a
linear term. In this work we will fit the data to both linear and
quadratic terms and use the series expansion for our main results.
There are alternatives to the $z$ expansion~\cite{mere-q2-expansion}.

In order to compare to lattice QCD calculations and previous
measurements, we will also compare the data to other
parametrizations of the form factor $q^2$ dependence.  A variety
of models have been traditionally used to parameterize the $q^2$
dependence. The most common, based on vector meson
dominance~\cite{BSW},
uses only the first term in the dispersion relation. In this \lq
\lq simple pole model" the $q^2$ dependence is given by
\begin{equation}
f_+(q^2) = \frac{\textstyle f_+(0)} {\textstyle (1-\frac{\textstyle q^2}{\textstyle M_{\rm pole}^2} ) }.
\end{equation}
\noindent Previous measurements of the $q^2$ spectrum in $D^0 \ra
K^- \ell^+ \nu_\ell$, the best measured charm semileptonic decay, find
a value of the pole mass many standard deviations from $M_{D_s^*}$
~\cite{Nadia,BABAR-06,CLEOIII,FOCUS-05,BELLE-06}. At low to medium values of
$q^2$ the $q^2$ spectrum is distorted compared to a simple pole,
suggesting contributions from a spectrum of poles above the pole
with the lowest mass.

The modified pole or Becirevic-Kaidalov (BK)
parametrization~\cite{modpole} attempts to address the
shortcoming of the simple pole model by keeping the first term in
the dispersion relation sum. The form factor is given by
\begin{equation}
f_+(q^2)=  \frac {\textstyle f_+(0)}{\textstyle (1- \frac{\textstyle q^2}{\textstyle M_{\rm pole}^2})(1- \alpha
\frac{\textstyle q^2}{\textstyle M_{\rm pole}^2}) },
\end{equation}
\noindent where $M_{\rm pole}$ is the spectroscopic pole mass
and $\alpha$, a free parameter, is an additional \lq \lq
effective" pole which represents the total contribution of all
additional poles.

In current data the $q^2$ evolution of form factors are indistinguishable from straight lines.
Therefore it is convenient to define the physical shape observables in terms of form factor slopes at $t=0$~\cite{Hill05, Hill06FPCP}

\begin{eqnarray}\label{eq:betadef}
{1\over \beta} &\equiv&
{m_D^2-m_P^2 \over f_+(0)} \left. df_{0} \over dt\right|_{t=0} \nonumber \\
\delta &\equiv&
1 - {m_D^2-m_P^2 \over f_+(0)}
\left( \left. df_{+} \over dt\right|_{t=0}  - \left. df_{0} \over dt\right|_{t=0} \right).
\end{eqnarray}
\noindent The quantities $\beta$ and $\delta$ depend on the masses of the mesons involved, and as they are physical quantities they are independent of the renormalization scale or scheme.

The BK parametrization requires several
assumptions to reduce the multiple parameters initially
present (Eq.~(\ref{eq:dispersion})) to one.
Specifically, it is assumed that $\beta$, which
measures scaling violations, is near unity, and $\delta$, which
measures spectator quark interactions, is near zero.
This sets the physical observable
\begin{equation}
1 + 1/\beta - \delta = \frac {m_D^2 -m_P^2}{f_+(0)}
\frac{df_+}{dq^2}  \sim 2  ~( {\rm at~} q^2=0),
\end{equation}
as noted in Ref.~\cite{Hill06FPCP},
corresponding to $\alpha \sim 1.75$
for $D \to K \ell^+\nu_\ell$ and 1.34 for $D \to \pi \ell^+\nu_\ell$.
Previous
experimental measurements of the $q^2$ spectrum in $D \ra K/ \pi
\ell^+ \nu_\ell$ do not agree with this value of
$\alpha$~\cite{Nadia,BABAR-06,CLEOIII,FOCUS-05,BELLE-06}.

Although the simple pole model and modified pole model are unable to
describe the $q^2$ spectrum of the data when the pole mass is fixed to the
relevant spectroscopic pole, or
$\alpha \sim 1.75$
for $D \to K \ell^+\nu_\ell$ and 1.34 for $D \to \pi \ell^+\nu_\ell$,
they do describe the data
well for values of the shape parameters many standard deviations from the expected values.

\subsection{Form Factor Calculations}

A variety of model dependent calculations of form factors exist.
In these models the form factors are evaluated at a fixed value of
$q^2$, {\it e.g.}, $q^2~=~0$ or $q^2~=~q^2_{{\rm max}} = (m_D -
m_P)^2$, and are extrapolated over the full range of $q^2$ using a
parametrization, such as those discussed above.

Quark model calculations estimate meson wave functions and use
them to compute the matrix elements that appear in the hadronic
current. There are a large variety of theoretical
calculations~\cite{survey-of-models}. Among them the ISGW
model~\cite{ISGW} has been widely used to simulate heavy hadron
semileptonic decays. This model is expected to be valid in the
vicinity of $q^2 =q^2_{\rm  max}$, the region of maximum overlap
between the initial and final meson wave functions.

In the ISGW model the form factors are assumed to have the form
\begin{equation}
f(q^2) = f(q^2_{\rm max})e^{- a (q^2_{\rm max} - q^2)}.
\end{equation}
The ISGW2 model~\cite{ISGW2}, an update of the ISGW model,
incorporates constraints from heavy quark symmetry. It uses a
dipole term for the form factor $q^2$ dependence expressed in
terms of the radius of a meson~($r$) rather than the mass of the
appropriate $(q_i \bar{q}_f)$ meson:
\begin{equation}
f_+( q^2 ) = f_+(q^2_{\rm max})  \left( 1+ \frac {r^2}{12}
\left(q_{\rm max}^2 - q^2 \right) \right)^{-2}.
\end{equation}
The ISGW2 model predicts $f_+^K(q^2_{\rm max}) =
1.23 $ and $r^K = 1.12 {\rm ~GeV}^{-1}$~\cite{ISGW2}.
Previous measurements (e.g., Refs.~\cite{CLEOIII,FOCUS-05,BELLE-06,BABAR-06}) do not agree with these values.

QCD sum rules~\cite{QCD-sum-rules-1,QCD-sum-rules-2}, are expected
to be valid at low $q^2$. For $D^0 \ra K^- \ell^+ \nu_\ell$, and using
a value of 150 MeV for the strange quark mass, one
obtains~\cite{QCD-sum-rules-2} $f^K_+(0) = 0.78(11)$ and
$\alpha_K = 0.07^{+0.15}_{-0.07}$ using the modified pole ansatz.
For $D \ra \pi \ell^+ \nu_\ell$~\cite{QCD-sum-rules-2} reports
$ f^\pi_+(0) = 0.65(11)$ and $\alpha_\pi =
0.01^{+0.11}_{-0.07}$.

The above models are based on theoretical assumptions and, in
consequence, introduce a difficult to quantify theoretical
uncertainty that is significantly larger than the presently
achievable experimental statistical and systematic uncertainties
combined. Therefore this limits the precision with which $\vcs$
and $\vcd$ can be determined from exclusive semileptonic
charm meson decays.

Lattice QCD computes $f_+(q^2)$ from first principles. Current
results must be extrapolated to physical values of light quark
masses and corrected for finite lattice size and discretization
effects. There have been several evaluations of $f_+(q^2)$ for
different values of the momentum transfer in the quenched
approximation~\cite{Flynn,Abada}. These results, which do not
include QCD vacuum polarization, have been combined~\cite{Flynn},
to give $f_+^K(0) = 0.73(7)$. LQCD calculations which incorporate
QCD vacuum polarization (unquenched calculations) have produced
results that agree with experiment to within a few percent for a
number of quantities~\cite{Davies}. The first unquenched LQCD
calculation~\cite{unquenched_LQCD} of form factors in $D \ra K e^+ \nu_e$
and $D \ra \pi e^+ \nu_e$
reports $f_+^K(0) = 0.73(3)(7)$, $\alpha_K  = 0.50(4)$,
$f_+^\pi(0)=0.64(3)(6)$, and $\alpha_{\pi}=0.44 (4) $ using the
modified pole ansatz to parameterize the $q^2$ dependence of the
form factor. Here the systematic uncertainty is dominated by the
effect of discretization. While the form factors are currently
calculated to a modest precision of ten percent, the
uncertainties are systematically improvable to a
precision that matches, or exceeds, the experimental measurements presented
here and in~\cite{Nadia}. Accordingly, we use~\cite{unquenched_LQCD} to extract values for $\vcs$ and
$\vcd$ in this work.

\section{Data Sample and the CLEO-\lowercase{c} Detector}

\label{dataAndDetector}

The data sample used in this analysis consists of  281~pb$^{-1}$
of $e^+ e^-$ annihilation data taken at the $\psi(3770)$, which is
about 40~MeV above the $D \bar{D}$ pair production threshold. (Throughout this paper $D$ is used to denote $D^0$ and $D^+$.) The
data include approximately  $1.0 \times 10^{6}$ $D^0 \bar{D}^0$
events and $0.8 \times 10^{6}$  $D^+ D^-$ events.

CLEO-c is a general-purpose solenoidal detector. The charged
particle tracking system covers a solid angle of 93\% of $4 \pi$
and consists of a small-radius six-layer low mass stereo wire
drift chamber concentric with and surrounded by a 47-layer
cylindrical drift chamber. The chambers operate in a 1.0 T
magnetic field and achieve a momentum resolution of $\sim$0.6\% at
$p=$1~GeV/$c$.  The main drift chamber provides
specific-ionization ($\dedx$) measurements that discriminate
between charged pions and kaons. Additional hadron identification
is provided by a Ring-Imaging Cherenkov (RICH) detector covering
approximately 80\% of $4 \pi$. Identification of positrons and
detection of neutral pions rely on an electromagnetic
calorimeter consisting of 7800 cesium iodide crystals and covering
95\% of $4 \pi$. The calorimeter achieves a photon energy
resolution of 2.2\% at $E_\gamma=$1~GeV and 5\% at 100~MeV. The
\cleoc detector is described in detail
elsewhere~\cite{cleo_detector}.

The response of the \cleoc detector was studied using a GEANT-based~\cite{GEANT}
Monte Carlo~(MC) simulation. To develop selection
criteria and test the analysis technique several MC simulations
are used.  $\psi(3770) \ra D \bar{D}$ events are generated using
EvtGen~\cite{EvtGen} and each $D$ meson is allowed to decay in accordance
with the best experimental and theoretical information.
We refer to this as \lq \lq generic MC".
The MC
sample generated corresponds to an integrated luminosity of about
$11~\rm{fb^{-1}}$ which is a factor 40 larger than the data.
Semileptonic signal decays are generated with the modified pole
model form factors~\cite{modpole} with parameters from the most
recent unquenched LQCD calculations~\cite{unquenched_LQCD}.

Due to the tagging technique employed in the analysis, backgrounds
from the non-$D \bar{D}$ processes $e^+ e^- \rightarrow q
\bar{q}$, where $q$ is a $u$, $d$, or $s$ quark, $e^+ e^-
\rightarrow \tau^+ \tau^-$, and $e^+ e^- \rightarrow \psi(2S)
\gamma$, are nearly absent. These non-$D \bar{D}$ processes are
also modeled using MC simulation and are scaled
absolutely according to their measured cross sections at the $\psi(3770)$.

A second type of MC sample, which we refer to as \lq \lq signal
MC", consists of several samples of $\psi(3770) \ra D \bar{D}$
events in which the $\bar{D}$ is allowed to decay to all possible
final states, and the $D$ decays to a specific semileptonic final
state.

\section{Event Reconstruction}

\label{sect_recon}

The reconstruction technique used in this analysis was first
applied by the Mark III collaboration~\cite{MkIII} at SPEAR. This
technique was used to measure $D$ semileptonic branching
fractions with a smaller data sample at
\cleoc\cite{DSemilBFs-2005}. That data sample was too small to
study charm semileptonic form factors, which are the focus of
studies reported in this paper.

The presence of two $D$ mesons in a $\psi(3770)$ event allows
a tag sample to be defined in which a $\bar{D}$
is reconstructed in a hadronic decay mode.
A sub-sample is then defined in which a positron and a set of hadrons, as a signature
of a semileptonic decay, are required in addition to the tag.
Tagging a $\bar{D}$ meson in a $\psi(3770)$ decay provides a $D$
with known four-momentum, allowing a semileptonic decay to be
reconstructed with no kinematic ambiguity, even though the
neutrino is undetected.

The tag yield can be expressed
as $N_{{\rm tag} } = 2N_{DD}  { \cal B}_{{\rm tag}} \epsilon_{{\rm tag}}$, where $N_{DD}$ is the produced
number of $D\bar{D}$ pairs, $ {\cal{B}}_{{\rm tag}}$ is the branching fraction of
hadronic modes used in the tag
sample, and $\epsilon_{{\rm tag}}$ is the tag efficiency.
The yield of tags with a semileptonic decay can be expressed as
$N_{{\rm tag, SL}} = 2N_{DD} {\cal{B}}_{{\rm tag}} {\cal{B}}_{{\rm SL}} \epsilon_{{\rm tag, SL}}$
where ${\cal{B}}_{{\rm SL}}$ is the semileptonic decay branching fraction,
including subsidiary branching fractions, and $\epsilon_{{\rm tag, SL}}$ is the efficiency of finding the tag and the semileptonic decay in the same event.
From the expressions for $N_{{\rm tag}}$ and  $N_{{\rm tag, SL}}$
we obtain
\begin{equation}
{\cal{B}}_{{\rm SL}} =  \frac {N_{ \rm tag, SL }} {N_{ \rm tag}}   \frac {\epsilon_{ \rm tag}} {\epsilon_{\rm tag, SL}} =
\frac {N_{ \rm tag, SL} / \epsilon} {N_{ \rm tag}},
\label{eq:master}
\end{equation}
where $\epsilon = \epsilon_{ \rm tag, SL}/\epsilon_{ \rm tag }$
is the effective signal efficiency.
The branching fraction determined by tagging is an absolute
measurement. It is independent of the integrated luminosity and
number of $D$ mesons in the data sample.
Due to the large solid angle acceptance
and high segmentation of the CLEO-c detector and the low multiplicity of the events
$\epsilon_{ \rm tag, SL} \approx \epsilon_{\rm tag} \epsilon_{\rm SL}$, where $\epsilon_{ \rm SL}$ is the semileptonic decay efficiency. Hence the ratio $  \epsilon_{\rm tag, SL }/\epsilon_{\rm tag}$
is insensitive to most systematic effects associated with the tag mode and the
absolute branching fraction determined with this procedure is nearly independent of the tag mode.
Below, we first describe the procedure
used for the reconstruction of tags followed by that for the
reconstruction of semileptonic decays~\cite{DSemilBFs-2005}.

\subsection{Tag Selection}

Hadronic tracks must have momenta above 50~MeV/$c$ and $|
\cos{\theta} | < 0.93$, where $\theta$ is the angle between the
track direction and the beam axis. Identification of hadrons is
based on measurements of specific ionization in the main drift
chamber and information from the RICH. Pion and kaon candidates
are required to have $dE/dx$ measurements within three standard
deviations (3$\sigma$) of the expected value. For tracks with
momenta greater than 700~MeV/$c$, RICH information, if available,
is combined with $dE/dx$.  The efficiencies ($95$\% or higher) and
misidentification rates (a few percent) are determined with
charged pion and kaon samples from hadronic $D$ decays.

We select $\piz$ candidates from pairs of photons, each having an
energy of at least 30~MeV, and a shower shape consistent with that
expected for a photon.  A kinematic fit is performed constraining
the invariant mass of the photon pair to the  known $\piz$ mass.
The candidate is accepted if the unconstrained invariant mass is
within 3$\sigma$, where $\sigma$ (typically 6 MeV/$c^2$) is
determined for that candidate from the kinematic fit, and the
kinematic parameters for the $\piz$ determined with the fit are
used in further reconstruction.

Candidate events are selected by reconstructing a $\bar{D}^0$ or
$D^-$ tag in the following hadronic final states: $K^+ \pi^-$,
$K^+ \pi^- \pi^0$, $K^+ \pi^- \pi^0 \pi^0$, $K^+ \pi^- \pi^-
\pi^+$, $K_S^0 \pi^0$, $K_S^0 \pi^- \pi^+$, $K_S^0 \pi^- \pi^+
\pi^0$, and $K^- K^+$ for neutral tags, and $K^0_S \pi^-$, $K^+
\pi^- \pi^-$, $K^0_S \pi^- \pi^0 $, $K^+ \pi^- \pi^- \pi^0$,
$K^0_S \pi^- \pi^- \pi^+$, and $K^- K^+ \pi^- $ for charged tags.
These modes constitute about 46\% and 28\% of all $\bar{D}^0$ and
$D^-$ decays, respectively. Tagged events are selected using two
variables: $\Delta E \equiv E_{D} - E_{\rm beam}$, the difference
between the energy of the tag candidate ($E_{D}$) and the beam
energy ($E_{\rm beam}$), and the beam-constrained mass $M_{\rm bc}
\equiv \sqrt{ E_{\rm beam}^2/c^4 - |\vec{p}_{D}|^2/c^2}$, where
$\vec{p}_{D}$ is the measured momentum of the tag candidate.
Note that the use of $E_{\rm beam}$ instead of $E_{D}$ improves
the resolution of $M_{\rm bc}$ by one order of magnitude, to about
$2~{\rm MeV/c^2}$, which is dominated by the beam energy spread. If
multiple candidates are present in the same tag mode, the one
candidate per tag flavor with the smallest $\Delta E$ is chosen.

\begin{figure}[htbp]
\epsfig{file=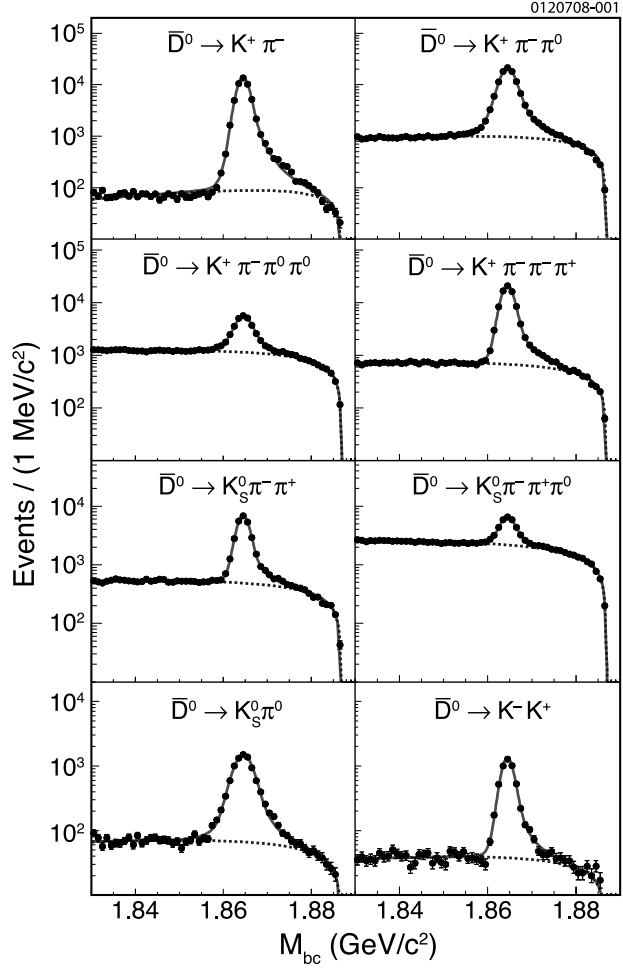,width=3.2in}
\caption{Fits~(solid line) to the $M_{\rm bc}$ distributions in
data for eight $\bar D^0$ tag modes. The backgrounds are shown by
the dashed line.} \label{mbcdistrs_d0}
\end{figure}

\begin{figure}[htbp]
\epsfig{file=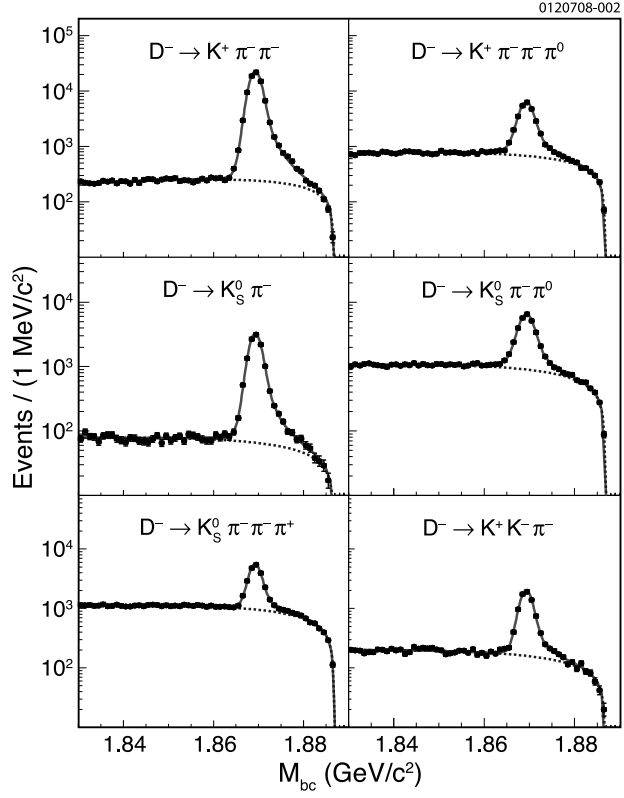,width=3.2in}
\caption{Fits~(solid line) to the $M_{\rm bc}$ distributions in
data for six $D^-$ tag modes. The backgrounds are shown by the
dashed line.} \label{mbcdistrs_dp}
\end{figure}

The number of tags reconstructed in each mode is obtained by
imposing a mode dependent requirement on $\Delta E$, counting the
number of events in the signal region of $M_{\rm bc}$, defined as
$-6.5~{\rm MeV}/c^2 < (M_{\rm bc} - m_D) < 9.5~{\rm MeV}/c^2$,
where $m_D$~\cite{PDG2004} is the known $D$ meson mass, and
subtracting  the background contribution from it. Fits to the
$M_{\rm bc}$ distributions, shown in Figs.~\ref{mbcdistrs_d0}
and~\ref{mbcdistrs_dp}, are made using the procedure described
in~\cite{cleoc-Dtagging}. We fit the $M_{\rm bc}$ distributions
to a signal shape and one or more background components. The
signal shape includes the effects of beam energy smearing, initial
state radiation, the line shape of the $\psi(3770)$, and
reconstruction resolution. The background is described by an ARGUS
function~\cite{ARGUS}, which models combinatorial contributions.
The background contribution in the signal region is estimated by
integrating this function. The yields of the eight neutral tag
modes and the six charged tag modes, and their reconstruction
efficiencies as determined with the generic MC simulation,
are given in Tables~\ref{tagsD0} and~\ref{tagsDPlus}.
There are approximately $3.1 \times 10^5$ and $1.6 \times 10^5$
neutral and charged tags, respectively.

\begin{table}[htbp]
\caption{Yields with statistical uncertainties and reconstruction
efficiencies of $\bar{D}^0$
         tags.}
\begin{center}
\begin{tabular}{lcc} \hline \hline
Tag Mode      \hspace{0.4cm}       & \hspace{0.4cm} $N_{\rm tag}$   \hspace{0.4cm}            &
\hspace{0.4cm}  $\epsilon_{\rm tag}$ (\%)  \hspace{0.4cm}   \\
\hline
\rule[-1mm]{-1mm}{4mm}
$\bar{D}^0 \to K^+ \pi^-$              & $51002(230)$   & $64.77(3)$ \\
$\bar{D}^0 \to K^+ \pi^- \pi^0$          & $98117(347)$   & $33.30(1)$ \\
$\bar{D}^0 \to K^+ \pi^- \pi^0 \pi^0$    & $23040(220)$   & $14.41(1)$ \\
$\bar{D}^0 \to K^+ \pi^- \pi^- \pi^+$    & $77641(303)$   & $45.46(2)$ \\
$\bar{D}^0 \to K^0_S \pi^- \pi^+$        & $24533(187)$   & $38.33(2)$ \\
$\bar{D}^0 \to K^0_S \pi^- \pi^+ \pi^0$  & $20355(260)$   & $17.81(5)$ \\
$\bar{D}^0 \to K^0_S \pi^0$              & $ 8175(99)$    & $31.01(5)$  \\
$\bar{D}^0 \to K^- K^+$                  & $ 4614(76)$    & $57.35(9)$  \\
\hline
All Neutral Tags  & $ 307478(657)$  &                                       \\
\hline \hline
\end{tabular}
\end{center}
\label{tagsD0}
\end{table}

\begin{table}[htbp]
\caption{Yields with statistical uncertainties and reconstruction
efficiencies of $D^-$
         tags .}
\begin{center}
\begin{tabular}{lcc} \hline \hline
Tag Mode     \hspace{0.4cm}        & \hspace{0.4cm}  $N_{\rm tag}$  \hspace{0.4cm}             &
\hspace{0.4cm}  $\epsilon_{\rm tag}$ (\%)  \hspace{0.4cm}   \\
\hline
$D^- \to K^+ \pi^- \pi^-$           & $79896(291)$   & $53.81(2)$ \\
$D^- \to K^+ \pi^- \pi^- \pi^0$     & $23740(196)$   & $25.23(2)$ \\
$D^- \to K^0_S \pi^-$               & $11456(113)$   & $45.14(5)$ \\
$D^- \to K^0_S \pi^- \pi^0$         & $25159(210)$   & $21.97(2)$ \\
$D^- \to K^0_S \pi^- \pi^- \pi^+$   & $16431(191)$   & $31.58(3)$ \\
$D^- \to K^- K^+ \pi^-$             & $ 6794(100)$   & $44.72(5)$ \\
\hline
All Charged Tags                   & $ 163476(477)$    &             \\
\hline \hline
\end{tabular}
\end{center}
\label{tagsDPlus}
\end{table}

\subsection{Selection of Semileptonic Decays}

After a tag is identified, we search for a positron and a set of
hadrons recoiling against the tag. (Muons are not used as $D$
semileptonic decays at the $\psi(3770)$ produce low momentum
leptons for which the \cleoc muon identification is not
efficient.) Positron candidates are required to have momenta of
at least 200~MeV/$c$ and to satisfy $| \cos{\theta} |$ $<$ 0.90,
where $\theta$ is the angle between the positron direction and the
beam axis. The efficiency for positron identification rises from
about $50\%$ at 200~MeV/$c$ to 95\% just above 300~MeV/$c$ and is
roughly constant thereafter.  The rate for misidentifying charged
pions and kaons as positrons averaged over the momentum range is
approximately 0.1\%. The energy lost by positrons to
bremsstrahlung photons is partially recovered by adding showers
that are within $5^\circ$ of the positron momentum and are not
matched to other particles. The selection of $\pi^-$, $\pi^0$,
$K^-$, and $K^0_S$ candidates is identical to that used for tags.

The tag and the semileptonic candidate are then combined. Events that
include tracks other than those of the tag and the semileptonic
candidate are vetoed~\cite{extra_track_cut}.
After all selection criteria are applied,
multiple candidates in the same event
are rare in all modes except $D^+ \to \pi^0 e^+ \nu_e$. For $D^+
\to \pi^0 e^+ \nu_e$, in the few percent of events with multiple
candidates, one combination is chosen per tag candidate based on the
proximity of the invariant masses of the $\pi^0$ candidates to the
expected mass.

Semileptonic decays are identified using the variable $U \equiv
E_{\rm miss} - c|\vec{p}_{\rm miss}|$, where $E_{\rm miss}$ and
$\vec{p}_{\rm miss}$ are the missing energy and momentum of the
$D$ meson decaying semileptonically, calculated using the
difference of the four-momentum of the tag and that of the observed
products of the semileptonic decay. If the decay products of the
semileptonic decay have been correctly identified, $U$ is expected
to be zero,  since only a neutrino is undetected. To improve the
resolution in $U$, the crossing angle of the beams~($\sim 3$~mrad)
is allowed for by recalculating all track momenta and shower
energies in the $\psi(3770)$ rest frame, and the four-momentum of
the tag is approximated by ($E_{\rm beam}/c, \sqrt{(E_{\rm
beam}/c)^2 - (c m_D)^2}  \hat{p}_D$), where $\hat{p}_D$
is the unit direction vector of the $D$ in the $\psi(3770)$
rest frame determined using the
direction of the $\bar{D}$ tag in the same frame. Due
to the finite resolution of the detector, the distribution in $U$
is approximately Gaussian, centered at $U=0$ with $\sigma \sim
12~{\rm MeV}$, for all modes except $D^+ \to \pi^0 e^+ \nu_e$, for
which $\sigma$ is approximately two times larger.

\begin{figure*}[htb]
\epsfig{file=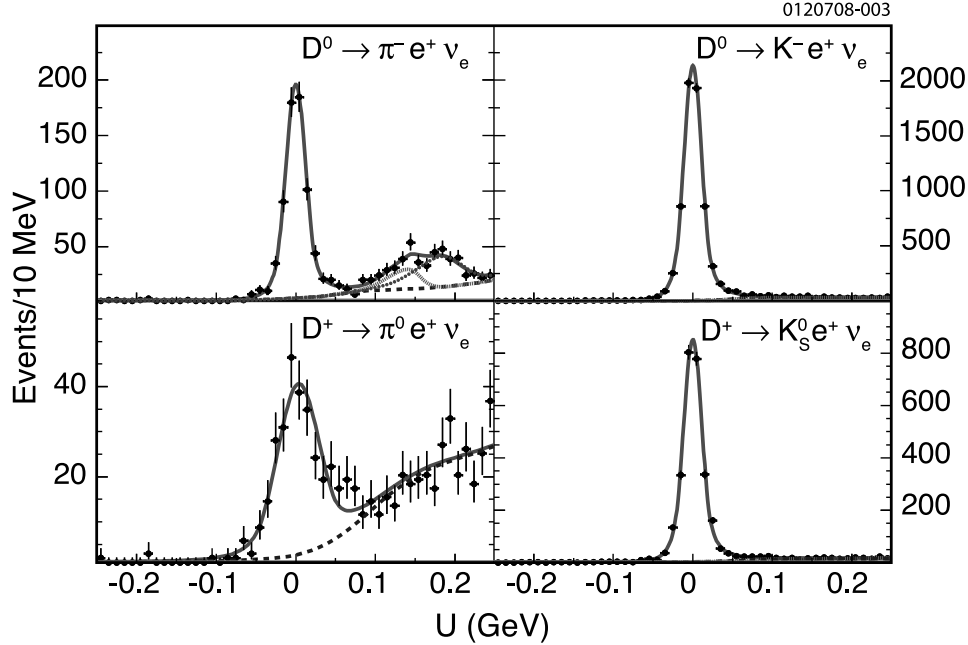,width=5in}
\caption{Fits~(solid line) to the $U$ distributions in
data~(points) for
$D^0 \to \pi^- e^+ \nu_e$, $D^0 \to      K^- e^+
\nu_e$, $D^+ \to \pi^0 e^+ \nu_e$, and $D^+ \to K^0_S e^+ \nu_e$. The
background contributions are represented by dotted or dashed lines.
In $D^0 \to \pi^- e^+ \nu_e$ the background peaks at positive $U$
are described in the text.}
\label{udistrs}
\end{figure*}

Using this procedure we obtain the $U$ distributions shown in
Fig.~\ref{udistrs}. For each mode a clear signal is evident
centered on $U=0$, while backgrounds are very
small near $U=0$. In $D^0 \ra \pi^- e^+ \nu_e$ the peak at
positive $U$ is from two sources: $D^0 \ra K^- e^+ \nu_e$ when a $K^-$
is misidentified as a $\pi^-$~(peak at 130 MeV) and from $D \ra
K^- \pi^+ \pi^0$ where the $K^-$ is mistaken for an electron and
the $\pi^0$ is unobserved~(peak at 180 MeV). This background is
present because each event is not required to have both a $D^0$
and a $\bar{D}^0$. Specifically, on the semileptonic side of the
event both $D^0 \ra \pi^- e^+ \nu_e$ and $\bar{D}^0 \ra \pi^+ e^-
\bar{\nu}_e$ are accepted, on the tag side for example both
$\bar{D}^0 \ra K^+ \pi^-$ and $D^0 \ra K^- \pi^+$ are accepted. The
kaon produced in the decay of the tag is not required to have the same
charge as the lepton produced in the semileptonic decay. If this
requirement were made the $D \ra K^- \pi^+ \pi^0$ background would
be removed, but decay sequences where the tag undergoes a doubly Cabibbo suppressed
decay such as
$D^0 \ra K^+ \pi^-,$  and $\bar{D}^0 \ra \pi^+ e^- \bar{\nu}_e$ would be
removed as well.

The yield for each semileptonic mode is determined from a fit to
the corresponding $U$ distribution, as shown in Fig.~\ref{udistrs}
with all tag modes combined. The yields are reported in Table~\ref{semilBFs_data}.
In each case the signal function
consists of a Gaussian to describe the core of the $U$
distribution and two power law tails to account for initial and
final state radiation~(ISR and FSR):

\begin{equation}
f = \left\{
\begin{array}{lr}
      a_1( \frac{n_1}{\alpha_1} - \alpha_1 + t )^{-n_1}  &  t >
\alpha_1  \\[4pt]
       e^{ -\frac{t^2}{2} }                                        &
-    \alpha_2 < t < \alpha_1    \\[4pt]
      a_2( \frac{n_2}{\alpha_2} - \alpha_2 - t )^{-n_2}  &  t < - \alpha_2
\end{array} \right. ,
\end{equation}

\noindent where
$t   \equiv (U - U_{\rm mean}) / \sigma_U$,
$a_1 \equiv \left( n_1/ \alpha_1 \right)^{n_1}  e^{ -{\alpha_1^2}/{2}}$, and
$a_2 \equiv \left( n_2/ \alpha_2 \right)^{n_2}  e^{ -{\alpha_2^2}/{2}}$.
The parameters describing the tails of the signal
function~($\alpha_1$, $\alpha_2$, $n_1$, and $n_2$) are always
fixed in fits to the data to the values found in signal MC simulation.
The $\sigma_U$ is fixed to the value predicted by the MC
simulation in the fit for $D^0 \rightarrow \pi^0 e^+ \nu_e$, which
has the smallest signal yield and the largest background level
among the four semileptonic modes, and allowed to float in the
fits for the other modes.

The background functions are determined from the generic MC simulation. The
backgrounds are small and arise mostly from misreconstructed
semileptonic decays with correctly reconstructed tags.
The background shape parameters are fixed, while the background
normalizations are allowed to float in all fits to the data.

\section{Absolute Branching Fraction Measurements}

\label{BFs}

\subsection{Determination of the Branching Fractions}

The absolute semileptonic branching fractions are obtained from
our tagged semileptonic yields $N_{\rm tag, SL}$, tag yields $N_{\rm tag}$,
and the efficiencies $\epsilon$, using Eq.~(\ref{eq:master}).
The simulation of each semileptonic mode employs the simple pole model
with $M_{\text {pole}}=2.0 ~{\rm GeV/c^2}$. The efficiency depends
weakly on $M_{\text {pole}}$; accordingly the efficiencies are
re-weighted to the value of $M_{\rm pole}$ measured in the data.
These efficiencies are then
weighted by the tag yields shown in Tables~\ref{tagsD0} and \ref{tagsDPlus}
to obtain the overall efficiency.
The absolute semileptonic branching fractions are obtained using
these weighted efficiencies.
Table~\ref{semilBFs_data} presents our absolute semileptonic branching
fraction measurements with statistical and systematic uncertainties.
A description of how the systematic uncertainties are obtained is provided in the next subsection.

The procedure for measuring semileptonic branching fractions is
tested using the generic MC sample.
In the test, the MC sample is treated
identically to the data.
In addition, the
procedure was separately tested for each combination of tag and
semileptonic mode. We find that the input and output branching
fractions are consistent within statistical uncertainties in all
cases.
The largest deviation is
observed for $D^+ \to \bar{K}^0 e^+ \nu_e$ with all tag modes
combined, where the discrepancy is less than one third of the statistical uncertainty on the measurement.

To check the consistency of the measurement of the semileptonic
branching fractions, we have also measured semileptonic branching
fractions for each tag mode separately for the two Cabibbo allowed
final states where there are adequate statistics in each tag mode.
We present the results in
Tables~\ref{semi_kenu_tag_data}
and~\ref{semi_ksenu_tag_data}.

We note that the effective semileptonic efficiency is
larger for tag modes with higher multiplicity. This happens
primarily because tag reconstruction efficiencies in events with
the second $D$ meson decaying hadronically are slightly
smaller compared to signal events with the second $D$ meson
decaying to a low multiplicity semileptonic final state.

We find that the branching fractions are consistent among tag
modes.  The results in Tables~\ref{semi_kenu_tag_data}
and~\ref{semi_ksenu_tag_data} also demonstrate consistency
between the weighted averages of the individual tag mode branching
fractions and the branching fractions obtained with all tag modes
combined.

\begin{table*}[htbp]
\caption{Signal efficiencies, yields, and branching fractions
         in this work~(first four columns) and, for comparison,
         the branching fractions measured using the
         first 56~pb$^{-1}$ \cleoc $\psi(3770)$ data sample
         ~\cite{DSemilBFs-2005}, and values from PDG-04~\cite{PDG2004}.
         The first uncertainty is statistical and the second
         systematic in the fourth and fifth columns,
         and statistical or total in the other columns.
       }
\begin{center}
\begin{tabular}{ l c c c c c}
\hline \hline

 Decay Mode &  \hspace{2mm} $\epsilon$ (\%) \hspace{2mm}  & \hspace{2mm} $N_{\rm tag, SL}$ \hspace{2mm}  &
 \hspace{2mm} $\mathcal{B}_{\rm SL}$ (\%) \hspace{2mm}  &   \hspace{2mm} $\mathcal{B}_{\rm SL}$ (\%) (56 pb$^{-1}$) \hspace{2mm}
 &   \hspace{2mm} $\mathcal{B}_{\rm SL}$ (\%) (PDG-04) \hspace{2mm}  \\

\hline\rule[-1mm]{-1mm}{4mm}
$D^0 \rightarrow  \pi^- e^+ \nu_e  $ & $72.54(11)$ & $699(28)$
                                      & $ 0.314(13)(4)$ & $ 0.262(25)(8)$  &  $0.36(6)$   \\
$D^+ \rightarrow  \pi^0 e^+ \nu_e  $ &  $44.72(13)$ & $281(19)$
                                      & $ 0.384(27)(23) $ & $ 0.44(6)(3) $  &  $0.31(15)$   \\
$D^0 \rightarrow  K^- e^+ \nu_e  $ & $61.06(7)$ & $6786(84)$
                                      &  $ 3.61(5)(5)$ &  $ 3.44(10)(10) $  &  $3.58(18)$      \\
$D^+ \rightarrow  \bar{K}^0 e^+ \nu_e  $ &  $ 20.01(4)$  & $ 2910(55) $
                                      & $ 8.90(17)(21) $  & $ 8.71(38)(37) $   &  $6.7(9)$      \\
\hline \hline
\end{tabular}
\end{center}
\label{semilBFs_data}
\end{table*}

\begin{table}[htbp]
\caption{ $\kenu$ semileptonic yields in data, the semileptonic
efficiency for each hadronic tag mode, $\epsilon$,
and the branching fraction measurement for each hadronic
tag mode. The last two lines show the weighted average of the
individual measurements and the result from the fit with all tag
modes combined.
}
\label{semi_kenu_tag_data}
\begin{center}
\begin{tabular}{lccc}
\hline \hline
                     Mode \hspace{0.4cm}  &
\hspace{0.4cm} $N_{\rm tag, SL}$      \hspace{0.4cm}        &
\hspace{0.4cm}$\epsilon$ (\%) \hspace{0.4cm}  &
\hspace{0.4cm} $\mathcal{B}_{\rm SL}$ (\%) \hspace{0.4cm}  \\ \hline
$K^+ \pi^-$               & $1088(34)$    & $59.36(15)$ & $3.60(11)$ \\
$K^+ \pi^- \pi^0$         & $2143(47)$    & $61.66(12)$ & $3.55(8)$ \\
$K^+ \pi^- \pi^0 \pi^0$   &  $593(25)$    & $67.11(30)$ & $3.84(15)$\\
$K^+ \pi^- \pi^- \pi^+$   & $1693(42)$    & $59.44(13)$ & $3.67(9)$ \\
$K^0_S \pi^- \pi^+$       &  $516(23)$    & $59.47(24)$ & $3.54(16)$ \\
$K^0_S \pi^- \pi^+ \pi^0$ &  $474(22)$    & $64.46(30)$ & $3.61(17)$\\
$K^0_S \pi^0$             &  $160(13)$    & $60.52(43)$ & $3.23(26)$ \\
$K^- K^+$                 &  $118(11)$    & $59.45(51)$ & $4.32(42)$\\
\hline
Average:         &                 &               & $3.61(5)$ \\
Combined Fit:    & $6786(84)$  & $61.06(7)$  & $3.61(5)$ \\
\hline\hline
\end{tabular}
\end{center}
\end{table}

\begin{table}[htbp]
\caption{ $\ksenu$ semileptonic yields in data,
the semileptonic
efficiency for each hadronic tag mode, $\epsilon$,
and the branching fraction measurement for each hadronic
tag mode. The semileptonic efficiency includes subsidiary branching fractions~\cite{subsidiarybf}.
The last two lines show the weighted average of the
individual measurements and the result from the fit with all tag
modes combined.
}
\label{semi_ksenu_tag_data}
\begin{center}
\begin{tabular}{lcccc} \hline \hline
Mode \hspace{0.4cm}    &
\hspace{0.4cm}  $N_{\rm tag, SL}$ \hspace{0.4cm}   &
\hspace{0.4cm}  $\epsilon$ (\%)  \hspace{0.4cm}        &
\hspace{0.4cm}  $\mathcal{B}_{\rm SL}$ (\%) \hspace{0.4cm} \\
\hline

$K^+ \pi^- \pi^-$          & $1437(39)$     & $ 19.88(5)$    & $9.04(25)$ \\
$K^+ \pi^- \pi^- \pi^0$    & $ 430(21)$     & $ 20.51(9) $   & $8.83(44)$ \\
$K^0_S \pi^-$              & $ 201(14)$     & $ 19.91(18) $   & $8.81(46)$ \\
$K^0_S \pi^- \pi^0$        & $ 443(22)$     & $ 20.17(9) $   & $8.73(43)$ \\
$K^0_S \pi^- \pi^- \pi^+$  & $ 272(17)$     & $ 19.82(11) $   & $8.35(53)$ \\
$K^- K^+ \pi^-$            & $ 130(12)$     & $ 19.97(16) $   & $9.59(88)$ \\

\hline
 Average :         &                &                     & $ 8.89(17)$ \\
Combined Fit:       & $ 2910(55)$   & $ 20.01(4)$    & $8.90(17)$ \\
\hline\hline
\end{tabular}
\end{center}
\end{table}

\subsection{Study of Systematic Uncertainties for Absolute Branching Fractions}

\label{BFSystErrs}

We have considered the following sources of systematic uncertainty
in the measurements of branching fractions and give our estimates
of their magnitudes in parentheses. The uncertainties associated
with the efficiency for finding a track (0.3\% for each pion,
kaon, or positron, combined in quadrature with an additional 0.6\%
for each kaon), for reconstructing a $\pi^0$ (4.3\%), and for
reconstructing a $K_S^0$~(1.8\%), are estimated using missing mass
techniques described in~\cite{cleoc-Dtagging}. The
uncertainty in the positron identification efficiency~(1.0\%) is
obtained using a comparison of the detector response to positrons
from radiative processes in the data and MC simulation. The effect of the event
complexity is incorporated by studying positrons both in isolation
and embedded in hadronic events. Uncertainties in the charged pion
and kaon identification efficiencies~(0.1\% per pion and 0.2\% per
kaon) are estimated using hadronic $D$ meson decays. The
uncertainty in the number of tags~(0.4\%) is estimated by using
alternative signal functions in the fits to the $M_{\rm bc}$
distributions and by varying the end point of the background
function~\cite{ARGUS}.
The uncertainty associated with the requirement that there be no
additional tracks in tagged semileptonic events~(0.3\%) is
estimated by comparing fully reconstructed $D \bar{D}$ events in
data and MC simulation. The uncertainty associated with the number of signal
events is estimated by using an alternative signal function (a
double Gaussian) in the fits and by counting events in the signal
region (4.2\% for $D^+ \to \pi^0 e^+ \nu_e$, 0.3\% for all other
modes). The uncertainty in the semileptonic reconstruction
efficiencies due to imperfect knowledge of the semileptonic form
factors~(0.0\% to 0.3\% depending on mode) is estimated by varying the form
factor shape parameters in the MC simulation within uncertainties
in their measurements reported in Sec.~\ref{ff_results_sec}.
The uncertainty associated with the simulation of FSR and
bremsstrahlung radiation in the detector material~(0.4\%) is
estimated by varying the amount of FSR modeled by the PHOTOS
algorithm~\cite{PHOTOS} and by repeating the analysis
without recovery of photons radiated by the positron and comparing to the standard results. The
uncertainty associated with the simulation of ISR ($e^+ e^- \rightarrow D \bar{D} \gamma$) is negligible.
There is a systematic uncertainty due to finite MC
statistics~(0.1\% to 0.3\% depending on mode).

\begin{table}[tbp]
\caption{Summary of systematic uncertainties considered in the measurements
         of absolute branching fractions of the four semileptonic modes. The modes are labeled by their final state hadrons.
         }
\begin{center}
\begin{tabular}{l l p{1cm}p{1cm}p{1cm}p{0.5cm}  }
\hline
\hline
  &                   &       \multicolumn{4}{l}{Systematic uncertainty (\%)} \\
\rule[0mm]{0mm}{4mm}Source &   &~$K^-$&~$\pi^-$ &$K_S^{0}$ & ~$\pi^0$   \\
\hline
Number of $D$ tags           & & 0.4 & 0.4 & 0.4 & 0.4 \\
Electron ID efficiency     & & 1.0 & 1.0 & 1.0 & 1.0 \\
Hadron ID efficiency       & & 0.2 & 0.1 & 0.0 & 0.0 \\
Track finding  efficiency  & & 0.8 & 0.6 & 0.9 & 0.3 \\
$\pi^0$ finding efficiency & & 0.0 & 0.0 & 0.0 & 4.3 \\
$K^0_S$ finding efficiency & & 0.0 & 0.0 & 1.8 & 0.0 \\
Unused tracks              & & 0.3 & 0.3 & 0.3 & 0.3 \\
Signal shape fit function  & & 0.3 & 0.3 & 0.3 & 4.2 \\
Simulation of FSR          & & 0.4 & 0.4 & 0.4 & 0.4 \\
Simulation of form factors & & 0.0 & 0.1 & 0.1 & 0.3 \\
Limited MC statistics      & & 0.1 & 0.2 & 0.2 & 0.3 \\
\hline
Total uncertainty           & & 1.5 & 1.4 &  2.4 & 6.1 \\
\hline
\hline
\end{tabular}
\end{center}
\label{allbfsyserrors}
\end{table}

Table~\ref{allbfsyserrors} is a summary of the systematic
uncertainties associated with the measurement of the four absolute
semileptonic branching fractions. These estimates of systematic
uncertainty are added in quadrature to obtain the total systematic
uncertainty: 1.4\%, 6.1\%, 1.5\%, and 2.4\% for $D^0 \to \pi^- e^+
\nu_e$,  $D^+ \to \pi^0 e^+ \nu_e$, $D^0 \to K^- e^+ \nu_e$, and
$D^+ \to \bar{K}^0 e^+ \nu_e$, respectively.

\subsection{Comparison to Previous Measurements}

The branching fraction measurements with all tag modes combined for each of the four semileptonic modes reported in Table~\ref{semilBFs_data}, are
in good agreement with previous \cleoc measurements
using the same technique~\cite{DSemilBFs-2005}
obtained with a smaller data sample, and
supersede them. In
Table~\ref{semilBFs_data} we also compare our measurements to PDG 2004~\cite{PDG2004}
averages.
We compare to PDG 2004 because subsequent PDG averages~\cite{PDG2006, PDG2008}
are dominated by our previous \cleoc measurements. In
Table~\ref{tab:BF-comp} we compare our measurements of $ {\cal B}(
D^0 \ra K^- e^+ \nu_e)$ and $ {\cal B}( D^0 \ra \pi^- e^+ \nu_e)$
to previous measurements and to theoretical predictions. Our measurements agree well with previous measurements including the \cleoc neutrino reconstruction analysis~\cite{Nadia}, which we denote by \lq \lq untagged" hereinafter.

\begin{table}[htbp]
\caption{Comparison of  ${\cal B} (D^0 \ra K^- \ell^+ \nu_{\ell})$
and ${\cal B} (D^0 \ra \pi^- \ell^+ \nu_{\ell})$ values among
different experiments and theoretical predictions. The first
uncertainty is statistical, the second is systematic. The third
uncertainty in the BABAR measurement is from the normalization to
${\cal B} (D^0 \ra K^- \pi^+)$.} \label{tab:BF-comp}
\begin{center}
\begin{tabular}{lcc}\hline\hline
& $K^- \ell^+ \nu_{\ell} (\%) $    & $\pi^- \ell^+ \nu_{\ell}$ (0.1\%)         \\
\hline
PDG (2004)~\cite{PDG2004}      & 3.58(18)        & 3.6(6)   \\
BES II ($e$)~\cite{BESII}   & 3.82(40)(27)    & 3.3(13)(3)   \\
LQCD~\cite{unquenched_LQCD}   & 3.77(29)(74)  & 3.16(25)(70)  \\
LQCD (Abada)~\cite{Abada}    & 2.99(45)      & 2.4(6)   \\
QCD SR (Ball)~\cite{QCD-sum-rules-1}   & 2.7(6) & 1.6(3)    \\
LCSR (KRWWY)~\cite{QCD-sum-rules-2}  & 3.6(14)    & 2.7(10)     \\
LCSR (WWZ)~\cite{WWZ}      & 3.9(1.2)  & 3.0(9)    \\
\cleoc ($e$)~\cite{DSemilBFs-2005}    & 3.44(10)(10)  & 2.62(25)(8)  \\
Belle ($e, \mu$)~\cite{BELLE-06}   & 3.45(7)(20)      & 2.55(19)(16)  \\
BABAR ($e$)~\cite{BABAR-06}  &3.522(27)(45)(65)&  --             \\
\cleoc (tagged, $e$)     & 3.61(5)(5)          & 3.14(13)(4)   \\
\cleoc (untagged, $e$)~\cite{Nadia}  & 3.56(3)(9)  & 2.99(11)(9)    \\
\hline\hline
\end{tabular}
\end{center}
\end{table}

The widths of the isospin conjugate exclusive semileptonic decay
modes of the $D^0$ and $D^+$ are related by isospin invariance of
the hadronic current. The ratio $\Gamma(D^0 \rightarrow K^-
e^+ \nu_e) / \Gamma(D^+ \rightarrow \bar{K}^0 e^+ \nu_e)$  is
expected to be unity, while the corresponding ratio for pions is expected to be two.
Using our results and the lifetimes
$\tau_{D^0}=410.3(1.5)\times 10^{-15} {\rm s}$
and $\tau_{D^+}= 1040(7)\times 10^{-15}{\rm s}$~\cite{PDG2004}, we obtain
\begin{equation}
 \frac{\Gamma(D^0
\rightarrow K^-e^+\nu_e)} {\Gamma(D^+ \rightarrow \bar{K}^0 e^+
\nu_e)} = 1.03(2)(2)
\end{equation}
and
\begin{equation}
\frac{\Gamma(D^0 \rightarrow \pi^-e^+\nu_e)} {2\Gamma(D^+
\rightarrow \pi^0 e^+ \nu_e)} = 1.04(9)(6),
\end{equation}
where correlated and uncorrelated systematic uncertainties are
taken into account. These ratios are consistent with isospin predictions,
and supersede the corresponding ratios in Ref.~\cite{DSemilBFs-2005}, which were measured with the same technique.
These ratios are also consistent with the \cleoc
untagged analysis~\cite{Nadia}, and two
less precise results: a measurement from BES II using the
same technique~\cite{BESII_ratio} and an indirect measurement from
\hbox{FOCUS}~\cite{FOCUS_ratio}.

As the data are consistent with isospin invariance, the precision
of each branching fraction can be improved by averaging the $D^0$
and $D^+$ results for isospin conjugate pairs. For the
isospin-averaged semileptonic decay widths, with correlations
among systematic uncertainties taken into account, we find
\begin{equation}
\Gamma(D  \rightarrow K e^+\nu_e) = 8.73(9)(15) \times10^{-2} {\rm ps^{-1}}
\end{equation}
\noindent and
\begin{equation}
\Gamma(D  \rightarrow \pi e^+\nu_e) = 0.76(3)(2)\times 10^{-2} {\rm ps^{-1}},
\end{equation}

\noindent where for the latter partial width we have used $\Gamma
(D^0 \rightarrow \pi^- e^+\nu_e) = 2 \Gamma(D^+ \rightarrow \pi^0
e^+\nu_e)$.  The measured ratio of decay widths for $D \rightarrow
\pi e^+ \nu_e$ and $D \rightarrow K e^+ \nu_e$ provides a test of the
LQCD charm semileptonic rate ratio
prediction~\cite{unquenched_LQCD}. Using the results obtained in
this analysis, we find
\begin{equation}
\frac{\Gamma(D^0 \rightarrow \pi^- e^+ \nu_e)} {\Gamma(D^0
\rightarrow K^- e^+ \nu_e)} = 0.0868(38)(4)
\end{equation}
 \noindent and
\begin{equation}
\frac{2\Gamma(D^+ \rightarrow \pi^0 e^+ \nu_e)}
{\Gamma(D^+ \rightarrow \bar{K}^0 e^+ \nu_e)} = 0.0863(64)(53).
\end{equation}
These results are consistent with LQCD~\cite{unquenched_LQCD} and
with previous measurements~\cite{Nadia,CLEOIII,pikenu_focus}. Finally, by
averaging the $D^0$ and $D^+$ results for isospin conjugate pairs
we obtain
\begin{equation}
\frac{ \Gamma(D \rightarrow \pi e^+ \nu_e)} {\Gamma(D \rightarrow K
e^+ \nu_e)} = 0.0868(33)(14),
\end{equation}
\noindent where we have again used $\Gamma (D^0 \rightarrow \pi^-
e^+\nu_e) = 2 \Gamma(D^+ \rightarrow \pi^0 e^+\nu_e)$. A complete
set of ratios of partial semileptonic decay widths measured in
this analysis is given in Table ~\ref{decayrate_results_ratios}.

\begin{table}[tbp]
\caption{ Ratios of semileptonic decay widths of $D^0$  and $D^+$
to the pseudoscalar mesons $\pi$ and $K$  (first four lines) and
the isospin averaged ratio of semileptonic decay widths (fifth
line). The uncertainties are statistical and systematic.}
\begin{center}
\begin{tabular}{lcc} \hline \hline
Ratios        & Measured values \\ \hline\rule[-1mm]{-1mm}{4.5mm}
$\Gamma(\kenu)/\Gamma(\ksenu)$   & $1.030(24)(20)$  \\
$\Gamma(\penu)/ 2 \Gamma(\p0enu)$   & $1.037(86)(57)$ \\
$\Gamma(\penu)/\Gamma(\kenu)$   & $0.0868(38)(4)$  \\
$2\Gamma(\p0enu)/\Gamma(\ksenu)$ & $0.0863(64)(53)$ \\
$\Gamma(D \to \pi e^+ \nu_e )/\Gamma( D \to K e^+ \nu_e )$ & $0.0868(33)(14)$ \\
\hline\hline
\end{tabular}
\end{center}
\label{decayrate_results_ratios}
\end{table}

\section{Study of Semileptonic Differential Decay Rates}

\label{decayRates}

\subsection{Measurement of the Differential Decay Rate}

We now describe how the efficiency-corrected
absolutely-normalized differential decay rate distributions are
obtained. Full event reconstruction allows a direct measurement of
the neutrino momentum with excellent resolution.
The invariant mass squared of the $e^+ \nu_e$ pair, $q^2$, is
calculated in the $\psi(3770)$ rest frame in the following way
(using as an example $D^0 \to K^- e^+ \nu_e$):

\begin{equation}
q^2 = ( E_{\rm beam} - E_K)^2 - (- \vec{p}_{\rm tag} - \vec{p}_K)^2,
\end{equation}
\begin{equation}
\vec{p}_{\rm tag} = \hat{p}_{\rm tag}  \sqrt{E_{\rm beam}^2 - m_D^2 },
\end{equation}

\noindent where $E_K$ and  $\vec{p}_K$ are the energy and three-momentum of the kaon.
The $q^2$ resolutions
($q^2_{\rm reconstructed}-q^2_{\rm generated}$)
averaged over the entire $q^2$ range are about
0.012~(GeV$/c^2)^2$ for $D^0 \to \pi^- e^+ \nu_e$, $D^0 \to K^-
e^+ \nu_e$ and  $D^+ \to \bar{K}^0 e^+ \nu_e$, and approximately
0.040~(GeV$/c^2)^2$ for $D^+ \to \pi^0 e^+ \nu_e$. For $D^+ \to
\pi^0 e^+ \nu_e$, the
$q^2_{\rm reconstructed}-q^2_{\rm generated}$
distribution is well described by a Gaussian. For
other semileptonic modes these distributions are consistent with a
double Gaussian with $\sigma$'s that differ by a factor of 2.5,
with the wider Gaussian mostly due to FSR.

As the $D$ mesons are produced almost at rest at the $\psi(3770)$,
and the \cleoc detector is nearly hermetic, the semileptonic
reconstruction efficiencies are almost constant across the $q^2$
range. In consequence the shape of the $q^2$ spectrum receives
only minor distortions due to detector acceptance. The excellent
$q^2$ resolution likewise leads to only minor distortions due to
$q^2$ smearing.

Events satisfying the reconstruction criteria of
Sec.~\ref{sect_recon} that lie in the $U$ signal region,
defined as $-60\;{\rm MeV} \le U  \le 60\;{\rm MeV}$, are sorted
into bins of $q^2$. Ten bins of equal size~($q^2_{\rm max}/10$)
are used for $D^0 \to K^- e^+ \nu_e$ and $D^+ \to \bar{K}^0 e^+
\nu_e$. Nine~(seven) bins are used for $D^0 \to \pi^- e^+
\nu_e$~($D^+ \to \pi^0 e^+ \nu_e$) with the last bin two~(four)
times wider than the other bins to allow for the smaller number of
events at large $q^2$ for these modes. The bin limits are given in
Table~\ref{batboldbinlimits}.

\begin{table*}[htbp]
\caption{The upper edge of each $q^2$ bin in units of ${\rm GeV}^2
/c^4$ for each semileptonic mode studied in this work. }
\begin{center}
\begin{tabular}{l | ccccc ccccc }
\hline \hline
     Mode &                 Bin 1& Bin 2& Bin 3& Bin 4& Bin 5
                             & Bin 6& Bin 7& Bin 8& Bin 9& Bin 10  \\
\hline \rule[-1mm]{-1mm}{4mm}
$D^0 \ra \pi^-  e^+ \nu_e$&0.30&0.60&0.89&1.19
                         &1.49&1.79&2.08&2.38&     $q^2_{\rm max}$& \\
$D^+ \ra \pi^0 e^+ \nu_e$&0.30&0.60&0.90&1.20
                         &1.50&1.80&$q^2_{\rm max}$& & & \\
$D^0 \ra K^-   e^+ \nu_e$&0.19&0.38&0.56&0.75
                         &0.94&1.13&1.32&1.50&1.69&$q^2_{\rm max}$ \\
$D^+ \ra \bar{K}^0 e^+ \nu_e$&0.19&0.38&0.56&0.75
                         &0.94&1.13&1.32&1.51&1.69&$q^2_{\rm max}$ \\
\hline \hline
\end{tabular}
\end{center}
\label{batboldbinlimits}
\end{table*}

\begingroup
\squeezetable
\begin{table*}[htbp]
\caption{ Numbers of events, estimated backgrounds and yields
          in $q^2$ bins for the four semileptonic modes. The uncertainty in parentheses is
          statistical. The $q^2$ bins are defined in Table~\ref{batboldbinlimits}.}
\begin{center}
\begin{tabular}{l | l l c c c c c c c c c c }
\hline \hline
     Mode &                      & & Bin 1& Bin 2& Bin 3& Bin 4& Bin 5
                             & Bin 6& Bin 7& Bin 8& Bin 9& Bin 10  \\
\hline \hline
& Number of events & & 130(11)   & 122(11)  & 99(10)  &
105(10)  & 76(9)  & 56(8)
                           & 66(8)   & 38(6)  & 19(4)  &    \\
$D^0 \ra \pi^- e^+ \nu_e$ & Background
                         & & 8.9(7)   & 8.3(7)  & 7.0(6)   & 6.2(5) &  4.7(4)  & 3.5(3)
                           & 2.8(2)    & 2.4(2)  & 2.9(2)  &    \\
& Yield                  & & 121(11)  & 114(11)  &  92(10)  &
99(10)  & 71(9)
                         & 52(8)  & 63(8)  & 36(6)  & 16(4)  &     \\
\hline
& Number of events & & 48(7)  & 46(7)  & 44(7) & 36(6) &
34(6)
                           & 30(6)  & 48(7)  &     &     &          \\
$D^+ \ra \pi^0 e^+ \nu_e$ & Background
                         & & 1.8(1)  & 1.6(1)  & 2.5(2) & 3.0(2) & 2.7(1)
                           & 3.1(2)  & 20.0(1.4)  &     &     &            \\
& Yield & & 46(7)  & 44(7)  & 42(7) & 33(6) & 31(6)
                           & 27(6)  &  28(7)  &     &     &  \\
\hline
& Number of events        & &  1239(35)  & 1169(34)
& 1006(31)  & 923(30)  & 821(29)
                            &  594(24)  & 464(22)  & 293(17)  & 139(12)  & 29(5)  \\
$D^0 \ra K^- e^+ \nu_e$ & Background
                          & &  6.7(6)  & 6.7(6)  & 8.1(7)  & 7.7(7) & 9.1(8) & 8.7(7)
                            &  5.3(5)  & 3.9(3)  & 3.1(3)  & 1.5(1)     \\
& Yield                   & &  1232(35)  & 1162(34)   & 998(32)  &
915(30)  & 811(29)
                            &  585(24)   & 459(22)  & 290(17)  & 136(12)  & 28(5)    \\
\hline
& Number of events & &   570(24)  & 502(22)  & 442(21)  &
379(19)  & 298(17)
                           &  255(16)  & 210(14)  & 112(11) & 64(8)  & 19(4)  \\
$D^+ \ra K^0_S e^+ \nu_e$  & Background
                            & & 2.4(3)  & 3.0(4) & 3.4(5)  & 3.8(5)  & 3.3(4)
                              & 3.5(5)  & 2.9(4) & 2.1(4)  & 1.8(2)  & 1.2(2)    \\
& Yield                     & & 568(24) & 499(22) & 439(21)&
375(19)  & 295(17)
                              & 251(16)  & 207(15) & 110(11)  & 62(8)  & 17(4) \\

\hline \hline
\end{tabular}
\end{center}
\label{q2rawyields}
\end{table*}
\endgroup

The number of events in the data, the estimated background, and
the background-subtracted yield in each bin of $q^2$ are provided
in Table~\ref{q2rawyields}.
To obtain $d \Gamma/d q^2$ for each semileptonic mode, the
background is subtracted from the observed $q^2$ distribution. The
number of signal events $N^{\rm tag, SL}$ in the $i$th bin is
given by
\begin{equation}
N^{\rm tag, SL}_i = \sum_{j} \epsilon^{ij} N^{\rm produced}_j,
\end{equation}
\noindent where $\epsilon^{ij}$ is the semileptonic efficiency
matrix which accounts for acceptance and resolution effects.  This
matrix equation is inverted to obtain $N^{\rm produced}_j$, a vector of
efficiency corrected signal events
with a $\bar{D}$ tag in the data. When properly normalized, the elements of
$N^{\rm produced}_j$ give the absolute decay rate in $q^2$ bins.
\noindent Efficiency matrices, $\epsilon^{ij}$, for each
semileptonic mode are obtained using signal MC samples. The
procedure for calculating the efficiency matrices is analogous to
that for $\epsilon $:
\beqn
\epsilon^{ij} =   \epsilon_{\rm tag, SL}^{ij}  / \epsilon_{\rm tag},
\eeqn
with  $\epsilon_{\rm tag, SL}^{ij} $  obtained as
\beqn
 \epsilon_{\rm tag, SL}^{ij} =
 \frac{N_{\rm signal}^{ij}}{N_{\rm total}^j},
\eeqn
where
$N_{\rm total}^j$ is the number of signal events generated
in the $j$th $q^2$ bin, and  $N_{\rm signal}^{ij}$ is the number
of signal events that are generated in the $j$th $q^2$ bin and
reconstructed in the $i$th $q^2$ bin. Efficiency matrices for each
of the four modes are given in Table~\ref{effmatrices}. These
efficiency matrices have been calculated for the simple pole
model, with the $q^2$ distribution re-weighted to the value of $M_{\rm pole}$
determined by the data for each mode,
and weighted by the tag yields given in Table~\ref{tagsD0} and~\ref{tagsDPlus}.
We note that  at the present
level of precision, due to the use of efficiency matrices combined
with the fine binning in $q^2$, the values we determine for
the shape and normalization parameters in the form factor fits are not sensitive to the model used to generate the
efficiency matrices.
The statistical uncertainty of the
background-subtracted and efficiency-corrected decay rate
distribution for each $q^2$ bin is given by
\begin{eqnarray}
 [\sigma_{N_{\rm produced}^i} ]^2 & = &
\sum_{j}
      ( [ \epsilon^{-1} ]_{ij}^2
       [ \sigma_{N_{\rm tag, SL}^j} ]^2 \nonumber \\
&&    \qquad +  [\sigma(\epsilon^{-1}) ]_{ij}^2
       [ N^j_{\rm tag, SL}]^2
       ).
\end{eqnarray}

\begingroup
\squeezetable
\begin{table*}[htbp]
\caption{ Abridged efficiency matrices for
the four semileptonic modes.
Matrix elements are in percent.
The semileptonic efficiency includes subsidiary branching fractions~\cite{subsidiarybf}.
Those that are not on or adjacent to a diagonal are null
and are not shown. The uncertainties are statistical. The $q^2$
bins are defined in Table~\ref{batboldbinlimits}. }
\begin{tabular}{l|lcccccccccc}
\hline \hline
&
$\epsilon(i,j)$       & $\epsilon(1,j)$ &
$\epsilon(2,j)$  & $\epsilon(3,j)$  & $\epsilon(4,j)$ &
$\epsilon(5,j)$ & $ \epsilon(6,j)$  & $\epsilon(7,j)$ &
$\epsilon(8,j)$ & $\epsilon(9,j)$   &  $\epsilon(10,j)$   \\
\hline \hline
&$ \epsilon(i,i-1) $ & 0.0(0)    &  2.08(5)
            & 1.98(5)  & 1.85(5)
                    & 1.67(5)  & 1.53(6)
                    & 1.23(6)  & 1.21(6)
                    & 0.85(7)  &  \\
$\dzpie$&$ \epsilon(i,i) $    & 62.81(26) & 64.13(28)
                     & 67.25(31) & 69.10(34)
                     & 69.48(37) & 70.36(42)
                     & 70.23(49) & 70.02(61)
                     & 68.17(75) & \\
&$\epsilon(i,i+1) $   & 2.23(5)  & 2.14(5)
                     & 2.00(6)  & 1.72(6)
                     & 1.58(6)  & 1.45(7)
                     & 1.21(8)  & 0.93(7)
                     &  0.0(0)  &
\\ \hline \hline

& $ \epsilon(i,i-1) $  &  0.0(0)   & 3.32(9)
                     & 3.68(10)  & 3.35(10)
                     & 2.82(10)  & 2.62(11)
                     & 2.50(12)  &
                     &                &    \\
$\dppie$ &$ \epsilon(i,i) $    & 36.00(29) & 35.48(30)
                     & 36.17(33) & 36.55(36)
                     & 35.84(39) & 35.67(44)
                     & 38.42(37) &
                     &                &  \\
& $ \epsilon(i,i+1) $  & 2.21(7)  & 2.07(8)
                     & 2.06(9)  & 1.87(9)
                     & 1.72(10)  & 0.81(5)
                     &  0.0(0)    &
                     &                &
\\ \hline \hline
 & $\epsilon(i,i-1)$
        & 0.0(0)      & 2.40(3)
     & 2.36(4)    & 2.24(4)      & 1.99(4)
        & 1.81(4)    & 1.48(4)     & 1.16(4)
     & 0.85(5)    & 0.47(2)    \\
$\dzke$ & $\epsilon(i,i)$ & 52.86(16)  & 52.78(17)     & 55.56(18)
                & 57.98(20)  &  59.49(23)
                & 59.72(25)  &  58.96(30)  & 57.43(36)
                & 53.78(50)   &  39.62(86)   \\
 & $\epsilon(i,i+1)$ & 2.70(4)   & 2.60(4)
    & 2.54(4)     & 2.37(5)   & 2.33(5)
        & 2.07(6)   &  1.86(7)   & 1.63(9)
        & 1.52(17)    &  0.0(0)
\\ \hline \hline

&$\epsilon(i,i-1)$    &  0.0(0)  & 0.82(2)
                     & 0.75(2)  & 0.73(2)
                     & 0.64(2)  & 0.62(2)
                     & 0.55(2)  & 0.49(2)
                     & 0.38(2)  & 0.21(2)
\\
$\dpke$ & $\epsilon(i,i)$      & 18.14(8) & 17.57(8)
                     & 18.12(8) & 18.64(9)
                     & 18.63(10) & 20.20(11)
                     & 19.01(13) & 19.47(17)
                     & 20.12(24) & 20.13(51)
\\
 & $\epsilon(i,i+1)$    & 0.84(2)  & 0.82(2)
                     & 0.82(2)  & 0.79(2)
                     & 0.72(2)  & 0.63(2)
                     & 0.66(3)  & 0.63(4)
                     & 0.78(10) &  0.0(0)
\\ \hline \hline

\end{tabular}
\label{effmatrices}
\end{table*}
\endgroup

Background-subtracted, efficiency-corrected and absolutely
normalized decay rate distributions for the four semileptonic
modes are given in
Table~\ref{decayrateinbins}. They constitute the main result of
this analysis and can be used to compare to other experimental
measurements and to theory without a need for knowledge of \cleoc
acceptance and resolution.

Table~\ref{decayrateinbins} includes statistical uncertainties and
the associated correlation matrices. As discussed in
Sec.~\ref{SystErrsdGdq2Section},
systematic
uncertainties are approximately fully correlated between $q^2$ bins across the
entire $q^2$ range. Therefore we include systematic uncertainties
for each $q^2$ bin for each semileptonic mode in
Table~\ref{decayrateinbins} without correlation matrices.

\begingroup
\squeezetable
\begin{table*}[htbp]
\caption{ Absolutely normalized decay rates in bins of $q^2$, with
statistical and systematic uncertainties in parentheses, are given
in the first row for each decay mode. These distributions are
background-subtracted and  efficiency-corrected. The truncated
statistical correlation matrices are shown in the last three or
five rows. The $q^2$ bins are defined in
Table~\ref{batboldbinlimits}.
} \label{decayrateinbins}
\begin{tabular}{lcccccccccc}
\hline \hline Bin: &   1 &   2 &   3 &   4  &   5 &   6 &   7 &  8
&   9 &  10 \\ \hline \hline

$\Gamma(\pi^- e^+ \nu_e)$ &
                $1.482$ & $1.325$ & $1.014$ & $1.087$ & $0.774$
          &     $0.560$ & $0.697$ & $0.388$ & $0.183$  & \\
$[{\rm ns}^{-1}]$ &
          $(149)(24)$ & $(142)(24)$ & $(122)(19)$ & $(121)(21)$ & $(103)(15)$
        & $(87)(11)$  & $(94)(13)$  & $(72)(7)$   & $(55)(4)$  & \\
\hline
$C_{ii-1}$ & --   & -0.068   & -0.063  & -0.056 & -0.049  & -0.045   &  -0.039  & -0.036  & -0.027 &   \\
$C_{ii}$   & 1.000 &  1.000   &  1.000  &  1.000 &  1.000  &  1.000   &   1.000  &  1.000  &  1.000 &   \\
$C_{ii+1}$ &-0.068 & -0.063   & -0.056  & -0.049 & -0.045  & -0.039   &  -0.036  & -0.027  &  --  &   \\
\hline \hline

\rule[-1mm]{-1mm}{4mm}
$2 \Gamma(\pi^0 e^+ \nu_e)$ &
          $1.432$ & $1.270$ & $1.167$ & $0.908$ &
          $0.917$ & $0.799$ & $0.786$ &  &  & \\
$[{\rm ns}^{-1}]$ &
          $(233)(97)$ & $(233)(77)$ & $(226)(63)$ & $(204)(44)$ &
          $(202)(39)$ & $(192)(30)$ & $(253)(25)$ &         &  & \\
\hline
$C_{ii-2}$ & --   &  --   &  0.020  &  0.019 &  0.015  &  0.011   &   0.002  &  &  &    \\
$C_{ii-1}$ & --   & -0.155   & -0.162  & -0.154 & -0.130  & -0.123   &  -0.078  &  &  &    \\
$C_{ii}$   & 1.000 &  1.000   &  1.000  &  1.000 &  1.000  &  1.000   &   1.000  &  &  &    \\
$C_{ii+1}$ &-0.155 & -0.162   & -0.154  & -0.130 & -0.123  & -0.078   &   --  &  &  &    \\
$C_{ii+2}$ & 0.020 &  0.019   &  0.015  &  0.011 &  0.002  &  --    &   --  &  &  &    \\
\hline \hline

$\Gamma(K^- e^+ \nu_e)$
             & $1.767 $ & $1.601$ & $1.303$ & $1.160$ &
               $1.015 $ & $0.725$ & $0.586$ & $0.383$ &
               $0.193 $ & $0.053$ \\
$[10 \times {\rm ns}^{-1}]$
         & $(53)(27) $  & $(52)(26)$  & $(46)(22)$  & $(42)(20)$ &
               $(39)(18) $  & $(33)(13)$  & $(29)(11)$  & $(24)(7)$ &
               $(18)(4)  $  & $(11)(1) $  \\
\hline
$C_{ii-1}$ & --  & -0.096    & -0.092  & -0.084 & -0.074  & -0.069   &  -0.059  & -0.051  & -0.043 &  -0.036 \\
$C_{ii}$   & 1.000  & 1.000   &  1.000  &  1.000 &  1.000  &  1.000   &   1.000  &  1.000  &  1.000 &   1.000 \\
$C_{ii+1}$ & -0.096 & -0.092  & -0.084  & -0.074 & -0.069  & -0.059   &  -0.051  & -0.043  & -0.036 &    -- \\
\hline \hline

\rule[-1mm]{-1mm}{4mm}
$\Gamma(\bar{K}^0 e^+ \nu_e)$ &
          $1.785$ & $1.540$ & $1.320$ & $1.105$ &
          $0.870$ & $0.757$ & $0.613$ & $0.312$ &
          $0.175$ & $0.050$ \\
$[10 \times {\rm ns}^{-1}]$ &
          $(79)(42)$ & $(77)(38)$ & $(70)(33)$ & $(63)(29)$ &
          $(55)(23)$ & $(52)(20)$ & $(46)(17)$ & $(33)(9)$ &
          $(24)(5)$  & $(13)(2)$ \\
\hline
$C_{ii-1}$ & --   & -0.093   & -0.088  & -0.084 & -0.077  & -0.072   &  -0.063  & -0.060  & -0.049 & -0.040  \\
$C_{ii}$   & 1.000 &  1.000   &  1.000  &  1.000 &  1.000  &  1.000   &   1.000  &  1.000  &  1.000 &  1.000  \\
$C_{ii+1}$ &-0.093 & -0.088   & -0.084  & -0.077 & -0.072  & -0.063   &  -0.060  & -0.049  & -0.040 &  --  \\
\hline \hline

\end{tabular}
\end{table*}
\endgroup

\begingroup
\squeezetable
\begin{table*}[htbp]
\caption{ Form factor distributions in bins of $q^2$ in the first
row, with statistical and systematic uncertainties in the second
row in parentheses for each decay mode. The $q^2$ bins are defined
in Table~\ref{batboldbinlimits}.
The entries for $\dppie$ have been scaled by the isospin factor $\sqrt{2}$.
}
\begin{tabular}{lcccccccccc}
\hline \hline

Bin: &   1 &   2 &   3 &   4  &   5 &   6 &   7 &  8
&   9 &  10 \\ \hline \hline

\rule[-1mm]{-1mm}{4mm}
$f^{\pi}_{+}(q^2)\vcd(\pi^- e^+ \nu_e)$ &
                $0.160$ & $0.175$ & $0.180$ & $0.222$ & $0.230$
          &     $0.249$ & $0.370$ & $0.398$ & $0.458$  & \\
    & $(8)(1)$ & $(9)(2)$ & $(11)(2)$ & $(12)(2)$ & $(15)(2)$
        & $(19)(2)$  & $(25)(4)$  & $(37)(38)$   & $(68)(5)$  & \\
\hline \hline

\rule[-1mm]{-1mm}{4mm}
$f^{\pi}_{+}(q^2)\vcd(\pi^0 e^+ \nu_e)$ &
          $0.156$ & $0.170$ & $0.191$ & $0.202$ &
          $0.249$ & $0.296$ & $0.357$ &  &  & \\
&         $(12)(5)$ & $(15)(5)$ & $(18)(5)$ & $(21)(5)$ &
          $(26)(5)$ & $(33)(6)$ & $(54)(6)$ &         &  & \\
\hline \hline
\rule[-1mm]{-1mm}{4mm}
$f^K_{+}(q^2)\vcs(K^- e^+
\nu_e)$
             & $0.759 $ & $0.806$ & $0.821$ & $0.887$ &
               $0.968 $ & $0.980$ & $1.098$ & $1.180$ &
               $1.268 $ & $1.519$ \\

         & $(12)(6) $  & $(14)(6)$  & $(15)(7)$  & $(17)(8)$ &
               $(20)(9) $  & $(24)(9)$  & $(30)(10)$ & $(40)(11)$ &
               $(63)(13)$  & $(159)(19) $  \\

\hline \hline

\rule[-1mm]{-1mm}{4mm}
$f^K_{+}(q^2)\vcs(\bar{K}^0 e^+ \nu_e)$ &
          $0.760$ & $0.788$ & $0.824$ & $0.862$ &
          $0.893$ & $0.997$ & $1.118$ & $1.061$ &
          $1.200$ & $1.437$ \\
&     $(18)(9)$ & $(21)(10)$ & $(23)(10)$ & $(26)(11)$ &
          $(30)(12)$ & $(36)(13)$ & $(44)(15)$ & $(58)(15)$ &
          $(87)(17)$  & $(202)(25)$ \\
\hline \hline

\end{tabular}
\label{formfactorinbins}
\end{table*}
\endgroup

\begin{figure*}[htb]
\begin{minipage}{6in}
\epsfig{file=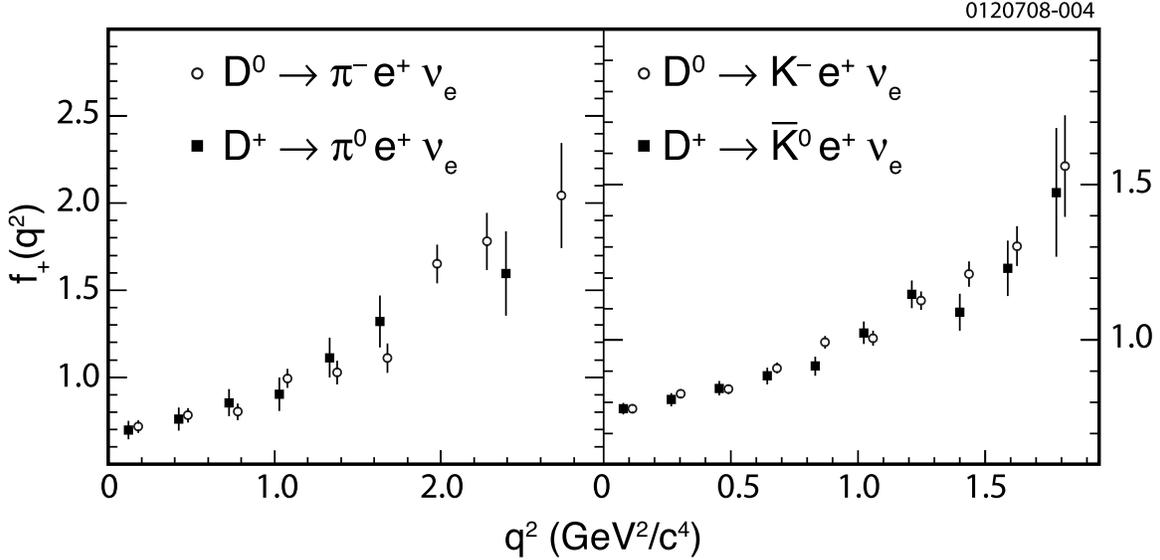,width=6in}
\end{minipage}
\caption{The data displayed as $f_+(q^2)$ for the four
semileptonic modes. In each case $f_+(q^2)$ is absolutely
normalized.
The $D^+$ and $D^0$ distributions are each offset
symmetrically in $q^2$ to facilitate display.}
\label{ffdist-isospin-plots}
\end{figure*}

The partial differential decay rates are expected to be identical
by isospin invariance. A powerful check of our understanding of
the data is therefore provided by comparing the
background-subtracted, efficiency-corrected rates in
Table~\ref{decayrateinbins} for $D^0$ and $D^+$.
We make the
comparison by removing the kinematic term and constants from the
differential rate to reveal  $f_+(q^2) |V_{cq}|$, where $q = d
{\rm ~or~} s$,
\begin{equation}
f_+ (q^2)|V_{cq}| =  \sqrt {  \frac{d \Gamma}{d q^2} \frac {
 24 \pi^3  } {G^2_F   p^{\prime 3}_{K,\pi} } },
\end{equation}  
where $d \Gamma/d q^2 $ is obtained by dividing the integrated rate in each $q^2$ interval by
the corresponding bin size, and $p^{\prime 3}_{K, \pi}$ in the $i$th $q^2$ bin
is given by
\begin{equation}
p^{\prime 3}_{K,\pi} (i) =
\frac{\displaystyle \int_{q^2_{\rm min} (i)}^{q^2_{\rm max} (i)}
p^{3}_{K,\pi} |f_+ (q^2)|^2 d q^2 }{|f_+(q^2_{{\rm center\: of\: bin}\: i})|^2
(q^2_{\rm max}(i) - q^2_{\rm min}(i))},
\end{equation}
where the form factor parameters are measured in the data using the
three parameter series parametrization~(see Sec.~\ref{FFFitingSection}).
For $D \ra K e^+ \nu_e$ ($ D \ra \pi e^+ \nu_e$),
$f_+ (q^2)|V_{cq}|$ varies by only a factor two (three) across the $q^2$ range.
Table~\ref{formfactorinbins} and Fig.~\ref{ffdist-isospin-plots}
show $f_+(q^2)|V_{cq}|$ and
$f_+(q^2)$
in data for all four semileptonic
modes.
The isospin conjugate distributions are
consistent.

\subsection{Fitting the Differential Decay Rate to Determine Form Factors }

\label{FFFitingSection}

We use the least squares method to fit the absolutely-normalized
efficiency-corrected and background-subtracted $q^2$
distributions. A $\chi^2$ is constructed from differences between
the number of efficiency corrected signal events with a $\bar{D}$ tag
in the $i$th $q^2$ bin,
$N^i_{\rm produced}$,
and the theoretically predicted number of events in the $i$th $q^2$ bin,
$N_{\rm predicted}^i$,
for a given set of form factor parameters,
where $N_{\rm predicted}^i$ is obtained using

\begin{eqnarray}
N_{\rm predicted}^i & = \hspace{5cm} \nonumber \\
N_{\rm tag} \tau_D & \displaystyle \int_{q^2_{\rm min} (i)}^{q^2_{\rm max} (i)}
 \frac{G^2_F |V_{cq}|^2 p^3_{K,\pi} } {
24 \pi^3} |f_+ (q^2, {\vec{\theta}})|^2 d q^2,
\end{eqnarray}
\noindent and where $\tau_D$ is the lifetime of the relevant $D$ meson,
and
$\vec{\theta}$ is the vector of form factor parameters that govern
the decay rate. Taking into account the correlations among the
bins and the correlations among the elements of the inverted
efficiency matrix~\cite{errors}, the $\chi^2$ is given by
\begin{eqnarray}
\chi^2 &= & \sum_{ij} ( [N_{\rm produced}^i] - [N_{\rm predicted}^i]) \nonumber \\
&& \qquad A_{ij}^{-1} ( [N_{\rm produced}^j] - [N_{\rm predicted}^j]),
\end{eqnarray}
where $A_{ij}$ is
\beqn
A_{ij} =
\sum_k \epsilon^{-1}_{i k} \epsilon^{-1}_{j k}
\sigma_{ N_{ \rm tag, SL}^k }^2.
\eeqn

Systematic uncertainties and correlations among them are not
included in the fit. Instead a systematic uncertainty from each
source is estimated separately. The fitting procedure has been
tested using ensembles of fits to 100~mock data samples that each
correspond to the same integrated luminosity as the data, for a
wide range of values of the form factor parameters. It has been
established that the statistical uncertainties from this fitting
procedure are consistent with the smallest statistical
uncertainties expected from a fit, estimated using the
Cramer~-~Rao inequality~\cite{eadie}, and that the fit is
consistent with being unbiased.

\begin{figure*}[htb]
\begin{minipage}{6in}
\epsfig{file=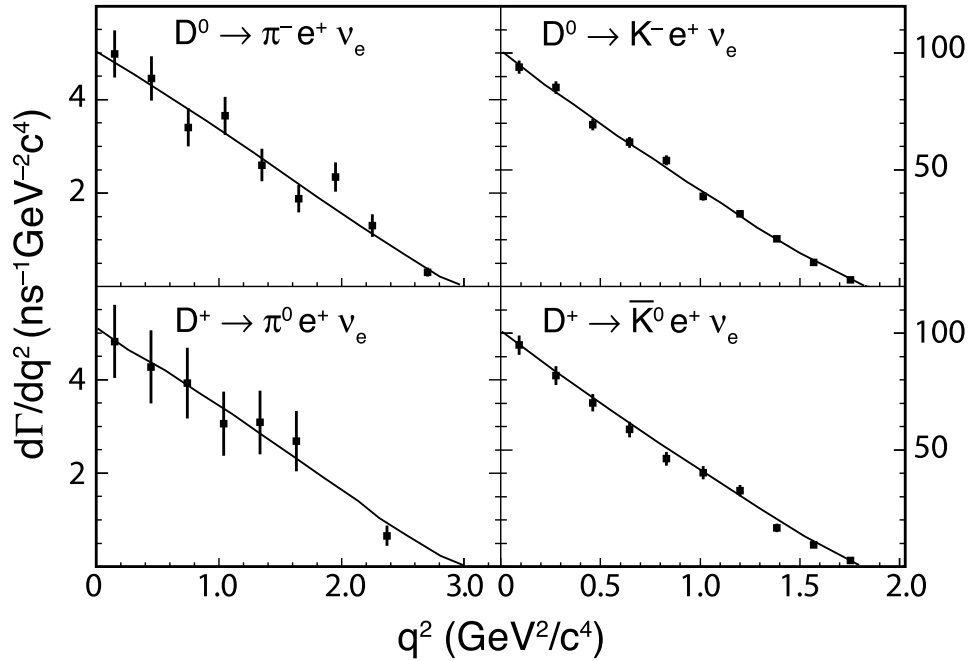,width=5in}
\end{minipage}
\caption{Simultaneous three parameter fits to the
background-subtracted, efficiency-corrected and
absolutely-normalized data derived decay rates for modes related by isospin
$\penu ; \p0enu$, and
$\kenu ; \ksenu$. The
$D^+ \to \pi^0  e^+ \nu_e$ distribution is scaled by a factor two
to account for isospin. The series parametrization with a quadratic term
is used
for these fits. The fit result is the line.} \label{fitsTodGdq2}
\end{figure*}

Fits to the data are made for two parameters related to the shape
and the normalization of the $f_+(q^2)$ form factors for the
series parametrization, the simple pole, the modified pole model,
and the ISGW2 model. For the series parametrization we also
present results of fits for three parameters, where the third
parameter is a second shape parameter. As an example,
Fig.~\ref{fitsTodGdq2} shows simultaneous fits to modes related
by isospin symmetry.
Before presenting numerical
results of the form factor measurements, we describe a study of
systematic uncertainties in the next section.

\subsection{\boldmath Study of Systematic Uncertainties for
${d \Gamma}/{d q^{2}}$ and Form Factor Measurements}

\subsubsection{Systematic Uncertainties for ${d \Gamma}/{d q^{2}}$ in $q^2$ bins}

\label{SystErrsdGdq2Section}

Each source contributing systematic uncertainty to the absolute
branching fractions also contributes systematic uncertainty to
measurements of the partial rate  ${d \Gamma}/{d q^{2}}$ in $q^2$
bins. Procedures identical to those used in the absolute branching
fraction measurements are employed to estimate systematic
uncertainties for ${d \Gamma}/{d q^{2}}$. In addition, there is a
systematic uncertainty associated with imperfect knowledge of
$D^0$ and $D^+$ meson lifetimes. Table~\ref{q2systerrors} reports
total systematic uncertainties, and the separated correlated and
uncorrelated components, for ${d \Gamma}/{d q^{2}}$
in $q^2$ bins for the four semileptonic modes.

\begin{table*}[htbp]
\caption{ Correlated, uncorrelated and total systematic uncertainties
          in ${d \Gamma}/{d q^{2}}$ for each $q^2$ bin for the four semileptonic modes.
          The $q^2$ bins are defined in Table~\ref{batboldbinlimits}. }
\begin{center}
\begin{tabular}{l | l l c c c c c c c c c c }
\hline\hline
          &                   & \multicolumn{10}{c}{Uncertainty (\%)}\\
     Mode &    Type                & & Bin 1& Bin 2& Bin 3& Bin 4& Bin 5
                             & Bin 6& Bin 7& Bin 8& Bin 9& Bin 10  \\
\hline

& Correlated  & & 1.5   & 1.7  & 1.8  & 1.9  & 1.8 & 1.8
                           & 1.8   & 1.8  & 2.0  &    \\
$D^0 \ra \pi^- e^+ \nu_e$ & Uncorrelated
                         & & 0.5   & 0.4  & 0.5   & 0.5 & 0.6  & 0.6
                           & 0.6    & 0.8  & 1.1  &    \\
& Total           & & 1.6  & 1.8  & 1.9  & 1.9  & 1.9
                                   & 1.9  & 1.9  & 1.9  & 2.3  &     \\
\hline

& Correlated  & & 6.8  & 6.1  & 5.4 & 4.8 & 4.2
                           & 3.6  & 3.1  &     &     &          \\
$D^+ \ra \pi^0 e^+ \nu_e$ & Uncorrelated
                         & & 0.7  & 0.8  & 0.8 & 0.9 & 0.9
                           & 1.0  & 0.9  &     &     &            \\
& Total & & 6.8  & 6.1  & 5.4 & 4.9 & 4.3
                           & 3.8  & 3.2  &     &     &  \\
\hline

& Correlated   & &  1.4  & 1.5  & 1.6  & 1.6 & 1.7
                             & 1.8  & 1.8  & 1.7  & 1.9  & 1.5  \\
$D^0 \ra K^- e^+ \nu_e$ & Uncorrelated
                          & &  0.4  & 0.4  & 0.4  & 0.4  & 0.4  & 0.4
                            &  0.5  & 0.7  & 0.9  & 1.9     \\
& Total          & & 1.5   & 1.6  & 1.7  & 1.7  & 1.8
                                  &  1.8   & 1.9  & 1.9  & 2.1  & 2.4    \\
\hline

& Correlated  & &  2.4  & 2.5  & 2.5  & 2.5  & 2.6
                           &  2.6  & 2.6  & 2.7  & 2.6  & 2.6   \\
$D^+ \ra \bar{K}^0 e^+ \nu_e$  & Uncorrelated
                            & & 0.4  & 0.4 & 0.5  & 0.5  & 0.6
                              & 0.6  & 0.7 & 0.9  & 1.2  & 2.2    \\
& Total     & & 2.4  & 2.5 & 2.5  & 2.6  & 2.6
                              & 2.7  & 2.7 & 2.9  & 2.9  & 3.4    \\
\hline
\hline
\end{tabular}
\end{center}
\label{q2systerrors}
\end{table*}

Systematic uncertainties associated with finding and identifying
the hadron~(positron) in the final state of a semileptonic decay
are measured in bins of hadron~(positron) momentum and propagated
to the $d \Gamma / d q^2$ distributions. In the rest frame of the
decaying $D$ meson, $q^2$ is determined by the momentum of the
final state hadron. Because $\psi(3770)$ decays produce $D$ mesons
with a small boost, $q^2$ is strongly correlated with the momentum
of the final state hadron measured in the laboratory frame.
Therefore systematic uncertainties measured in hadron momentum
bins, when propagated to $d \Gamma / d q^2$, lead to uncertainties
that are mostly uncorrelated between $q^2$ bins. The correlation
between $q^2$ and the positron momentum in the laboratory frame is
less pronounced due to additional degrees of freedom associated
with the undetected neutrino. Systematic uncertainties in positron
momentum bins are therefore averaged over a range in $q^2$ and
their net effect is to produce uncertainties that are nearly
constant and fully correlated between $q^2$ bins.

To simplify the estimation of the systematic uncertainties, we
assume that a given systematic uncertainty is either fully
correlated or uncorrelated between $q^2$ bins as discussed in the
remainder of this section.

Studies of the momentum dependence of the systematic uncertainty
associated with track finding efficiencies are performed in three
momentum bins covering the entire momentum range accessible in $D$
meson decays at the $\psi(3770)$.  Efficiencies for positively and
negatively charged pions and kaons are measured separately. We
assume that track finding efficiencies for positrons are identical
to those for positively charged pions. A systematic uncertainty
from track finding efficiencies in a $q^2$ bin is calculated by
weighting charged hadron~(positron) spectra with the efficiency
uncertainties measured in the hadron~(positron) momentum bins and
summing contributions from different momentum bins.
Due to the coarse
binning used in the tracking studies and because positron and
charged hadron track finding uncertainties are combined in each
$q^2$ bin, systematic uncertainties associated with track finding
efficiency are strongly correlated between $q^2$ bins. We
assume that they are fully correlated.

Systematic uncertainties associated with charged hadron
identification in $D^0 \rightarrow K^- e^+ \nu_e$ $(D^0
\rightarrow \pi^- e^+ \nu_e)$ are obtained by weighting charged
hadron spectra with the statistical uncertainties associated with
charged hadron identification measured in 100 (80)
~MeV/$c$~--~wide momentum bins and summing contributions from
different momentum bins in quadrature. Because the hadron momentum
is strongly correlated with $q^2$, these systematic uncertainties
are largely independent for well-separated values of $q^2$.
We therefore assume that the systematic uncertainties associated
with hadron identification are uncorrelated between $q^2$ bins.

The systematic uncertainty in the $\pi^0$ reconstruction
efficiency varies from  1.3\% for low $\pi^0$ momenta to 6.3\%
for high $\pi^0$ momenta and is found to be fully correlated
between $q^2$ bins.
Systematic uncertainties associated with the $K^0_S$
reconstruction are found to be independent of the $K^0_S$ momentum
and
are fully correlated between $q^2$ bins.

Systematic uncertainties due to simulation of ISR and FSR are
strongly correlated between $q^2$ bins. Systematic uncertainties
from FSR are assigned based on differences between the main
results and results of fits with efficiency matrices obtained
using a subset of signal MC events without FSR.
To evaluate systematic uncertainties associated with ISR, we repeated
the analysis with two alternative efficiency matrices: one using
signal MC events
with soft ISR photons~($E_\gamma \le 25$~keV) and the other from
the remainder of the signal MC events. Comparing results of fits
with these two efficiency matrices, we conclude that systematic
uncertainties due to ISR
are negligible.

The background is modeled using the generic MC sample. Systematic
uncertainties associated with the modeling of background are
obtained by varying the composition of the background sample
according to uncertainties in the branching fractions of processes
producing background, and by the statistical uncertainties in the
normalization for each background component. In addition, in cases
where a background component arises from misidentified hadrons or
leptons, background normalizations are varied according to the
uncertainty in the relative misidentification rates between the
data and MC simulation.

Systematic uncertainties from imperfect knowledge of the $D^0$ (0.4\%) and
$D^+$ (0.7\%) meson lifetimes, the number of tags (0.4\%), and unused
tracks~(0.3\%) are fully correlated between $q^2$ bins. Systematic
uncertainties due to the limited size of the MC samples used to
measure the efficiency matrices are statistical in origin and are
therefore uncorrelated between $q^2$ bins.

Three systematic uncertainties for each $q^2$ bin are presented in
Table~\ref{q2systerrors}. These are the combined sum in
quadrature of all correlated and all uncorrelated
contributions, and the total systematic uncertainty. The magnitude
of the systematic uncertainty in each $q^2$ bin is significantly
smaller than the corresponding statistical uncertainty, and
the relative size of the uncorrelated systematic uncertainty is
small compared to the correlated systematic uncertainty in nearly
all bins. (Note, in the last $q^2$ bin the uncorrelated systematic
uncertainty is dominated by uncertainty due to the limited size of
the MC sample, and for $D \ra K e^+ \nu_e$ it is comparable to the
correlated systematic uncertainty.)
For comparison to theory and for the form factor measurements
presented here we assume that systematic uncertainties are fully
correlated between $q^2$ bins.

\subsubsection{Systematic Uncertainties for Measurements of $f_+(0)$ and
Form Factor Shape Parameters}

The normalization parameter,~$f_+(0)|V_{cq}|$,
and form factor shape parameters are
determined from simultaneous two parameter fits to $d \Gamma / d
q^2 ( q^2) $ for each isospin conjugate semileptonic mode.
In each
case the correlation coefficient between the form factor shape
parameter and the normalization parameter is found to be small.

Systematic uncertainties associated with the absolute form factor
normalization, $f_+(0)$, for each semileptonic mode, are one half
the systematic uncertainties in the branching fraction
measurements presented in Sec.~\ref{BFSystErrs} combined in
quadrature with the small uncertainties associated with the
knowledge of $D^0$~(0.4\%) and $D^+$~(0.7\%)~\cite{PDG2004}
lifetimes and the CKM matrix elements $|V_{cs}|$~(0.1\%) and
$|V_{cd}|$~(1.3\%) obtained from the unitarity constraints of
the CKM matrix.

Systematic uncertainties for form factor shape parameters are obtained
from one parameter fits with absolute normalizations fixed. In the rest
of this section, we describe sources of systematic uncertainty for
form factor shape parameters and how they are estimated.

A systematic uncertainty associated with the fit procedure is
assigned by examining the pull distributions resulting from fits
to ensembles of mock data samples.
The studies determine that the fit has good fidelity, and place an
upper limit on the existence of bias at 15\% of the statistical
uncertainty in the measurement on data, which we take as a
systematic uncertainty associated with the fit
method~(Table~\ref{syspolemass}).

As discussed in the previous section, most sources of systematic
uncertainty are $q^2$ independent
and, consequently, do not contribute a systematic uncertainty in
the form factor shape parameter measurement.
Accordingly, to assign systematic uncertainties for tracking
efficiency, $K_S^0$ and $\pi^0$ finding, and hadron and electron
identification, a correlation with particle momentum consistent
with our knowledge of each systematic effect is introduced. This
is achieved by constructing a model according to which each
systematic uncertainty varies linearly as a function of the
particle momentum. The slope for each systematic uncertainty is
determined by the precision with which each systematic effect is
known. We fit the data using efficiency matrices modified
according to this model. We also construct a set of mock data
samples for each source using the model and fit them to obtain
systematic uncertainties for the form factor shape parameters. We
find that systematic uncertainties measured by these two methods
are consistent.

Systematic uncertainties associated with the simulation of FSR and
from background estimation are obtained as described in the
previous section.  The total systematic uncertainty, given in
Table~\ref{syspolemass}, ranges from 19\% to 53\% of the
statistical uncertainty.
The ratio of the systematic to statistical uncertainties for shape
parameters are found to be consistent for all
parametrizations.

\begin{table}[htbp]
\caption{ Systematic uncertainties for form factor shape parameters 
in units of the statistical uncertainty of the measurement with
data. The modes are labeled by their final state hadrons.
} \label{syspolemass}
\begin{center}
\begin{tabular}{lc p{1cm}p{1cm}p{1cm}p{0.5cm}}
\hline \hline
  &               &      \multicolumn{4}{l}{Systematic uncertainty ($\sigma_{stat}$)} \\
\rule[-1mm]{-1mm}{4.5mm}
Sources & & ~$K^-$& $K_S^{0}$ & ~$\pi^-$ & ~$\pi^0$ \\ \hline
Track finding   &&   $0.03$     & $0.00$            & $0.02$    &  $0.00$             \\
$K_S^0$ finding &&   $0.00$     & $0.20$            & $0.00$    &  $0.00$             \\
$\pi^0$ finding &&   $0.00$     & $0.00$            & $0.00$    &  $0.38$             \\
Hadron ID       &&   $0.11$     & $0.00$            & $0.05$    &  $0.00$             \\
Electron ID     &&   $0.02$     & $0.01$            & $0.01$    &  $0.01$             \\
FSR             &&   $0.20$     & $0.16$            & $0.07$    &  $0.04$   \\
Background      &&   $0.05$     & $0.07$            & $0.06$    &  $0.34$             \\
MC size         &&   $0.10$     & $0.10$            & $0.05$    &$0.05$             \\
Fitter          &&   $0.15$     & $0.15$            & $0.15$   & $0.15$             \\ \hline
Total  && $0.30$  & $0.32$                           & $0.19$  & $0.53$     \\ 
\hline \hline
\end{tabular}
\end{center}
\end{table}

\subsection{Form Factor Measurement Results}

\label{ff_results_sec}

The fit described in Sec.~\ref{FFFitingSection} is applied to
the decay rates in Table~\ref{decayrateinbins} for each
semileptonic mode and to pairs of modes related by isospin. Five fits
are carried out per mode. In each case the
normalization parameter $f_+(0)|V_{cq}|$ and one or more form
factor shape parameters are determined. Specifically the shape
parameters are $M_{\rm pole}$~(simple pole model),
$\alpha$~(modified pole model), and $r$ (ISGW2). For the series
parametrization we map the data to the variable $z$. The quantity
$P(z) \phi(z) f_+(z)$ is, by convention, constrained to unity at
$z = z_{\rm max}$, which corresponds to $q^2 = 0$. We fit to the
distribution $ d\Gamma/dq^2 = (G_F^2/ 24 \pi^3) p^3_{K,\pi}|V_{cq}|^2 a_0^2
F_+^2(z)$, where: $F_+(z) = [P(z)\phi(z)]^{-1} \times (1 + r_1 z +
r_2 z^2)$
with $r_1 = a_1/a_0$, $r_2=a_2/a_0$, and $f_+(q^2) = a_0 F_+(z)$.

The fit returns the normalization parameter $f_+(0)|V_{cq}|$
and either $r_1$ or $r_1$ and $r_2$. We test the
sensitivity of the data to the number of parameters and the
convergence of the series. For the series parametrization
the slope at the intercept, $1 + 1/\beta - \delta$, is also reported. Results
of fits to each parametrization are given in
Table~\ref{ffResultsTable-2para} and
Table~\ref{ffResultsTable-3para}.

\begin{figure*}[htb]
\begin{minipage}{6in}
\epsfig{file=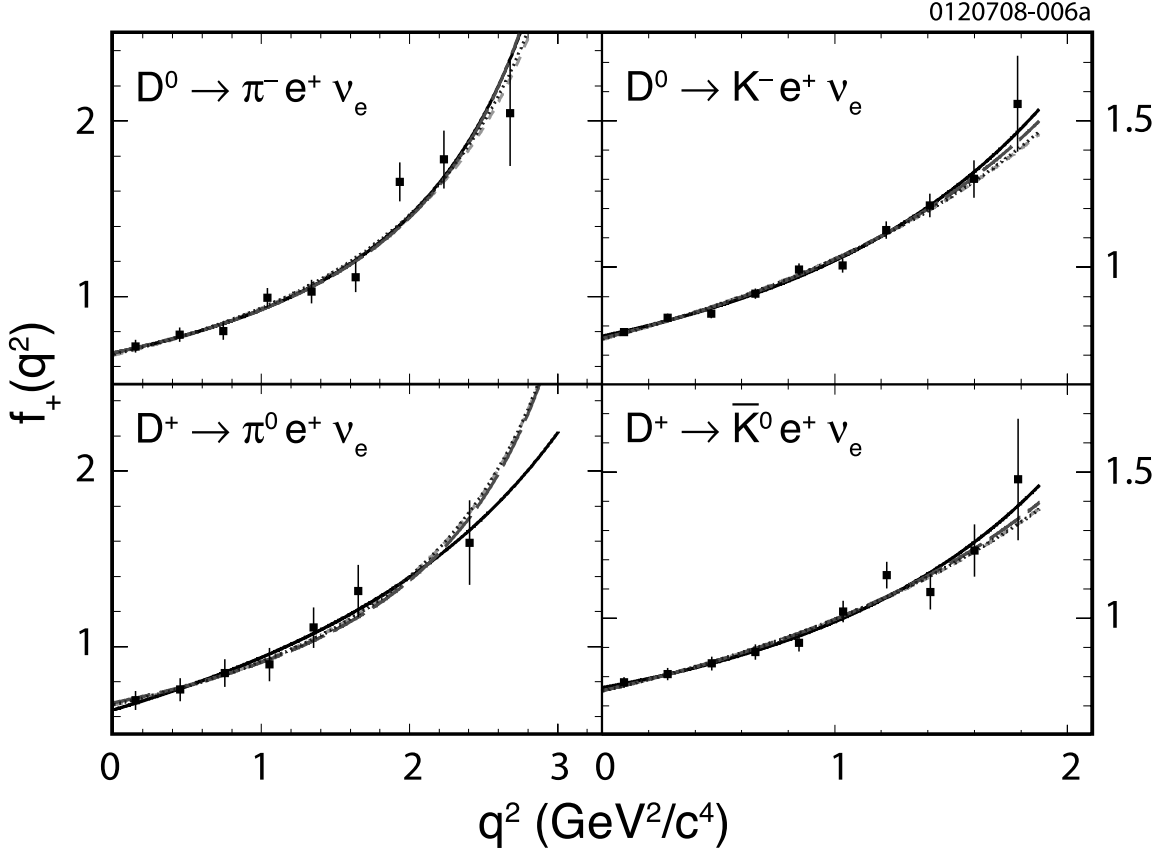,width=6in}
\end{minipage}
\caption{Projections of fits onto $f_+(q^2)$ for each semileptonic
mode.
In each case $f_+(q^2)$ is absolutely normalized.
The data are shown as points with error bars.
The lines are fits
to the simple pole model (long dash),
the modified pole model (short dash), the series
parametrization with two free parameters (dot), and the series
parametrization with three free parameters (solid).
}
\label{ffdistplots}
\end{figure*}

\begin{figure*}[htb]
\begin{minipage}{6in}
\epsfig{file=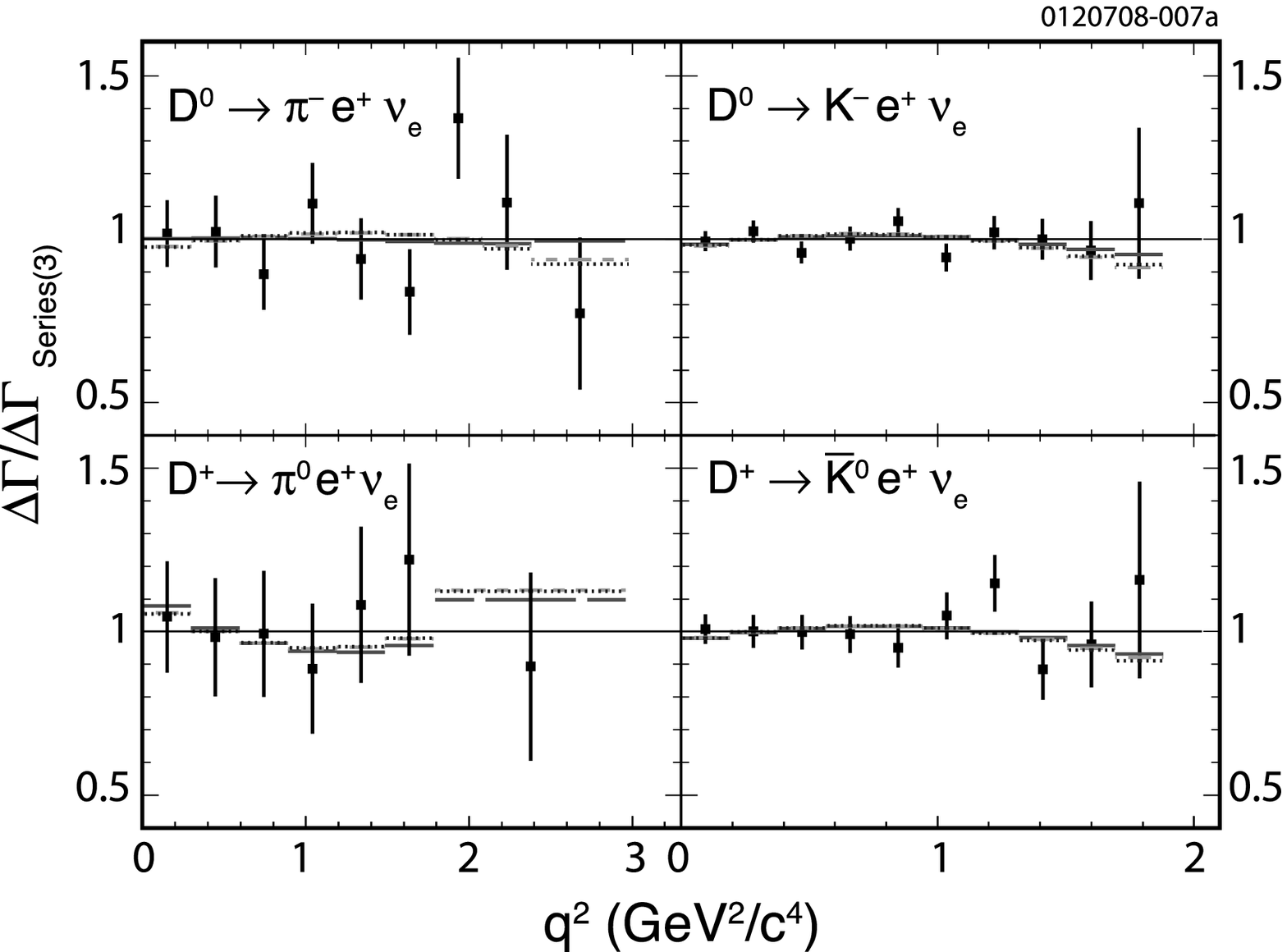,width=6in}
\end{minipage}
\caption{Form factor fit comparison for each semileptonic mode.
All data (squares) and fits (histograms) are normalized to the
relevant three parameter fit result (Series (3) solid line at unity).
The simple pole, modified pole, and two parameter series fit
(Series (2)) are represented by
long dash, short dash, and dotted
histograms, respectively.
}
\label{dGdq2Diffplots}
\end{figure*}

\begin{figure*}[htb]
\begin{minipage}{6in}
\epsfig{file=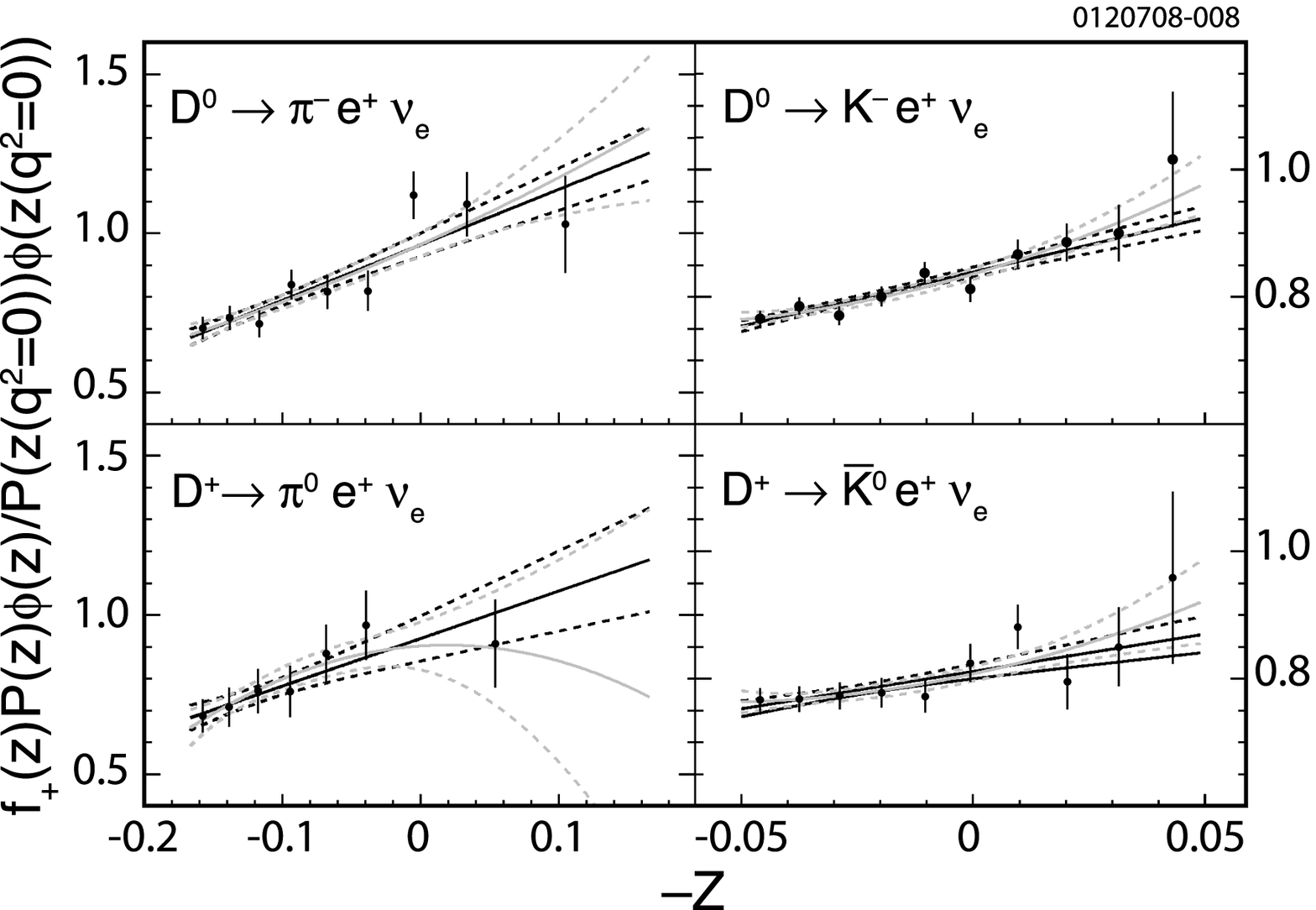,width=6in}
\end{minipage}
\caption{Series parametrization form factor fit comparison to the
data for two and three expansion parameters for each semileptonic
mode. Series(2) (black line) with $\pm 1 \sigma$ uncertainty (black dashed
lines). Series (3) (gray line) with $\pm 1 \sigma$ uncertainty
(gray dashed lines).
} \label{ffphipidistplots}
\end{figure*}

Comparisons of four of the five fits to the
data for each of the four modes are shown in
Fig.~\ref{ffdistplots}~(ISGW2 is excluded). To facilitate a comparison, in
Fig.~\ref{dGdq2Diffplots} we normalize each fit to the result of
the three parameter series fit. It can be seen that each of these
parametrizations provides an adequate, and almost identical,
description of the data when the shape parameter is allowed to be free.
To illustrate the difference between the linear and
quadratic $z$-expansion fits,
Fig.~\ref{ffphipidistplots} shows  $P(z) \phi(z)
f_+(z)/P(z(q^2=0))\phi(z(q^2=0))$ for both as a function of $z$.

\begin{figure*}[htb]
\epsfig{file=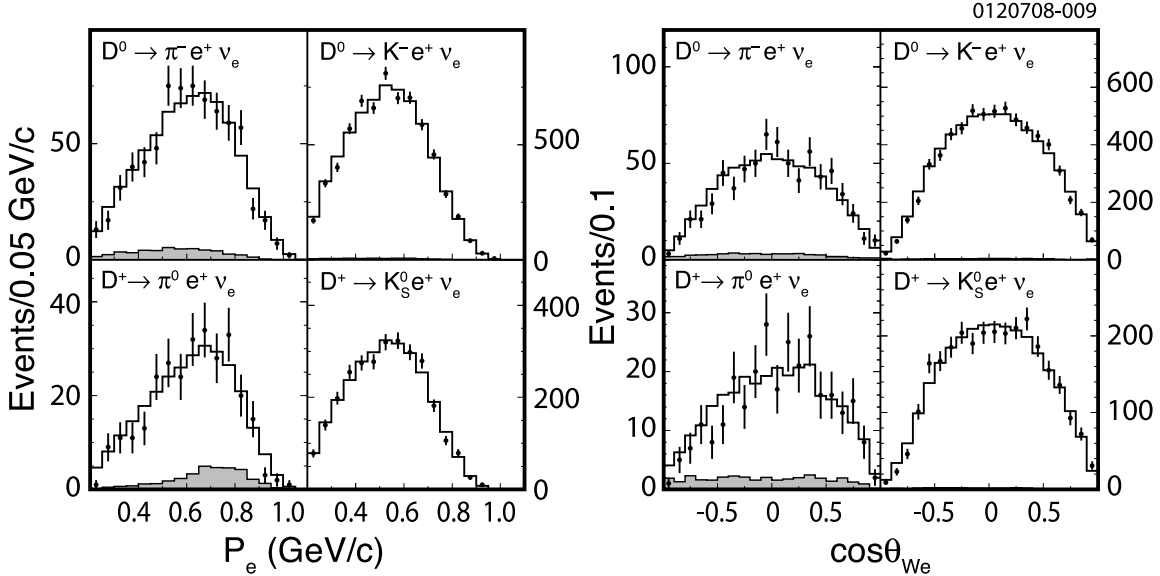,width=6.0in}
\caption{Distributions of positron momenta~(left) and
$\cos{\theta_{We}}$~(right) in data~(points with error bars) and
the projection of the fit~(solid line) for each semileptonic mode.
The
background contributions are the hatched histograms.
}
\label{elecmomentumplots}
\end{figure*}

An independent assessment of the quality of the fits to the data
is obtained from the ability of the fit to describe distributions
in the data in two variables that are not used to constrain the
fit. The first variable is the angle between the $W^+$ in the $D$
meson frame and the positron in the $W^+$ frame, $\theta_{We}$.
The second variable is the laboratory momentum of the
positron, $|\vec{p}_e|$. Figure~\ref{elecmomentumplots} shows
distributions for $\cos{\theta_{We}}$ and $|\vec{p}_e|$ in data and the projections of the
fit, where the background contributions are shown as
hatched histograms. The fits describe the distributions
in these two variables well.

\begingroup
\squeezetable
\begin{table*}[htpb]
\caption{Summary of results of form factor fits to the data: the
normalization parameters $f^K_+(0)|V_{cs}|$ or
$f^{\pi}_+(0)|V_{cd}|$, shape parameters~($M_{\rm pole}$,
$\alpha$, $r$, and $r_1$), and the correlation coefficient~$\rho$
between the normalization and shape parameters for each fit. The
last column gives the $\chi^2$ per degree of freedom for each fit.
} \label{ffResultsTable-2para}
\begin{center}
\begin{tabular}{lccccc}
\hline \hline
\rule[-1mm]{-1mm}{4mm}
Simple pole                     & $f_+(0)|V_{cq}|$  & $M_{\rm pole}$ & $\rho$ & $\chi^2$ per d.o.f. \\
\hline\rule[-1mm]{-1mm}{4mm}
$D^0 \rightarrow \pi^- e^+ \nu_e$     &  0.152(4)(1)        & 1.94(4)(1)  & 0.68   &  1.26       \\
$D^+ \rightarrow \pi^0 e^+ \nu_e$     &  0.153(7)(5)        &1.99(10)(5) & 0.68   &  0.37       \\
$D^0 \rightarrow K^- e^+ \nu_e$       &  0.736(7)(6)        & 1.95(4)(1)  & 0.78   &  0.81       \\
$D^+ \rightarrow \bar{K}^0 e^+ \nu_e$ &  0.733(11)(9)       & 2.02(6)(2)  & 0.78   &  0.80       \\\hline

$D \rightarrow \pi e^+ \nu_e$         &  0.152(4)(1)        & 1.95(4)(2)  & 0.68   &  0.82       \\
$D \rightarrow K e^+ \nu_e$           &  0.735(7)(5)        & 1.97(3)(1)  & 0.78   &  1.00       \\
\hline \hline
\rule[-1mm]{-1mm}{4mm}
Mod.pole                              & $f_+(0)|V_{cq}|$  & $\alpha$ & $\rho$ & $\chi^2$ per d.o.f.     \\
\hline\rule[-1mm]{-1mm}{4mm}
$D^0 \rightarrow \pi^- e^+ \nu_e$     &  0.150(6)(1)        & 0.18(12)(4) & -0.83    &  1.26      \\
$D^+ \rightarrow \pi^0 e^+ \nu_e$     &  0.151(9)(4)        &0.09(22)(12) & -0.80   & 0.35    \\
$D^0 \rightarrow K^- e^+ \nu_e$       &  0.733(8)(6)        & 0.25(6)(2)  & -0.83    &  0.94       \\
$D^+ \rightarrow \bar{K}^0 e^+ \nu_e$ &  0.732(12)(9)       & 0.12(10)(4) & -0.82    &  0.83       \\\hline

$D \rightarrow \pi e^+ \nu_e$         &  0.150(5)(2)        & 0.16(10)(5)  & -0.81   &  0.78       \\
$D \rightarrow K e^+ \nu_e$           &  0.733(7)(6)        & 0.21(5)(2)   & -0.83   &  1.01       \\
\hline \hline
\rule[-1mm]{-1mm}{4mm}
ISGW2                                & $f_+(0)|V_{cq}|$  & $r(ISGW2)$       & $\rho$ & $\chi^2$ per d.o.f. \\
\hline\rule[-1mm]{-1mm}{4mm}
$D^0 \rightarrow \pi^- e^+ \nu_e$     &  0.147(5)(1)        & 1.98(12)(2) & -0.80   &  1.37      \\
$D^+ \rightarrow \pi^0 e^+ \nu_e$     &  0.149(8)(4)        & 1.85(22)(12) & -0.77   &  0.26       \\
$D^0 \rightarrow K^- e^+ \nu_e$       &  0.730(8)(6)        & 1.56(4)(1)  & -0.81   &  1.07       \\
$D^+ \rightarrow \bar{K}^0 e^+ \nu_e$ &  0.723(12)(9)       & 1.48(7)(2)  & -0.81   &  0.91       \\\hline

$D \rightarrow \pi e^+ \nu_e$         &  0.147(4)(2)        & 1.95(10)(5)  & -0.80   &  0.80       \\
$D \rightarrow K e^+ \nu_e$           &  0.730(7)(6)        & 1.53(4)(2)   & -0.81   &  1.09       \\
\hline \hline
\rule[-1mm]{-1mm}{4mm}
Series (2 param.)                         & $f_+(0)|V_{cq}|$ & $r_1$  & $1+1/\beta - \delta$ & $\rho$ & $\chi^2$ per d.o.f. \\
\hline\rule[-1mm]{-1mm}{4mm}
$D^0 \rightarrow \pi^- e^+ \nu_e$     &  0.150(6)(1)        & -1.80(27)(5)  & 1.00(11)(2) & 0.84   &  1.26      \\
$D^+ \rightarrow \pi^0 e^+ \nu_e$     &  0.151(9)(4)        & -1.57(49)(26)  & 0.92(17)(9) & 0.81   &  0.36       \\
$D^0 \rightarrow K^- e^+ \nu_e$       &  0.734(8)(6)        & -1.96(28)(8)  & 0.89(5)(2) & 0.83   &  0.88      \\
$D^+ \rightarrow \bar{K}^0 e^+ \nu_e$ &  0.733(12)(9)       & -1.40(44)(14)  & 0.79(7)(3) & 0.82   &  0.81       \\\hline

$D \rightarrow \pi e^+ \nu_e$     &  0.151(5)(2)            & -1.75(11)(23)  & 0.99(9)(4) & 0.83   &  0.78      \\
$D \rightarrow K e^+ \nu_e$       &  0.734(7)(6)            & -1.78(24)(10)  & 0.86(4)(2) &  0.83   &  0.98      \\
\hline \hline
\end{tabular}
\end{center}
\end{table*}
\endgroup

\begingroup
\squeezetable
\begin{table*}[htbp]
\caption{Summary of the results of form factor fits for the series parametrization with three
parameters to the data: the normalization parameters
$f^K_+(0)\vcs$ or $f^{\pi}_+(0)\vcd$, the shape
parameters $r_1$, and $r_2$, and correlation
coefficients~$\rho_{ij}$ between the parameters determined by the
fit. The last column gives the $\chi^2$ per degree of freedom for
each fit.
} \label{ffResultsTable-3para}
\begin{center}
\begin{tabular}{lcccccccc}
\hline \hline
\rule[-1mm]{-1mm}{4mm}
Decay                                 &
$f_+(0)|V_{cq}|$   & $r_1$       &  $r_2$     & $1+1/\beta - \delta$
                                      &  $\rho_{01}$     & $\rho_{02}$ & $\rho_{12}$ &  $\chi^2$ per d.o.f. \\
\hline\rule[-1mm]{-1mm}{4mm}

$D^0 \rightarrow \pi^- e^+ \nu_e$     &  0.152(8)(1)     &
-2.0(6)(1) & 1.6(4.2)(0.8)    & 0.91(24)(5)
                                      &  -0.36            &  0.67       & -0.91        & 1.45      \\
$D^+ \rightarrow \pi^0 e^+ \nu_e$     &  0.144(12)(4)    &
-0.3(1.8)(1.0) & -7.9(10.5)(5.6)  & 1.36(43)(23)
                                      &  -0.46            & 0.68        & -0.96        & 0.29       \\

$D^0 \rightarrow K^- e^+ \nu_e$       &  0.745(12)(6)    &
-2.4(5)(2) & 15.6(12.8)(3.8)  & 0.70(15)(5)
                                      &  -0.26           &  0.71       &  -0.82       & 0.79     \\
$D^+ \rightarrow \bar{K}^0 e^+ \nu_e$ &  0.744(17)(9)    &
-1.9(7)(2) & 16.6(19.3)(6.2)  & 0.61(21)(7)
                                      &  -0.22            & 0.71        &  -0.80       & 0.82     \\\hline

$D \rightarrow \pi e^+ \nu_e$         &  0.151(7)(2)     &
-1.8(6)(3) & 0.3(3.9)(1.7)  & 0.98(23)(10)
                                      &  -0.38            &  0.67       & -0.92        & 0.84      \\
$D \rightarrow K e^+ \nu_e$           &  0.744(10)(6)    &
-2.2(4)(2) & 16.9(11.4)(4.7)  & 0.69(12)(5)
                                      &  -0.25            &  0.71       &  0.81       & 0.93     \\

\hline \hline
\end{tabular}
\end{center}
\end{table*}
\endgroup

Using the ISGW2 parametrization we determine the isospin
conjugate average values of the meson radius to be
\beqn
r^K = 1.53(4)(2) {\rm ~GeV}^{-1},
\eeqn
\beqn
r^\pi = 1.95(10)(5) {\rm ~GeV}^{-1}.
\eeqn
They are the most precise measurements of these quantities to
date, and are $18 \sigma$ and $5 \sigma$ from the ISGW2 expected
values $r^K=1.12{\rm ~GeV^{-1}}$ and $r^\pi=1.410 {\rm
~GeV^{-1}}$, respectively. We have assigned no uncertainty to the
theoretical prediction, and assume here and in what follows, that
the experimental uncertainties derived from the fit are Gaussian
distributed. A comparison to other recent measurements is given in
Table~\ref{tab:ISGUR-compare}. The measurements by BABAR and Belle
disagree by $3.8 \sigma$. Our measurement of $r^K$ is over $4 \sigma$ smaller than Belle, and $ 1.6 \sigma$ smaller than BABAR.

\begin{table}[htbp]
\caption{Compilation of recent measurements  of the ISGW2
parameters $r^K$ and $r^\pi$ in units of GeV$^{-1}$. }
\begin{center}
\begin{tabular}{lcc}\hline\hline\rule[-1mm]{-1mm}{4.5mm}
               & $r^K$  & $r^\pi$                        \\\hline
Belle~\cite{BELLE-06}     &   2.47(15)(15) &  2.68(45)(40)     \\
BABAR~\cite{BABAR-06}     &   1.645(36)(44)   &  --           \\
\cleoc (tagged) &  1.53(4)(2) &   1.95(10)(5)     \\
\hline\hline
\end{tabular}
\end{center}
\label{tab:ISGUR-compare}
\end{table}

Using the simple pole model, we determine the isospin conjugate
average pole masses to be
\beqn
M_{\rm pole}^K  =  1.97(3)(1) {\rm ~GeV}/c^{2},
\eeqn
\beqn
M_{\rm pole}^\pi  = 1.95(4)(2)  {\rm ~GeV}/c^{2}.
\eeqn
These values differ by  $4.6 \sigma$ and $1.3 \sigma$ from the
well-measured masses: $M_{D_s^{*+}}=2112.0 \pm 0.6 ~{\rm MeV}/c^2$
and $M_{D^{*+}}= 2010.0 \pm 0.4~{\rm MeV}/c^2$~\cite{PDG2006},
respectively. Comparison to previous measurements are given in
Table~\ref{tab:mpole-comp-kenu}~and
Table~\ref{tab:mpole-comp-pienu}. For $D \ra K \ell^+ \nu_{\ell}$
all of the
more recent measurements are below the mass of the $D_s^*$ meson.
Our measurement of
$M_{\rm pole}^K$ is in excellent agreement with Ref.~\cite{Nadia}, but is $2.3\sigma$ larger than the BABAR~\cite{BABAR-06} measurement, and $2.5 \sigma$ larger than the Belle~\cite{BELLE-06} measurement.
For $D \ra \pi \ell^+ \nu_{\ell}$ all measurements are much less
precise and are in reasonable agreement, albeit within large uncertainties.
All measurements are below the mass of the $D^*$ meson.

\begin{table}[htbp]
\caption{Compilation of measurements  of $M_{\rm pole}$ in $ D \ra K
\ell \nu_{\ell} $. \cleoc (tagged) is the isospin averaged
value; \cleoc (untagged) is for $ D^0 \ra K^- e^+ \nu_{e}$. }
\begin{center}
\begin{tabular}{lc}\hline\hline\rule[-1mm]{-1mm}{4.5mm}
               & $M_{\rm pole}^K$ ${\rm GeV/c^2}$                       \\\hline
Mark III~\cite{MARKIII-91}      & 1.80$_{-0.20}^{+0.50}$(25)   \\
E691~\cite{E691-89}            &   2.10$_{-0.20}^{+0.40}$(20)    \\
CLEO~\cite{CLEO-91}        &   2.10$_{-0.20}^{+0.40}$(25)   \\
CLEOII~\cite{CLEOII-93}         &   2.00(12)(18)        \\
E687 (Tag)~\cite{E687-95}         &   1.97$_{-0.22}^{+0.43}$(7)   \\
E687 (Incl)~\cite{E687-95}        &   1.87$_{-0.08}^{+0.11}$(7)   \\
CLEO~\cite{CLEOIII}        &   1.89(5)$^{+0.04}_{-0.03}$          \\
FOCUS~\cite{FOCUS-05}        &  1.93(5)(3)           \\
Belle~\cite{BELLE-06}     &     1.82(4)(3)           \\
BABAR~\cite{BABAR-06}     &     1.884(12)(15)           \\
\cleoc (tagged) &     1.97(3)(1)         \\
\cleoc (untagged)~\cite{Nadia}     &   1.97(3)(1)           \\
\hline\hline
\end{tabular}
\end{center}
\label{tab:mpole-comp-kenu}
\end{table}

\begin{table}[htbp]
\caption{Compilation of measurements of $M_{\rm pole}$ in $D \ra \pi
\ell \nu_{\ell}$. \cleoc (tagged) is the isospin average value;
\cleoc (untagged) is for $ D^0 \ra \pi^- e^+ \nu_{e}$.}

\begin{center}
\begin{tabular}{lcc}\hline\hline\rule[-1mm]{-1mm}{4mm}
               & $M_{\rm pole}^\pi$ ${\rm GeV/c^2}$          \\\hline
CLEO (2004)~\cite{CLEOIII}        &  1.86$_{-0.06}^{+0.10}$(5) \\
FOCUS (2004~\cite{FOCUS-05})        &   1.91$_{-0.15}^{+0.30}$(7) \\
Belle (2006)~\cite{BELLE-06}     &     1.97(8)(4)  \\
\cleoc (tagged) &    1.95(4)(2)  \\
\cleoc (untagged)~\cite{Nadia}    &   1.87(3)(1)  \\
\hline\hline
\end{tabular}
\end{center}
\label{tab:mpole-comp-pienu}
\end{table}

Using the modified pole model, we determine the isospin conjugate
average shape parameters to be
\beqn
\alpha^K  =  0.21(5)(2),
\eeqn
\beqn
\alpha^\pi  = 0.16(10)(5).
\eeqn
The values of $\alpha^K$ and $\alpha^\pi$ are $ 27 \sigma$ and $11
\sigma$, respectively, from the values of $\sim 1.75$ and $1.34$ required by
the BK parametrization. A comparison to previous measurements is
given in Table~\ref{tab:alpha-comp}. For $ K e^+ \nu_e$ there is
excellent agreement between this result and
\cleoc (untagged)~\cite{Nadia}, good agreement with previous measurements
by CLEO III~\cite{CLEOIII}, and
FOCUS~\cite{FOCUS-05}, and QCD sum rules~\cite{QCD-sum-rules-1},
but our result is lower than BABAR~\cite{BABAR-06} by $2.6
\sigma$, lower than Belle~\cite{BELLE-06} by $2.7 \sigma$  and lower than the LQCD
fit~\cite{unquenched_LQCD} by $4.2 \sigma$. The significance of the discrepancy
between our result and the LQCD fit cannot be quantified
rigorously, as the covariance matrix for the LQCD form factor is
lost during the chiral
extrapolation~\cite{unquenched_LQCD}. For $\pi e^+ \nu_e$ there is
reasonable agreement with \cleoc (untagged)~\cite{Nadia} and other previous measurements, albeit within large uncertainties. Our measurement of
$\alpha^\pi$ is $2.4 \sigma$ smaller than the LQCD fit.

\begin{table}[htbp]
\caption{Compilation of measurements and theoretical predictions
for $\alpha^K$ and $\alpha^\pi$. \cleoc (tagged) are the
isospin average values; \cleoc (untagged) is for $ D^0 \ra K^- e^+
\nu_{e}$ and $ D^0 \ra \pi^- e^+ \nu_{e}$ respectively.}

\begin{center}
\begin{tabular}{lcc}\hline\hline\rule[-1mm]{-1mm}{4.5mm}
               & $\alpha^K $            & $\alpha^{\pi} $           \\\hline
FOCUS~\cite{FOCUS-05}   &0.28(8)(7)  &       --                       \\
CLEO III~\cite{CLEOIII}       &0.36(10)(5)  & 0.37(25)(15)    \\
Belle~\cite{BELLE-06}   &0.52(8)(6)  &0.10(21)(10)    \\
BABAR~\cite{BABAR-06}   &0.377(23)(29) &     --                          \\
LQCD~\cite{unquenched_LQCD}           &0.50(4)   &0.44(4)     \\
LCSR~\cite{QCD-sum-rules-1} & $ 0.07^{+0.15}_{-0.07} $  & $0.01^{+0.11}_{-0.07}$  \\
CQM~\cite{CQM-alpha} & 0.24 & 0.30 \\
\cleoc (tagged)   &0.21(5)(2) &0.16(10)(5)   \\
\cleoc (untagged)~\cite{Nadia} &0.21(5)(3)  &0.37(8)(3)   \\
\hline
\end{tabular}
\end{center}
\label{tab:alpha-comp}
\end{table}

Fits to the data using the first two terms of the $z$ expansion
are reported in Table~\ref{ffResultsTable-2para}.
Fits using the first three terms are given in
Table~\ref{ffResultsTable-3para} and shown in
Fig.~\ref{ffphipidistplots}. The expansion parameters are not
predicted. The central value of the ratio of expansion parameters
$r_2$ is an order of magnitude larger than $r_1$, however the
statistical uncertainty is of similar magnitude to the central
value, and therefore no statement can be made about the
convergence of the expansion. Moreover, the data lack the
precision, even in the copious $D \ra K e^+ \nu_e$ mode, to
determine $r_2$. For this reason there is no appreciable
difference between the probability of the $\chi^2$ between the two
parameter series expansion and three parameter series expansion
fits for any mode. The compatibility of the data with linear
dependence is consistent with the modified pole ansatz for
$f_+(q^2)$.
Recently BABAR~\cite{BABAR-06} using a data sample of
75,000 $\dzke$ events, found
$r_1 = -2.5 (2)(2)$ and $ r_2= 0.6(6)(5)$, and that the differential rate is well-described
by the $z$ expansion with only a linear term.
The results reported here for $r_1$ and $r_2$
are in excellent agreement with
\cleoc (untagged)~\cite{Nadia} and agree with BABAR~\cite{BABAR-06} to better than $2 \sigma$ with the precise level depending on the correlation coefficient for the BABAR $r_1$ and $r_2$ parameters.

The quadratic series expansion fit returns
isospin conjugate average values for $1 + 1/\beta - \delta$ of
0.69(12)(5) and  0.98(23)(10)
for $D \ra K e^+ \nu_e$ and $D \ra \pi e^+ \nu_e$, respectively.
These values are $ 10 \sigma$ and $4 \sigma$  from the value of
$\sim 2$ required by the BK parametrization, and are consistent
with the results in~\cite{Nadia} given in
Table~\ref{tab:ratio-expansion-slope-comp}.

\begin{table}[htbp]
\caption{Compilation of measurements of  $1 + 1/\beta - \delta$ from
\cleoc tagged and untagged~\cite{Nadia}.}

\begin{center}
\begin{tabular}{lcc}\hline\hline
 &  \multicolumn{2}{c}{$ 1 + 1/\beta - \delta $} \\
 &   Tagged  & Untagged~\cite{Nadia}
\\\hline\rule[-1mm]{-1mm}{4mm}
$D^0 \ra \pi^- e^+ \nu_e$   &0.91(24)(5)  & 1.30(37)(12) \\
$D^+ \ra \pi^0 e^+ \nu_e$   &1.36(43)(23)  & 1.58(60)(13) \\
$D^0 \ra K^- e^+ \nu_e $ & 0.70(15)(6)  & 0.62(13)(4)  \\
$D^+ \ra \bar{K}^0 e^+ \nu_e$ & 0.61(21)(7)  &  0.51(20)(4) \\\hline
$D \ra \pi e^+ \nu_e$ &  0.98(23)(10) & -- \\
$D \ra K e^+ \nu_e$ & 0.69(12)(5) & -- \\
 \hline
\end{tabular}
\end{center}
\label{tab:ratio-expansion-slope-comp}
\end{table}

When the shape parameters are not fixed the $q^2$
parametrizations of the simple pole model, the modified pole
model, the ISGW2 model, and the series expansion with two and
three parameters are functionally almost identical over the $q^2$
range accessible in $D$ meson semileptonic decay.  For this reason
each parametrization is able to describe the data with a
comparable $\chi^2$ probability.

Measurements of  $f_+(0) |V_{cq}|$ are given in
Table~\ref{ffResultsTable-2para} for the ISGW2, simple pole,
modified pole, and two parameter series parametrization, and in
Table~\ref{ffResultsTable-3para} for the three parameter series
parametrization. As each parametrization is able to describe the
data, measurements of $f_+(0) |V_{cq}|$ are very similar
among parametrizations. For $ D \ra K e^+ \nu_e$ the values of
$f^K_+(0) |V_{cs}|$ span about one half of a statistical sigma
between the pole model, modified pole model, and series expansion
(linear). However, the fit to the series expansion including a quadratic
term returns a value of $f^K_+(0) |V_{cs}|$ one statistical
sigma larger than for the series expansion using a linear term.
The statistical uncertainty is also increased by one third.

For $ D \ra \pi e^+ \nu_e$ the values of $f^{\pi}_+(0) |V_{
cd}|$ span a statistical sigma among the pole model, modified pole
model, and the series expansion (linear). The fit to the series
expansion including a quadratic term returns a value of
$f^{\pi}_+(0) |V_{cd}|$ that only differs in the least
significant digit from the value obtained  for the series
expansion using a linear term, but the statistical uncertainty is
increased by one third.

Using $|V_{cs}| = 0.97334 \pm 0.00023$ and $ |V_{cd}| =
0.2256 \pm 0.0010$
obtained using CKM unitarity constraints~\cite{PDG2008}, we
calculate $f_+(0)$ for each semileptonic mode separately and also
for isospin averages. These are presented in
Table~\ref{f0resultsTable} and compared to previous measurements
in Table~\ref{tab:fzero-comp}~\cite{note:newVcx}.
The measurement of $f_+^\pi(0)$
presented here is the most precise to date.

\begingroup
\squeezetable
\begin{table*}[htbp]
\caption{ Results for  $f_+(0)$  obtained from fits to five form
factor parametrizations.
          The first uncertainty is statistical, the second is systematic and the third
          uncertainty is from the relevant CKM matrix element.
}. \label{f0resultsTable}
\begin{center}
\begin{tabular}{lcccccccc}
\hline \hline Mode                                  & Simple Pole
& Mod. Pole   &  ISGW2
                                      & Series (2~param.)&  Series (3~param.) \\
\hline\rule[-1mm]{-1mm}{4mm}
$D^0 \rightarrow \pi^- e^+ \nu_e$     &  0.676(20)(6)(3) & 0.666(25)(6)(3) & 0.651(23)(6)(3) & 0.667(26)(6)(3) & 0.675(34)(6)(3)  \\
$D^+ \rightarrow \pi^0 e^+ \nu_e$     &  0.678(32)(17)(3) & 0.670(39)(17)(3) & 0.660(35)(17)(3) & 0.672(40)(17)(3) & 0.640(57)(16)(3) \\
$D^0 \rightarrow K^- e^+ \nu_e$       &  0.756(7)(6)(0)& 0.753(8)(6)(0) & 0.750(8)(6)(0) & 0.755(8)(6)(0)  & 0.765(12)(7)(0) \\
$D^+ \rightarrow \bar{K}^0 e^+ \nu_e$ &  0.753(11)(9)(0)& 0.752(12)(9)(0)& 0.748(12)(9)(0)& 0.753(13)(9)(0) & 0.764(18)(10)(0) \\
\hline
$D \rightarrow \pi e^+ \nu_e$     &  0.676(17)(7)(3) & 0.667(21)(7)(3) & 0.653(19)(7)(3) & 0.668(21)(7)(3) & 0.669(29)(7)(3)  \\
$D \rightarrow K e^+ \nu_e$           &  0.756(6)(6)(0)& 0.753(7)(6)(0) & 0.750(7)(6)(0) & 0.754(7)(6)(0)  & 0.764(10)(6)(0) \\

\hline \hline
\end{tabular}
\end{center}
\end{table*}
\endgroup

\begin{table}[htbp]
\caption{Compilation of measurements and theoretical predictions
for $f_+^K(0)$ and $f_+^\pi(0)$. For the experimental measurements the first uncertainty is statistical, and the second is systematic. For this work the third
          uncertainty is from the relevant CKM matrix element. For BABAR the third uncertainty includes
contributions from ${\cal B}(D^0 \rightarrow K^- \pi^+)$, $\tau_{D^0}$ and $|V_{cs}|$.
}

\begin{center}
\begin{tabular}{lcc}\hline\hline\rule[-1mm]{-1mm}{4.5mm}
                               &$f_+^K(0)$                   & $f_+^\pi(0)$ \\\hline
LQCD1~\cite{Abada}                   &0.66(4)(1)    & 0.57(6)(2)     \\
QCD SR~\cite{QCD-sum-rules-1}    &0.60(2)           & 0.50(1)       \\
LCSR1~\cite{QCD-sum-rules-2} &0.785(11)        & 0.65(11)      \\
LCSR2~\cite{WWZ}           &0.67(20)       & 0.67(19)      \\
ISGW2~\cite{ISGW2}             &1.23                         & --    \\
LQCD2~\cite{unquenched_LQCD}       &0.73(3)(7)   & 0.64(3)(6)     \\
Belle~\cite{BELLE-06}           &0.695(7)(22) & 0.624(20)(30) \\
BABAR~\cite{BABAR-06}           &0.727(7)(5)(7) & -- \\
\cleoc (tagged)             &0.764(10)(6)(0)&0.669(29)(7)(3) \\
\cleoc {average}            & 0.763(7)(6)(0) & 0.629(22)(7)(3) \\
\hline\hline
\end{tabular}
\end{center}
\label{tab:fzero-comp}
\end{table}

\begingroup
\squeezetable
\begin{table*}[htbp]
\caption{ Results for  $|V_{cs}|$ and $|V_{cd}|$ obtained from
fits to five form-factor parametrizations.
          The first uncertainty is statistical, the second is systematic
          and the third is from the $f_+(0)$ LQCD prediction.}
\begin{center}
\begin{tabular}{lcccccccc}
\hline \hline Decay               & Simple Pole& Mod. Pole & ISGW2
                    & Series (2 param.)  & Series (3 param.) \\

\hline\rule[-1mm]{-1mm}{4mm}
$D^0 \rightarrow \pi^- e^+ \nu_e$     &  0.238(7)(2)(25)  &
0.235(9)(2)(25) & 0.230(8)(2)(24) & 0.235(9)(2)(25)
                                      &  0.238(12)(2)(25) \\
$D^+ \rightarrow \pi^0 e^+ \nu_e$     &  0.239(11)(6)(25) &
0.236(14)(6)(25) & 0.233(12)(6)(24) & 0.236(14)(6)(25)
                                      &  0.226(20)(6)(24) \\
$D^0 \rightarrow K^- e^+ \nu_e$       &  1.008(10)(9)(105)&
1.004(11)(9)(105) & 1.000(11)(9)(104) & 1.006(11)(8)(105)
                                      &  1.020(16)(9)(106) \\
$D^+ \rightarrow \bar{K}^0 e^+ \nu_e$ &  1.004(15)(13)(104)&
1.003(17)(13)(104)&0.997(16)(13)(104)& 1.004(17)(13)(105)
                                      &  1.019(24)(13)(106) \\
\hline
$D \rightarrow \pi e^+ \nu_e$     &  0.238(6)(2)(25)  &
0.235(7)(3)(24) & 0.230(7)(3)(24)  & 0.234(8)(2)(25)
                                      &  0.236(10)(2)(25) \\
$D \rightarrow K e^+ \nu_e$       &  1.007(8)(8)(105)&
1.004(9)(8)(105) & 0.999(9)(8)(104) & 1.006(9)(8)(105)
                                      &  1.019(13)(9)(106) \\
\hline \hline
\end{tabular}
\end{center}
\label{VcqResultsTable}
\end{table*}
\endgroup

\section{\boldmath Determination of $|V_{cs}|$ and $|V_{cd}|$}

\label{overview}

Using recent unquenched LQCD calculations of the form factor
normalizations~\cite{unquenched_LQCD}
we obtain $|V_{cq}|$ for each of the four
semileptonic modes and for the isospin averages. These are
presented in Table~\ref{VcqResultsTable} for both pole models, the
ISGW2 model and the series expansion with two and three
parameters.

As the data do not support the physical interpretation of the
shape parameter in the ISGW2, simple pole, and modified pole
parametrizations we choose the value of $|V_{cq}|$ obtained
with the series expansion as our main result. Although the
$|V_{cq}|$ statistical uncertainty is one third larger when
data is fit to the series expansion with three parameters, we
choose this rather than the results obtained with the fit to two
parameters to facilitate comparison with~\cite{Nadia}.

We find $ |V_{cd}|= 0.238(12)(2)(25)$  for $D^0 \ra \pi^- e^+
\nu_e$, and $|V_{cd}|= 0.226 (21)(6)(24)$ for $D^+ \ra \pi^0
e^+ \nu_e$. We find $|V_{cs}| = 1.020(16)(9)(106)$ for $D^0
\ra K^- e^+ \nu_e$, and $|V_{cs}| = 1.019(24)(13)(106)$ for
$D^+ \ra \bar{K}^0 e^+ \nu_e$. In each case the third
uncertainty in the determination of the CKM matrix element is from
theory.  Averaging the $D^0$ and $D^+$ results and taking into
account correlated and uncorrelated uncertainties we find
\beqn
|V_{cd}| = 0.236 \pm 0.010 \pm 0.002 \pm 0.025
\eeqn
\noindent and
\beqn
|V_{cs}| = 1.019 \pm 0.013 \pm 0.009 \pm 0.106,
\eeqn
where the uncertainties are statistical, systematic and
theoretical, respectively. The theoretical uncertainty dominates
and is expected to be reduced soon.  We compare our measurements
to other determinations in Table~\ref{tab:Vcs-comp} and
Table~\ref{tab:Vcd-comp}. Our determination of $|V_{cs}|$ is
consistent with previous measurements, is in good agreement with~\cite{Nadia},
and is the most precise to date. Our determination of
$\vcd$ is in good agreement with the result derived from
neutrino-nucleon scattering, it is consistent with~\cite{Nadia}
and is the most precise determination from $D$ meson semileptonic
decay to date.

\begin{table}[htbp]
\caption{Compilation of determinations of $|V_{cs}|$. For the
$\Gamma(K \ell \nu_{\ell})$ determination we use
PDG2000~\cite{PDG2000}, as PDG2004~\cite{PDG2004} does not quote a
value from this technique and subsequent PDG determinations~\cite{PDG2006, PDG2008} include a
result obtained from an earlier \cleoc measurement, the data
sample for which is a subset of the data used in this work.}

\begin{center}
\begin{tabular}{lcc}\hline\hline
                               & $|V_{cs}|$                  \\\hline
$\Gamma(K \ell \nu_{\ell})$ PDG2000~\cite{PDG2000}  & $1.04 \pm 0.16$       \\
Charm tagged $W$ decay~\cite{PDG2006}    & $0.94^{+0.32}_{-0.26}\pm 0.14$      \\
$\Gamma(K \ell \nu_{\ell})$BESII~\cite{BESIIVcs}  & $1.00 \pm0.05\pm 0.11$       \\
\cleoc (tagged) &  $1.019 \pm 0.013 \pm 0.009 \pm 0.106$     \\
\cleoc (untagged)  &  $1.015 \pm 0.010 \pm 0.011 \pm 0.106 $ \\
\cleoc {average} & $1.018 \pm 0.010 \pm 0.008 \pm 0.106$ \\
 \hline \hline
\end{tabular}
\end{center}
\label{tab:Vcs-comp}
\end{table}

\begin{table}[htbp]
\caption{Comparison of determinations of $\vcd$. }

\begin{center}
\begin{tabular}{lcc}\hline\hline
                               & $|V_{cd}|$                  \\\hline
$\nu d  \ra c d$~\cite{PDG2006}    & $0.22\pm 0.011$      \\
\cleoc (tagged) & $0.236 \pm 0.010 \pm 0.002 \pm 0.025$     \\
\cleoc (untagged)  &  $0.217 \pm 0.009 \pm 0.004 \pm 0.023 $ \\
\cleoc {average} & $0.222\pm 0.008 \pm 0.003 \pm 0.023$ \\
 \hline \hline
\end{tabular}
\end{center}
\label{tab:Vcd-comp}
\end{table}

We also extract the ratio $|V_{cd}|/|V_{cs}|$ from the
ratio of measured form factors. From the simultaneous quadratic
$z$ expansion fits to isospin conjugate pairs
we obtain:
\beqn
\frac{|V_{cd}|f_+^\pi(0)}{\rule[-1mm]{-1mm}{4.5mm} |V_{cs}|f_+^K(0)} = 0.203 \pm 0.009
\pm 0.003,
\eeqn
where the uncertainties are statistical and systematic,
respectively, and correlations have been taken into account. We
can compare this result to calculations of $f_+^\pi(0)/f_+^K(0)$
to obtain the ratio of CKM elements. A recent light cone sum rules
(LCSR) calculation obtains~\cite{Patricia}  $f_+^\pi(0)/f_+^K(0) =
0.84 \pm 0.04$, from which we find
\beqn
  \frac{|V_{cd}|}{|V_{cs}|} = 0.242 \pm 0.011 \pm  0.004
  \pm 0.012,
\eeqn
where the third uncertainty is from LCSR. This value is in reasonable agreement with~\cite{Nadia}.

\section{CLEO-\lowercase{c} Averages}

\label{average}

In this section we compute average values of the measurements of
branching fractions, form factors and $\vcs$ and $\vcd$, obtained
in this work (tagged), with previous untagged \cleoc measurements of the same
quantities~\cite{Nadia}. These average values represent the best
determinations of the branching fractions, form factors, and $\vcs$
and $\vcd$ with the \cleoc 281~pb$^{-1}$ data set.

The analysis of the data, both in this work and in
Ref.~\cite{Nadia}, does not support the physical interpretation of the shape
parameter in the ISGW2, simple pole, and modified pole parametrization.
Accordingly, both here and in Ref.~\cite{Nadia}, the values of $|V_{cq}|$ obtained with
the series expansion with a quadratic term are chosen as the primary
results. Therefore, in this section we present averages of $|V_{cq}|$ and the
shape parameters only for the series expansion with a quadratic term.

To allow external use of the set of partial branching fractions
presented in this paper and in Ref.~\cite{Nadia}, we determine the full
statistical and systematic uncertainty correlation matrices and present
them in Appendix A. These matrices allow for simultaneous fits of the
results in this work and in Ref.~\cite{Nadia}
to any form factor parametrization to obtain form factor
parameters. They also allow for simultaneous fits with other
experimental results.

The two analyses use the same data set. The untagged analysis has a significantly higher efficiency, resulting in signal yields
$\sim 2.5$ times greater than the tagged analysis,  but
also has larger backgrounds. Most of the signal events found by
the tagged analysis are also found by the untagged
analysis, and so the measurements produced by the
two analyses are highly correlated.

To compute averages
we use error matrices to take into account
the correlations between measurements made by the two techniques.
The statistical covariance matrix between the two analyses has  a 2 $\times$ 2 block form.
The diagonal blocks are obtained from the untagged and tagged analyses, respectively.
The off-diagonal blocks arise from correlations between the two analyses.
As the covariance matrix is symmetric, only one off-diagonal block needs to be determined.

The off-diagonal blocks are computed using a bootstrap~\cite{bootstrap} MC simulation,
where 185 data-sized MC samples are constructed.
Each sample is created by randomly selecting events from the
generic MC sample. Each event cannot appear more than
once in a given sample. Each analysis runs on each bootstrap sample
and the statistical correlation
between the analyses can be measured and the off-diagonal block of
the statistical covariance matrix
computed.

A systematic correlation matrix is constructed
by taking each systematic uncertainty
as either 100\% correlated or uncorrelated between the two analyses.
The complete four-block
combined statistical and systematic correlation matrix is used to obtain
the averages.
A comparison of correlation matrices calculated by this technique
to those determined by each analysis is made and good agreement is found.

Consider first the determination of the average branching
fractions. The untagged analysis determines the branching fractions
from the sum of the partial branching fractions in each $q^2$ bin. To
treat each analysis similarly
the derived branching fractions are computed for the tagged analysis
from the partial branching fractions in each $q^2$ bin, using the
partial rates in Table~\ref{decayrateinbins}, corrected for the
lifetime factors. The derived branching fractions are reported in
Table~\ref{tab:derived-bfractions-average}, and are consistent
with the branching fractions found in Sec.~\ref{BFs}.

\begin{table}[htbp]
\caption{The derived absolute branching fractions (in percent)
obtained by summing the partial branching fractions in each $q^2$
bin, for \cleoc tagged, untagged~\cite{Nadia}, and the average absolute branching
fraction.} \label{tab:derived-bfractions-average}
\begin{center}{\small
{
\begin{tabular}{lccc}\hline\hline
  & Tagged  & Untagged   & Average   \\\hline

$\pi^- e^+ \nu_e$     & 0.308(13)(4)  & 0.299(11)(8) & 0.304(11)(5)   \\
$\pi^0 e^+ \nu_e$     & 0.379(27)(23) & 0.373(22)(13)& 0.378(20)(12) \\
$K^- e^+ \nu_e $      & 3.60(5)(5)    & 3.56(3)(9)   & 3.60(3)(6)     \\
$\bar{K}^0 e^+ \nu_e$ & 8.87(17)(21)  & 8.53(13)(23) & 8.69(12)(19)   \\

 \hline\hline
\end{tabular}
}}
\end{center}
\end{table}

To extract the statistical correlations between the two analyses,
both analyses have run on the bootstrap samples.
The analysis procedures in each case are identical to those used with data.
For each semileptonic mode, and each bootstrap sample,
we obtain the partial branching fractions in each $q^2$ bin. We
sum individual $q^2$ bins to obtain the total branching fractions.
These are used to calculate the statistical covariance matrices.

Table~\ref{tab:bf_corr_matrix} gives the complete 8 $ \times$ 8 statistical correlation matrix.
The tagged internal block is diagonal as
the branching fractions are
uncorrelated with each other. The untagged internal block is obtained from the untagged analysis, where
the correlations between the total branching fractions have been
computed from those between the individual $q^2$ bins.
The off-diagonal elements in the off-diagonal blocks are small, so we set them to zero.
In so doing, we neglect the
statistical correlation from using common tag yields.
Performing the averaging procedure with or without these elements produces a negligible difference in the final results.

\begin{table}[htbp]
\caption{The complete branching fraction statistical correlation matrix. Untagged $q^2$ intervals are in columns and tagged $q^2$ intervals are in rows.  The modes are labeled by their final state hadrons.}
\begin{center}
\begin{tabular}{cc|rrrr|rrrr} \hline \hline
\multicolumn{2}{c|}{} &  \multicolumn{4}{c|}{Untagged} & \multicolumn{4}{c}{Tagged}    \\
\multicolumn{2}{c|}{} & $\pi^-$ & $K^-$ & $\pi^0$ & $\bar{K}^0$
                     & $\pi^-$ & $K^-$ & $\pi^0$ & $\bar{K}^0$  \\\hline

   &$\pi^-$     & 1.00 & -0.04 & -0.02 & -0.01 &  0.60 &  0.00 &  0.00 &  0.00\\
Untagged &$K^-$       &      &  1.00 &  0.00 & -0.02 &  0.00 &  0.49 &  0.00 &  0.00\\
   &$\pi^0$     &      &       &  1.00 & -0.02 &  0.00 &  0.00 &  0.36 &  0.00\\
   &$\bar{K}^0$ &      &       &       &  1.00 &  0.00 &  0.00 &  0.00 &  0.43\\\hline
   &$\pi^-$     &      &       &       &       &  1.00 &  0.00 &  0.00 &  0.00\\
Tagged&$K^-$       &      &       &       &       &       &  1.00 &  0.00 &  0.00\\
   &$\pi^0$     &      &       &       &       &       &       &  1.00 &  0.00\\
   &$\bar{K}^0$ &      &       &       &       &       &       &       &  1.00\\

\hline \hline
\end{tabular}
\end{center}
\label{tab:bf_corr_matrix}
\end{table}

To be conservative, all of the systematic uncertainties in the tagged analysis are taken to be fully correlated with the corresponding
systematic uncertainties in the untagged analysis. The untagged analysis has additional systematic uncertainties
which have no analog in the tagged analysis. Accordingly, these are taken to be uncorrelated between the two analyses.
A covariance matrix for each systematic uncertainty
is then constructed.
A 39\% correlation between the numbers of charged $D$ and neutral $D$ pairs
in the untagged analysis is also taken into account.  The complete branching fraction systematic correlation matrix is given in Table~\ref{tab:bf_rho_syst}.

\begin{table}[htbp]
\caption{The complete branching fraction systematic correlation matrix.
The modes are labeled by their final state hadrons.
}
\begin{center}
\begin{tabular}{cc|rrrr|rrrr} \hline \hline
\multicolumn{2}{c|}{} &  \multicolumn{4}{c|}{Untagged} & \multicolumn{4}{c}{Tagged}    \\
\multicolumn{2}{c|}{} & $\pi^-$ & $K^-$ & $\pi^0$ & $\bar{K}^0$
                     & $\pi^-$ & $K^-$ & $\pi^0$ & $\bar{K}^0$  \\\hline

   &$\pi^-$     & 1.00 &  0.85 &  0.32 &  0.31 &  0.48 &  0.46 &  0.08 &  0.23\\
Untagged&$K^-$       &      &  1.00 &  0.39 &  0.37 &  0.48 &  0.49 &  0.08 &  0.24\\
   &$\pi^0$     &      &       &  1.00 &  0.62 &  0.24 &  0.24 &  0.23 &  0.18\\
   &$\bar{K}^0$ &      &       &       &  1.00 &  0.38 &  0.38 &  0.10 &  0.59\\\hline
   &$\pi^-$     &      &       &       &       &  1.00 &  0.98 &  0.33 &  0.60\\
Tagged&$K^-$       &      &       &       &       &       &  1.00 &  0.31 &  0.61\\
   &$\pi^0$     &      &       &       &       &       &       &  1.00 &  0.21\\
   &$\bar{K}^0$ &      &       &       &       &       &       &       &  1.00\\
\hline \hline

\end{tabular}
\end{center}
\label{tab:bf_rho_syst}
\end{table}

We fit the 8 branching fractions with the four-block statistical
covariance matrix. The fit returns $\chi^2/{\rm d.o.f.} = 6.7/4$.
We then repeat the fit with
the four-block
combined statistical and systematic correlation matrix to obtain
the central values for the averages and the combined statistical and systematic uncertainty.
This fit returns $\chi^2/{\rm d.o.f.} = 2.1/4$.
The quadrature difference between the uncertainties obtained in the two fits is used to compute the systematic uncertainty.
The results of the fits are reported in Table~\ref{tab:derived-bfractions-average}.
The averaged branching fractions are more
precise than those measured by either the tagged or untagged analysis.

To determine the form factor
parameters we perform a
simultaneous fit to the ($N_1 + N_2$) partial branching fractions, where $N_1$ and $N_2$ are the numbers of $q^2$ bins in the untagged and tagged analyses,
respectively. To improve precision, we also simultaneously fit pairs of modes related by isospin.
The corresponding correlation matrix has dimensions ($N_1 + N_2$) $\times$ ($N_1 + N_2) = (56 \times 56$) as there are 5 bins for each mode in the untagged analysis and
(9+7+10+10) bins in the tagged analysis.
We give averaged results for the series expansion.

\begin{table*}[!htb]
\caption{Measurements of $f_+(0)|V_{cq}|$, $r_1$ and $r_2$ in \cleoc tagged, untagged~\cite{Nadia} and
the \cleoc average.
}
\begin{center}
\begin{tabular}{cc|ccc} \hline \hline
 & &   Tagged  & Untagged   & Average   \\\hline

 & $f_+(0)|V_{cq}|$           & 0.152(8)(1)  & 0.140(7)(3)  & 0.142(6)(2)        \\
 $D^0\to \pi^-\enu$& $r_1$    & -2.0(6)(1)    & -2.1(7)(3)  & -2.1(5)(1)          \\
     & $r_2$                 & 1.6(4.2)(0.8)  & -1.2(4.8)(1.7)&-0.6(3.5)(0.9)     \\\hline

 & $f_+(0)|V_{cq}|$       &  0.745(12)(6)   & 0.747(9)(9)  & 0.745(8)(6)       \\
$D^0\to K^-\enu$ & $r_1$     &  -2.4(5)(2)     & -2.4(4)(1)   & -2.4(4)(1)         \\
     & $r_2$                & 15.6(12.8)(3.8)  & 21(11)(2)  & 17.9(10.0)(1.9)     \\\hline

 & $f_+(0)|V_{cq}|$       &  0.144(12)(4)   & 0.138(11)(4)   & 0.140(9)(2)     \\
$D^+\to \pi^0 \enu$& $r_1$  & -0.3(1.8)(1.0) & -0.2(1.5)(4)   & -0.2(1.3)(0.2) \\
     & $r_2$                & -7.9(10.5)(5.6) & -9.8(9.1)(2.1) & -9.4(7.5)(1.1)  \\\hline

 & $f_+(0)|V_{cq}|$         & 0.744(17)(9)  & 0.733(14)(11)   & 0.733(13)(9)    \\
$D^+\to \bar{K}^0 \enu$& $r_1$ & -1.9(7)(2) & -2.8(6)(2)         & -2.5(6)(1)     \\
     & $r_2$                   & 16.6(19.3)(6.2) & 32(18)(4)  & 26.8(15.4)(2.6)    \\\hline\hline

 & $f_+(0)|V_{cq}|$        & 0.151(7)(2)  &   & 0.142(5)(2)       \\
 $D\to \pi\enu$& $r_1$   & -1.8(6)(3)    &      & -1.7(5)(2)      \\
     & $r_2$                & 0.3(3.9)(1.7)  &  & -1.9(3.2)(1.1)    \\\hline

 & $f_+(0)|V_{cq}|$       &  0.744(10)(6)   &   & 0.743(7)(6)      \\
$D\to K\enu$ & $r_1$     &  -2.2(4)(2)     &      & -2.4(3)(1)     \\
    & $r_2$                & 16.9(11.4)(4.7)  &   & 20.8(8.3)(2.0)    \\\hline\hline

\end{tabular}
\end{center}
\label{tab:FFresults}
\end{table*}

For the fits to the $q^2$ distributions for each mode the untagged and tagged diagonal blocks
are again taken from the analysis-determined correlation coefficients
from the data yield fits.
The bootstrap method determines the off-diagonal block statistical
correlation matrix.
As the $q^2$ binning differs between the two analyses we take into account
both the correlations between $q^2$ intervals that overlap in the two analyses and
the correlations between non-overlapping tagged$-$untagged bins.
The off-diagonal block of the ($56 \times 56$)
statistical correlation matrix may be found in Tables~\ref{tab:ff_rho_stat_ut1} and \ref{tab:ff_rho_stat_ut2}.
The diagonal untagged block is given in~\cite{Nadia}.
The systematic covariance matrix is constructed from the systematic uncertainties in each analysis. The off-diagonal block may be found in Tables~\ref{tab:ff_rho_syst_ut1} and \ref{tab:ff_rho_syst_ut2}, the untagged diagonal block is given in~\cite{Nadia}, and the tagged diagonal block may be found in Tables~\ref{tab:ff_rho_syst_tt1} and \ref{tab:ff_rho_syst_tt2}.

We first perform the fit with the statistical covariance matrix only,
then repeat it with the combined statistical and systematic covariance matrix.
The quadrature difference between the uncertainties obtained in the two fits
is used to compute the systematic uncertainty.
The central values for the averages are taken from the combined statistical and
systematic fit and are given in Table~\ref{tab:FFresults},
and the parameter correlations from the fit with statistical and systematic uncertainties
are given in Table~\ref{tab:FFparcorr}.
Finally, to improve precision, isospin constraints are imposed and we determine form factor parameters for $D\to K\enu$
and $D\to \pi\enu$ from simultaneous fits to the respective isospin conjugate modes.  Results and parameter correlations are shown in Tables~\ref{tab:FFresults}
and \ref{tab:FFparcorr_iso}, respectively.
Using $\vcs$ and $\vcd$ values constrained by CKM unitarity~\cite{PDG2008},
the $f^K_+(0)$ and $f^\pi_+(0)$ averages are calculated and reported in Table ~\ref{tab:fzero-comp}.

\begingroup
\squeezetable
\begin{table*}[!hbt]
\caption{The parameter correlation coefficients for $f_+(0)|V_{cq}|$, $r_1$ and $r_2$ between the tagged and untagged~\cite{Nadia} \cleoc analyses. The modes are labeled by their final state hadrons.
}
\begin{center}
\begin{tabular}{cc|cr@{\hspace{5mm}}r|cr@{\hspace{5mm}}r|cr@{\hspace{5mm}}r|cr@{\hspace{5mm}}r} \hline \hline\rule[-1mm]{-1mm}{4mm}
& & \multicolumn{3}{c|}{$ \pi^-$}& \multicolumn{3}{c|}{$ K^-$ }
& \multicolumn{3}{c|}{$ \pi^0 $}& \multicolumn{3}{c}{$ \bar{K}^0 $}\\
   &    & $f_+(0)|V_{cq}|$ & $r_1$\hspace{1mm} & $r_2$\hspace{1mm} & $f_+(0)|V_{cq}|$ & $r_1$\hspace{1mm} & $r_2$\hspace{1mm}
   & $f_+(0)|V_{cq}|$ & $r_1$\hspace{1mm} & $r_2$\hspace{1mm} & $f_+(0)|V_{cq}|$ & $r_1$\hspace{1mm} & $r_2$\hspace{1mm}\\\hline

 & $f_+(0)|V_{cq}|$         &  1.00 & -0.38 &  0.65 &  0.11 &  0.01 &  0.00 &  0.02 &  0.00 &  0.00 &  0.07 & -0.01 &  0.01 \\
 $ \pi^-$& $r_1$     &              &  1.00 & -0.93 &  0.01 &  0.00 &  0.00 &  0.01 & -0.03 &  0.02 &  0.01 &  0.00 &  0.00  \\
     & $r_2$           &            &       &  1.00 &  0.00 &  0.00 &  0.01 &  0.00 &  0.01 & -0.01 &  0.01 & -0.01 &  0.01    \\\hline
 & $f_+(0)|V_{cq}|$        &        &       &       &  1.00 & -0.23 &  0.59 &  0.04 &  0.00 &  0.00 &  0.16 &  0.01 & -0.01   \\
$ K^-$ & $r_1$       &              &       &       &       &  1.00 & -0.83 &  0.00 & -0.01 &  0.01 &  0.03 & -0.03 &  0.03  \\
     & $r_2$             &          &       &       &       &       &  1.00 &  0.00 & -0.01 &  0.01 & -0.01 &  0.03 & -0.01    \\\hline
 & $f_+(0)|V_{cq}|$        &        &       &       &       &       &       &  1.00 & -0.43 &  0.65 &  0.03 &  0.00 &  0.00   \\
$ \pi^0 $& $r_1$     &              &       &       &       &       &       &       &  1.00 & -0.96 &  0.01 & -0.02 &  0.01  \\
     & $r_2$              &         &       &       &       &       &       &       &       &  1.00 & -0.01 &  0.01 & -0.01    \\\hline
 & $f_+(0)|V_{cq}|$        &        &       &       &       &       &       &       &       &       &  1.00 & -0.24 &  0.62   \\
$ \bar{K}^0 $& $r_1$  &             &       &       &       &       &       &       &       &       &       &  1.00 & -0.81  \\
     & $r_2$              &         &       &       &       &       &       &       &       &       &       &       &  1.00    \\ \hline\hline
\end{tabular}
\end{center}
\label{tab:FFparcorr}
\end{table*}
\endgroup

\begin{table}[!hbt]
\caption{The parameter correlation coefficients for $f_+(0)|V_{cq}|$, $r_1$ and $r_2$ between the tagged and untagged~\cite{Nadia} \cleoc analysis
with isospin constraints imposed. The modes are labeled by their final state hadrons.
}
\begin{center}{\footnotesize
\begin{tabular}{cc|cr@{\hspace{5mm}}r|cr@{\hspace{5mm}}r} \hline \hline
& & \multicolumn{3}{c|}{$\pi$}& \multicolumn{3}{c}{$K$ }\\
   &    & $f_+(0)|V_{cq}|$ & $r_1$\hspace{1mm} & $r_2$\hspace{1mm} & $f_+(0)|V_{cq}|$ & $r_1$\hspace{1mm} & $r_2$\hspace{1mm}\\\hline

 & $f_+(0)|V_{cq}|$   &1.00 & -0.42 &  0.66 &  0.13 &  0.00 &  0.01   \\
$\pi$& $r_1$    &           &  1.00 & -0.94 &  0.01 &  0.00 &  0.00   \\
     & $r_2$    &           &       &  1.00 &  0.00 &  0.00 &  0.01  \\\hline
 & $f_+(0)|V_{cq}|$   &     &       &       &  1.00 & -0.21 &  0.55    \\
$K$& $r_1$    &             &       &       &       &  1.00 & -0.81  \\
     & $r_2$&               &       &       &       &       &  1.00   \\ \hline\hline
\end{tabular}}
\end{center}
\label{tab:FFparcorr_iso}
\end{table}

The \cleoc averages are
\beqn
|V_{cd}| = 0.222\pm 0.008 \pm 0.003 \pm 0.023
\eeqn
\noindent and
\beqn
|V_{cs}| = 1.018 \pm 0.010 \pm 0.008 \pm 0.106,
\eeqn
where the uncertainties are statistical, systematic, and
theoretical, respectively.   We compare these averages
to the untagged and tagged standalone determinations in Table~\ref{tab:Vcs-comp} and
Table~\ref{tab:Vcd-comp}. In each case the average value is more
precise than that obtained with either analysis.
This is the most precise determination
of $|V_{cs}|$ to date, and the most precise determination of
$|V_{cd}|$ from $D$ meson semileptonic decay to date.

\section{Summary}

\label{summary}

In this paper we have presented precise measurements of the
absolute branching fractions of $D^0$ decays to $K^- e^+ \nu_e$
and $\pi^- e^+ \nu_e$ and $D^+$ decays to $\bar{K}^0 e^+ \nu_e$
and $\pi^0 e^+ \nu_e$, that agree well with world averages~\cite{PDG2004} and Ref.~\cite{Nadia}.
We have combined these measurements to
demonstrate that $D$ meson exclusive semileptonic decays to
pseudoscalar final states are consistent with isospin invariance.

From the $q^2$ spectrum of all four decay modes
studied, we have made the most precise determinations
of $r^\pi$ and $r^K$. Our measurement of $r^K$ is over $4 \sigma$ smaller
than Belle~\cite{BELLE-06}, and  $1.6 \sigma$ smaller than BABAR~\cite{BABAR-06}.
Our measurement of
$M_{\rm pole}^K$ ($\alpha^K$) is in excellent agreement with Ref.~\cite{Nadia}, but is  $2.3\sigma$ larger ($2.6 \sigma$ smaller)
than the BABAR~\cite{BABAR-06} measurement, and  $2.5 \sigma$ larger ($2.7 \sigma$ smaller) than the Belle~\cite{BELLE-06} measurement.
Our determinations of $r^\pi$, $M_{\rm pole}^\pi$ and $\alpha^\pi$ are in
reasonable agreement with Ref~\cite{Nadia} and other previous measurements.
Our measurement of $\alpha^K$ ($\alpha^\pi$)
is more than  $4 \sigma$ ($2.4 \sigma$) smaller than the LQCD fit. However, the discrepancy with LCQD is difficult to quantify because the covariance matrix for the LQCD form factors is lost
during the chiral extrapolation procedure for the published analysis~\cite{unquenched_LQCD}.
The results reported here for the series expansion parameters $r_1$ and $r_2$
are in excellent agreement with
Ref.~\cite{Nadia} and agree with BABAR~\cite{BABAR-06} to better than $2 \sigma$ with the precise level depending on the correlation coefficient for the BABAR $r_1$ and $r_2$ parameters.

Our data, and other recent measurements (e.g., Refs.~\cite{CLEOIII,FOCUS-05,BELLE-06,BABAR-06}) do not
support the physical interpretation of the shape parameter in the ISGW2,
simple pole, and modified pole parametrizations.
Accordingly, the $f_+(0)|V_{cq}|$ values
obtained when the data is fit with the quadratic series expansion
were selected as our primary normalization results. We combined these values with the unitarity of the CKM matrix
to make a precise determination of the form factor absolute magnitude $f_+^K(0)$
and the most precise determination of $f_+^\pi(0)$. Using unquenched LQCD predictions for $f_+^K(0)$ and
$f_+^\pi(0)$ we have made the most precise determination of $|V_{cs}|$, and the most precise determination of $|V_{cd}|$ from
$D$
semileptonic decays to date. The results agree
well with previous measurements using semileptonic decays including Ref~\cite{Nadia}, and agree well with
charm-tagged $W$ decay measurements of $|V_{cs}|$ and neutrino based
determinations of $|V_{cd}|$.

To allow external use of the set of partial branching fractions presented in this paper and in Ref.~\cite{Nadia}
we determined the full statistical and systematic uncertainty correlation matrices.
These matrices allow for simultaneous fits of the results of this work and Ref~\cite{Nadia} to any form factor parametrization to obtain form factor parameters.
They also allow for simultaneous fits with other experimental results.

Finally, we averaged values of the measurements obtained
in this work with Ref.~\cite{Nadia}.
These averages
represent the best determinations of the branching fractions, form
factors and $\vcs$ and $\vcd$ with the \cleoc 281~pb$^{-1}$ data
set~\cite{nompoleave}. They are the most
precise measurements of the absolute branching fractions of $D^0$
decays to $K^- e^+ \nu_e$ and $\pi^- e^+ \nu_e$ and $D^+$ decays
to $\bar{K}^0 e^+ \nu_e$ and $\pi^0 e^+ \nu_e$,
and the most precise direct
determination of $\vcs$ and the most precise determination of
$\vcd$ from $D$ semileptonic decay.

CESR has recently collected a
larger $\psi(3770)$ data sample.  It is expected that this sample
will result in a further improvement in measurements of $D^{0}$
and $D^{+}$ semileptonic branching fractions, measurements of the
decay form factors, and the CKM matrix elements $\vcs$ and
$\vcd$~\cite{cleoc}.

\section{Acknowledgments}

We gratefully acknowledge the effort of the CESR staff in
providing us with excellent luminosity and running conditions. D.
Cronin-Hennessy and A. Ryd thank the A.P. Sloan Foundation. This
work was supported by the National Science Foundation, the U.S.
Department of Energy, and the Natural Sciences and Engineering
Research Council of Canada.

\appendix

\renewcommand{\thetable}{A.\arabic{table}}\setcounter{table}{0}
\section{Correlation Matrices}

To allow external use of the set of partial branching
fractions presented in this paper and~\cite{Nadia}
this Appendix contains the statistical
and systematic uncertainty correlation matrices. These
matrices allow for simultaneous fits of
these results with other experimental results to obtain
form factor parameters. Sec.~\ref{average} describes the procedures
that have been used to obtain these matrices.

\begingroup
\squeezetable
\begin{table*}[htbp]
\caption{The untagged-tagged block of the statistical correlation matrix obtained from the bootstrap procedure
. Untagged $q^2$ intervals are in columns and tagged $q^2$ intervals are in rows.  The lines indicate mode boundaries. The modes are labeled by their final state hadrons. Within each submode, the $q^2$ intervals are ordered from lowest to highest (part I).}
\begin{center}
\begin{tabular}{cc|rrrrrrrrr|rrrrr rrrrr} \hline\hline
& & \multicolumn{19}{c}{Tagged} \\
& & \multicolumn{9}{c|}{$\pi^-$} & \multicolumn{10}{c}{$K^-$} \\\hline

   &            &  0.46 &  0.20 & -0.01 &  0.00 &  0.00 &  0.00 &  0.00 &  0.00 &  0.00 & -0.01 & -0.02 &  0.00 &  0.00 &  0.00 &  0.00 &  0.00 &  0.00 &  0.00 &  0.00 \\
   &            &  0.00 &  0.35 &  0.30 & -0.02 &  0.00 &  0.00 &  0.00 &  0.00 &  0.00 &  0.00 &  0.00 &  0.00 &  0.00 &  0.00 &  0.00 &  0.00 &  0.00 &  0.00 &  0.00 \\
   &$\pi^-$     &  0.00 & -0.02 &  0.28 &  0.53 &  0.01 & -0.01 &  0.00 &  0.00 &  0.00 &  0.00 &  0.00 &  0.00 &  0.00 &  0.00 &  0.00 &  0.00 &  0.00 &  0.00 &  0.00 \\
   &            &  0.00 &  0.00 & -0.01 & -0.02 &  0.43 &  0.20 &  0.00 &  0.00 & -0.01 &  0.00 &  0.00 &  0.00 &  0.00 &  0.00 &  0.00 &  0.00 &  0.00 &  0.00 &  0.00 \\
   &            &  0.00 &  0.00 &  0.00 &  0.00 & -0.01 &  0.35 &  0.26 &  0.15 &  0.35 &  0.00 &  0.00 &  0.00 &  0.00 &  0.00 &  0.00 &  0.00 &  0.00 &  0.00 &  0.00 \\\cline{2-21}
   &            & -0.03 & -0.01 &  0.00 &  0.00 &  0.00 &  0.00 &  0.00 &  0.00 &  0.00 &  0.25 &  0.27 &  0.04 & -0.02 & -0.01 &  0.00 &  0.00 &  0.00 &  0.00 &  0.00 \\
   &            &  0.00 &  0.00 &  0.00 &  0.00 &  0.00 &  0.00 &  0.00 &  0.00 &  0.00 &  0.00 &  0.00 &  0.44 &  0.37 &  0.08 & -0.03 & -0.01 &  0.00 &  0.00 &  0.00 \\
   &$K^-$       &  0.00 &  0.00 &  0.00 &  0.00 &  0.00 &  0.00 &  0.00 &  0.00 &  0.00 &  0.00 &  0.00 & -0.03 & -0.03 &  0.28 &  0.48 &  0.13 & -0.03 & -0.01 &  0.00 \\
   &            &  0.00 &  0.00 &  0.00 &  0.00 &  0.00 &  0.00 &  0.00 &  0.00 &  0.00 &  0.00 &  0.00 &  0.00 &  0.00 & -0.02 & -0.03 &  0.28 &  0.47 &  0.17 & -0.02 \\
Untagged &           &  0.00 &  0.00 &  0.00 &  0.00 &  0.00 &  0.00 &  0.00 &  0.00 &  0.00 &  0.00 &  0.00 &  0.00 &  0.00 &  0.00 &  0.00 & -0.03 & -0.04 &  0.25 &  0.23 \\\cline{2-21}
   &            &  0.00 &  0.00 &  0.00 &  0.00 &  0.00 &  0.00 &  0.00 &  0.00 &  0.00 &  0.00 &  0.00 &  0.00 &  0.00 &  0.00 &  0.00 &  0.00 &  0.00 &  0.00 &  0.00 \\
   &            &  0.00 &  0.00 &  0.00 &  0.00 &  0.00 &  0.00 &  0.00 &  0.00 &  0.00 &  0.00 &  0.00 &  0.00 &  0.00 &  0.00 &  0.00 &  0.00 &  0.00 &  0.00 &  0.00 \\
   &$\pi^0$     &  0.00 &  0.00 &  0.00 &  0.00 &  0.00 &  0.00 &  0.00 &  0.00 &  0.00 &  0.00 &  0.00 &  0.00 &  0.00 &  0.00 &  0.00 &  0.00 &  0.00 &  0.00 &  0.00 \\
   &            &  0.00 &  0.00 &  0.00 &  0.00 &  0.00 &  0.00 &  0.00 &  0.00 &  0.00 &  0.00 &  0.00 &  0.00 &  0.00 &  0.00 &  0.00 &  0.00 &  0.00 &  0.00 &  0.00 \\
   &            &  0.00 &  0.00 &  0.00 &  0.00 &  0.00 & -0.04 & -0.03 & -0.02 & -0.04 &  0.00 &  0.00 &  0.00 &  0.00 &  0.00 &  0.00 &  0.00 &  0.00 &  0.00 &  0.00 \\\cline{2-21}
   &            &  0.00 &  0.00 &  0.00 &  0.00 &  0.00 &  0.00 &  0.00 &  0.00 &  0.00 &  0.00 &  0.00 &  0.00 &  0.00 &  0.00 &  0.00 &  0.00 &  0.00 &  0.00 &  0.00 \\
   &            &  0.00 &  0.00 &  0.00 &  0.00 &  0.00 &  0.00 &  0.00 &  0.00 &  0.00 &  0.00 &  0.00 &  0.00 &  0.00 &  0.00 &  0.00 &  0.00 &  0.00 &  0.00 &  0.00 \\
   &$\bar{K}^0$ &  0.00 &  0.00 &  0.00 &  0.00 &  0.00 &  0.00 &  0.00 &  0.00 &  0.00 &  0.00 &  0.00 &  0.00 &  0.00 &  0.00 &  0.00 &  0.00 &  0.00 &  0.00 &  0.00 \\
   &            &  0.00 &  0.00 &  0.00 &  0.00 &  0.00 &  0.00 &  0.00 &  0.00 &  0.00 &  0.00 &  0.00 &  0.00 &  0.00 &  0.00 &  0.00 &  0.00 &  0.00 &  0.00 &  0.00 \\
   &            &  0.00 &  0.00 &  0.00 &  0.00 &  0.00 & -0.02 & -0.01 & -0.01 & -0.02 &  0.00 &  0.00 &  0.00 &  0.00 &  0.00 &  0.00 &  0.00 &  0.00 & -0.02 & -0.01 \\
\hline\hline

\end{tabular}
\end{center}
\label{tab:ff_rho_stat_ut1}
\end{table*}
\endgroup

\begingroup
\squeezetable
\begin{table*}[htbp]
\caption{The untagged-tagged block of the statistical correlation matrix obtained from the bootstrap procedure
. Untagged $q^2$ intervals are in columns and tagged $q^2$ intervals are in rows.  The lines indicate mode boundaries. The modes are labeled by their final state hadrons. Within each submode, the $q^2$ intervals are ordered from lowest to highest (part II).}
\begin{center}
\begin{tabular}{cc|rrrrrrr|rrrrr rrrrr} \hline\hline
& & \multicolumn{17}{c}{Tagged} \\
& & \multicolumn{7}{c|}{$\pi^0$} & \multicolumn{10}{c}{$\bar{K}^0$}\\\hline

   &            &  0.00 &  0.00 &  0.00 &  0.00 &  0.00 &  0.00 &  0.00 &  0.00 &  0.00 &  0.00 &  0.00 &  0.00 &  0.00 &  0.00 &  0.00 &  0.00 &  0.00   \\
   &            &  0.00 &  0.00 &  0.00 &  0.00 &  0.00 &  0.00 &  0.00 &  0.00 &  0.00 &  0.00 &  0.00 &  0.00 &  0.00 &  0.00 &  0.00 &  0.00 &  0.00   \\
   &$\pi^-$     &  0.00 &  0.00 &  0.00 &  0.00 &  0.00 &  0.00 &  0.00 &  0.00 &  0.00 &  0.00 &  0.00 &  0.00 &  0.00 &  0.00 &  0.00 &  0.00 &  0.00   \\
   &            &  0.00 &  0.00 &  0.00 &  0.00 &  0.00 &  0.00 &  0.00 &  0.00 &  0.00 &  0.00 &  0.00 &  0.00 &  0.00 &  0.00 &  0.00 &  0.00 &  0.00   \\
   &            &  0.00 &  0.00 &  0.00 &  0.00 &  0.00 & -0.03 & -0.04 &  0.00 &  0.00 &  0.00 &  0.00 &  0.00 &  0.00 &  0.00 &  0.00 & -0.02 & -0.01   \\\cline{2-19}
   &            &  0.00 &  0.00 &  0.00 &  0.00 &  0.00 &  0.00 &  0.00 &  0.00 &  0.00 &  0.00 &  0.00 &  0.00 &  0.00 &  0.00 &  0.00 &  0.00 &  0.00   \\
   &            &  0.00 &  0.00 &  0.00 &  0.00 &  0.00 &  0.00 &  0.00 &  0.00 &  0.00 &  0.00 &  0.00 &  0.00 &  0.00 &  0.00 &  0.00 &  0.00 &  0.00   \\
   &$K^-$       &  0.00 &  0.00 &  0.00 &  0.00 &  0.00 &  0.00 &  0.00 &  0.00 &  0.00 &  0.00 &  0.00 &  0.00 &  0.00 &  0.00 &  0.00 &  0.00 &  0.00   \\
   &            &  0.00 &  0.00 &  0.00 &  0.00 &  0.00 &  0.00 &  0.00 &  0.00 &  0.00 &  0.00 &  0.00 &  0.00 &  0.00 &  0.00 &  0.00 &  0.00 &  0.00   \\
Untagged    &        &  0.00 &  0.00 &  0.00 &  0.00 &  0.00 &  0.00 &  0.00 &  0.00 &  0.00 &  0.00 &  0.00 &  0.00 &  0.00 &  0.00 &  0.00 & -0.02 & -0.02   \\\cline{2-19}
   &            &  0.28 &  0.18 & -0.02 &  0.00 &  0.00 &  0.00 &  0.00 &  0.00 &  0.00 &  0.00 &  0.00 &  0.00 &  0.00 &  0.00 &  0.00 &  0.00 &  0.00   \\
   &            &  0.00 &  0.31 &  0.27 & -0.03 &  0.00 &  0.00 &  0.00 &  0.00 &  0.00 &  0.00 &  0.00 &  0.00 &  0.00 &  0.00 &  0.00 &  0.00 &  0.00   \\
   &$\pi^0$     &  0.00 & -0.03 &  0.12 &  0.36 &  0.00 & -0.01 &  0.00 &  0.00 &  0.00 &  0.00 &  0.00 &  0.00 &  0.00 &  0.00 &  0.00 &  0.00 &  0.00   \\
   &            &  0.00 &  0.00 & -0.01 & -0.03 &  0.41 &  0.13 & -0.02 &  0.00 &  0.00 &  0.00 &  0.00 &  0.00 &  0.00 &  0.00 &  0.00 &  0.00 &  0.00   \\
   &            &  0.00 &  0.00 &  0.00 &  0.00 & -0.03 &  0.22 &  0.30 &  0.00 &  0.00 &  0.00 &  0.00 &  0.00 &  0.00 &  0.00 &  0.00 & -0.02 & -0.01   \\\cline{2-19}
   &            &  0.00 &  0.00 &  0.00 &  0.00 &  0.00 &  0.00 &  0.00 &  0.34 &  0.28 &  0.04 & -0.01 & -0.01 &  0.00 &  0.00 &  0.00 &  0.00 &  0.00   \\
   &            &  0.00 &  0.00 &  0.00 &  0.00 &  0.00 &  0.00 &  0.00 &  0.00 &  0.00 &  0.26 &  0.21 &  0.09 & -0.02 & -0.01 &  0.00 &  0.00 &  0.00   \\
   &$\bar{K}^0$ &  0.00 &  0.00 &  0.00 &  0.00 &  0.00 &  0.00 &  0.00 &  0.00 &  0.00 & -0.02 & -0.01 &  0.22 &  0.39 &  0.20 & -0.02 & -0.01 &  0.00   \\
   &            &  0.00 &  0.00 &  0.00 &  0.00 &  0.00 &  0.00 &  0.00 &  0.00 &  0.00 &  0.00 &  0.00 & -0.02 & -0.03 &  0.23 &  0.26 &  0.17 & -0.03   \\
   &            &  0.00 &  0.00 &  0.00 &  0.00 &  0.00 & -0.01 & -0.02 &  0.00 &  0.00 &  0.00 &  0.00 &  0.00 &  0.00 & -0.02 & -0.03 &  0.33 &  0.27   \\
\hline\hline

\end{tabular}
\end{center}
\label{tab:ff_rho_stat_ut2}
\end{table*}
\endgroup

\begingroup
\squeezetable
\begin{table*}[htbp]
\caption{The untagged-tagged block of the systematic correlation matrix. Untagged $q^2$ intervals are in columns and tagged $q^2$ intervals are in rows.  The lines indicate mode boundaries. The modes are labeled by their final state hadrons. Within each submode, the $q^2$ intervals are ordered from lowest to highest (part I).}
\begin{center}
\begin{tabular}{cc|rrrrrrrrr|rrrrr rrrrr} \hline\hline
& & \multicolumn{19}{c}{Tagged} \\
& & \multicolumn{9}{c|}{$\pi^-$} & \multicolumn{10}{c}{$K^-$} \\\hline

   &            & 0.34 &  0.34 &  0.33 &  0.33 &  0.33 &  0.32 &  0.32 &  0.31 &  0.29 &  0.29 &  0.32 &  0.33 &  0.33 &  0.33 &  0.33 &  0.32 &  0.31 &  0.30 &  0.21 \\
   &            & 0.44 &  0.47 &  0.47 &  0.47 &  0.47 &  0.47 &  0.47 &  0.44 &  0.44 &  0.32 &  0.39 &  0.42 &  0.44 &  0.45 &  0.45 &  0.45 &  0.44 &  0.44 &  0.28 \\
   &$\pi^-$     & 0.39 &  0.40 &  0.39 &  0.39 &  0.38 &  0.38 &  0.38 &  0.36 &  0.35 &  0.33 &  0.36 &  0.37 &  0.38 &  0.38 &  0.38 &  0.38 &  0.36 &  0.36 &  0.24 \\
   &            & 0.31 &  0.32 &  0.31 &  0.31 &  0.31 &  0.31 &  0.31 &  0.29 &  0.28 &  0.25 &  0.28 &  0.30 &  0.31 &  0.31 &  0.31 &  0.31 &  0.30 &  0.29 &  0.20 \\
   &            & 0.26 &  0.26 &  0.26 &  0.26 &  0.25 &  0.25 &  0.25 &  0.24 &  0.23 &  0.21 &  0.24 &  0.25 &  0.25 &  0.25 &  0.25 &  0.25 &  0.24 &  0.24 &  0.16 \\\cline{2-21}
   &            & 0.37 &  0.38 &  0.37 &  0.37 &  0.37 &  0.36 &  0.36 &  0.35 &  0.33 &  0.37 &  0.39 &  0.39 &  0.39 &  0.40 &  0.39 &  0.39 &  0.38 &  0.37 &  0.27 \\
   &            & 0.33 &  0.34 &  0.33 &  0.33 &  0.33 &  0.32 &  0.32 &  0.31 &  0.29 &  0.33 &  0.35 &  0.35 &  0.35 &  0.35 &  0.35 &  0.34 &  0.34 &  0.33 &  0.24 \\
   &$K^-$       & 0.33 &  0.33 &  0.33 &  0.33 &  0.32 &  0.32 &  0.32 &  0.31 &  0.29 &  0.34 &  0.35 &  0.35 &  0.35 &  0.35 &  0.35 &  0.34 &  0.34 &  0.32 &  0.24 \\
   &            & 0.27 &  0.27 &  0.26 &  0.26 &  0.26 &  0.25 &  0.25 &  0.25 &  0.23 &  0.28 &  0.29 &  0.29 &  0.29 &  0.29 &  0.28 &  0.28 &  0.27 &  0.26 &  0.20 \\
 Untagged&            & 0.26 &  0.27 &  0.27 &  0.27 &  0.26 &  0.26 &  0.26 &  0.25 &  0.23 &  0.26 &  0.27 &  0.28 &  0.28 &  0.28 &  0.28 &  0.27 &  0.27 &  0.26 &  0.19 \\\cline{2-21}
   &            & 0.19 &  0.18 &  0.18 &  0.17 &  0.17 &  0.16 &  0.16 &  0.16 &  0.14 &  0.21 &  0.21 &  0.20 &  0.20 &  0.19 &  0.19 &  0.18 &  0.17 &  0.16 &  0.12 \\
   &            & 0.14 &  0.15 &  0.16 &  0.16 &  0.16 &  0.15 &  0.15 &  0.15 &  0.14 &  0.14 &  0.15 &  0.16 &  0.16 &  0.16 &  0.16 &  0.16 &  0.14 &  0.14 &  0.09 \\
   &$\pi^0$     & 0.09 &  0.10 &  0.10 &  0.10 &  0.10 &  0.10 &  0.10 &  0.10 &  0.09 &  0.08 &  0.09 &  0.10 &  0.10 &  0.10 &  0.10 &  0.10 &  0.09 &  0.09 &  0.06 \\
   &            & 0.23 &  0.23 &  0.23 &  0.23 &  0.22 &  0.22 &  0.22 &  0.21 &  0.20 &  0.19 &  0.21 &  0.21 &  0.21 &  0.22 &  0.22 &  0.21 &  0.20 &  0.20 &  0.14 \\
   &            & 0.10 &  0.11 &  0.11 &  0.12 &  0.11 &  0.12 &  0.12 &  0.11 &  0.11 &  0.07 &  0.09 &  0.10 &  0.11 &  0.11 &  0.11 &  0.11 &  0.11 &  0.12 &  0.08 \\\cline{2-21}
   &            & 0.24 &  0.24 &  0.23 &  0.23 &  0.22 &  0.21 &  0.21 &  0.21 &  0.19 &  0.27 &  0.27 &  0.26 &  0.26 &  0.26 &  0.25 &  0.25 &  0.23 &  0.22 &  0.17 \\
   &            & 0.29 &  0.29 &  0.28 &  0.28 &  0.28 &  0.27 &  0.27 &  0.26 &  0.24 &  0.29 &  0.31 &  0.30 &  0.30 &  0.30 &  0.30 &  0.29 &  0.28 &  0.27 &  0.20 \\
   &$\bar{K}^0$ & 0.30 &  0.30 &  0.30 &  0.30 &  0.29 &  0.29 &  0.29 &  0.28 &  0.26 &  0.29 &  0.31 &  0.32 &  0.32 &  0.32 &  0.31 &  0.31 &  0.30 &  0.29 &  0.21 \\
   &            & 0.31 &  0.31 &  0.31 &  0.31 &  0.30 &  0.30 &  0.30 &  0.29 &  0.27 &  0.30 &  0.32 &  0.32 &  0.33 &  0.33 &  0.32 &  0.32 &  0.31 &  0.30 &  0.21 \\
   &            & 0.23 &  0.24 &  0.23 &  0.23 &  0.23 &  0.22 &  0.22 &  0.21 &  0.20 &  0.22 &  0.24 &  0.24 &  0.24 &  0.24 &  0.24 &  0.23 &  0.22 &  0.22 &  0.15 \\
\hline\hline

\end{tabular}
\end{center}
\label{tab:ff_rho_syst_ut1}
\end{table*}
\endgroup

\begingroup
\squeezetable
\begin{table*}[htbp]

\caption{The untagged-tagged block of the systematic correlation matrix. Untagged $q^2$ intervals are in columns and tagged $q^2$ intervals are in rows.  The lines indicate mode boundaries. The modes are labeled by their final state hadrons. Within each submode, the $q^2$ intervals are ordered from lowest to highest (part II).}
\begin{center}
\begin{tabular}{cc|rrrrrrr|rrrrr rrrrr} \hline\hline
& & \multicolumn{17}{c}{Tagged} \\
& & \multicolumn{7}{c|}{$\pi^0$} & \multicolumn{10}{c}{$\bar{K}^0$}\\\hline

   &            & 0.05 &  0.07 &  0.08 &  0.10 &  0.11 &  0.13 &  0.16 &  0.16 &  0.19 &  0.20 &  0.20 &  0.21 &  0.21 &  0.21 &  0.22 &  0.20 &  0.17 \\
   &            & 0.05 &  0.09 &  0.12 &  0.14 &  0.16 &  0.20 &  0.25 &  0.14 &  0.22 &  0.24 &  0.26 &  0.27 &  0.28 &  0.28 &  0.31 &  0.28 &  0.23 \\
   &$\pi^-$     & 0.05 &  0.08 &  0.10 &  0.12 &  0.13 &  0.16 &  0.19 &  0.16 &  0.21 &  0.22 &  0.22 &  0.23 &  0.24 &  0.24 &  0.25 &  0.22 &  0.19 \\
   &            & 0.03 &  0.05 &  0.07 &  0.08 &  0.10 &  0.12 &  0.15 &  0.11 &  0.15 &  0.17 &  0.17 &  0.18 &  0.19 &  0.19 &  0.20 &  0.18 &  0.15 \\
   &            & 0.02 &  0.04 &  0.05 &  0.07 &  0.08 &  0.10 &  0.12 &  0.09 &  0.12 &  0.13 &  0.13 &  0.14 &  0.14 &  0.14 &  0.15 &  0.14 &  0.11 \\\cline{2-19}
   &            & 0.05 &  0.07 &  0.09 &  0.11 &  0.12 &  0.15 &  0.18 &  0.17 &  0.21 &  0.21 &  0.22 &  0.23 &  0.23 &  0.23 &  0.23 &  0.21 &  0.18 \\
   &            & 0.05 &  0.07 &  0.08 &  0.10 &  0.11 &  0.13 &  0.16 &  0.15 &  0.18 &  0.19 &  0.20 &  0.20 &  0.21 &  0.21 &  0.21 &  0.19 &  0.16 \\
   &$K^-$       & 0.05 &  0.07 &  0.08 &  0.09 &  0.11 &  0.13 &  0.16 &  0.16 &  0.19 &  0.20 &  0.20 &  0.21 &  0.21 &  0.21 &  0.21 &  0.19 &  0.16 \\
   &            & 0.04 &  0.05 &  0.06 &  0.07 &  0.08 &  0.10 &  0.12 &  0.13 &  0.15 &  0.16 &  0.16 &  0.16 &  0.17 &  0.17 &  0.17 &  0.15 &  0.13 \\
 Untagged&            & 0.03 &  0.05 &  0.06 &  0.07 &  0.08 &  0.10 &  0.13 &  0.09 &  0.12 &  0.12 &  0.13 &  0.13 &  0.14 &  0.14 &  0.14 &  0.13 &  0.11 \\\cline{2-19}
   &            & 0.23 &  0.23 &  0.24 &  0.24 &  0.23 &  0.22 &  0.20 &  0.18 &  0.18 &  0.17 &  0.17 &  0.17 &  0.17 &  0.17 &  0.16 &  0.15 &  0.13 \\
   &            & 0.16 &  0.17 &  0.18 &  0.18 &  0.18 &  0.18 &  0.18 &  0.13 &  0.15 &  0.15 &  0.15 &  0.15 &  0.16 &  0.15 &  0.15 &  0.14 &  0.12 \\
   &$\pi^0$     & 0.17 &  0.18 &  0.18 &  0.18 &  0.18 &  0.17 &  0.15 &  0.07 &  0.09 &  0.09 &  0.09 &  0.09 &  0.10 &  0.09 &  0.10 &  0.09 &  0.08 \\
   &            & 0.32 &  0.33 &  0.33 &  0.34 &  0.33 &  0.32 &  0.29 &  0.13 &  0.15 &  0.15 &  0.16 &  0.16 &  0.17 &  0.17 &  0.17 &  0.15 &  0.13 \\
   &            & 0.24 &  0.25 &  0.25 &  0.25 &  0.24 &  0.23 &  0.19 &  0.04 &  0.06 &  0.07 &  0.08 &  0.08 &  0.08 &  0.08 &  0.09 &  0.08 &  0.07 \\\cline{2-19}
   &            & 0.07 &  0.08 &  0.09 &  0.11 &  0.12 &  0.14 &  0.16 &  0.52 &  0.52 &  0.51 &  0.51 &  0.50 &  0.50 &  0.49 &  0.47 &  0.46 &  0.39 \\
   &            & 0.07 &  0.09 &  0.11 &  0.12 &  0.14 &  0.16 &  0.20 &  0.51 &  0.53 &  0.52 &  0.52 &  0.52 &  0.52 &  0.51 &  0.49 &  0.48 &  0.40 \\
   &$\bar{K}^0$ & 0.07 &  0.09 &  0.11 &  0.13 &  0.15 &  0.17 &  0.20 &  0.50 &  0.51 &  0.51 &  0.52 &  0.51 &  0.51 &  0.51 &  0.49 &  0.47 &  0.40 \\
   &            & 0.08 &  0.10 &  0.12 &  0.13 &  0.15 &  0.18 &  0.21 &  0.46 &  0.49 &  0.49 &  0.49 &  0.49 &  0.49 &  0.48 &  0.47 &  0.45 &  0.38 \\
   &            & 0.06 &  0.07 &  0.09 &  0.10 &  0.12 &  0.14 &  0.17 &  0.34 &  0.36 &  0.36 &  0.36 &  0.36 &  0.36 &  0.35 &  0.34 &  0.33 &  0.28 \\
\hline\hline

\end{tabular}
\end{center}
\label{tab:ff_rho_syst_ut2}
\end{table*}
\endgroup

\begingroup
\squeezetable
\begin{table*}[htbp]
\caption{The tagged block of the systematic correlation matrix. The lines indicate mode boundaries. The modes are labeled by their final state hadrons. Within each submode, the $q^2$ intervals are ordered from lowest to highest (part I).}
\begin{center}
\begin{tabular}{cc|rrrrrrrrr|rrrrr rrrrr} \hline\hline
& & \multicolumn{19}{c}{Tagged} \\
& & \multicolumn{9}{c|}{$\pi^-$} & \multicolumn{10}{c}{$K^-$} \\\hline

   &            &  1.00 &  0.93 &  0.91 &  0.90 &  0.89 &  0.88 &  0.88 &  0.85 &  0.80 &  0.85 &  0.91 &  0.92 &  0.92 &  0.92 &  0.91 &  0.90 &  0.87 &  0.84 &  0.60  \\
   &            &       &  1.00 &  0.94 &  0.94 &  0.92 &  0.92 &  0.92 &  0.88 &  0.84 &  0.81 &  0.89 &  0.92 &  0.93 &  0.94 &  0.93 &  0.92 &  0.90 &  0.87 &  0.61  \\
   &            &       &       &  1.00 &  0.93 &  0.92 &  0.91 &  0.91 &  0.87 &  0.83 &  0.79 &  0.88 &  0.91 &  0.92 &  0.93 &  0.93 &  0.91 &  0.89 &  0.87 &  0.60  \\
   &            &       &       &       &  1.00 &  0.92 &  0.92 &  0.92 &  0.88 &  0.84 &  0.77 &  0.87 &  0.91 &  0.92 &  0.93 &  0.93 &  0.91 &  0.89 &  0.87 &  0.60  \\
   &$\pi^-$     &       &       &       &       &  1.00 &  0.90 &  0.90 &  0.86 &  0.83 &  0.76 &  0.85 &  0.89 &  0.90 &  0.91 &  0.91 &  0.90 &  0.88 &  0.86 &  0.59  \\
   &            &       &       &       &       &       &  1.00 &  0.90 &  0.86 &  0.83 &  0.75 &  0.84 &  0.88 &  0.90 &  0.91 &  0.91 &  0.90 &  0.88 &  0.86 &  0.58  \\
   &            &       &       &       &       &       &       &  1.00 &  0.86 &  0.83 &  0.75 &  0.84 &  0.89 &  0.90 &  0.91 &  0.91 &  0.90 &  0.88 &  0.86 &  0.58  \\
   &            &       &       &       &       &       &       &       &  1.00 &  0.79 &  0.73 &  0.82 &  0.86 &  0.87 &  0.88 &  0.87 &  0.86 &  0.84 &  0.82 &  0.56  \\
Tagged&            &       &       &       &       &       &       &       &       &  1.00 &  0.66 &  0.75 &  0.80 &  0.81 &  0.82 &  0.83 &  0.82 &  0.80 &  0.79 &  0.53  \\\cline{2-21}
   &            &       &       &       &       &       &       &       &       &       &  1.00 &  0.94 &  0.91 &  0.90 &  0.88 &  0.86 &  0.85 &  0.82 &  0.77 &  0.62  \\
   &            &       &       &       &       &       &       &       &       &       &       &  1.00 &  0.96 &  0.95 &  0.94 &  0.92 &  0.91 &  0.89 &  0.84 &  0.64  \\
   &            &       &       &       &       &       &       &       &       &       &       &       &  1.00 &  0.97 &  0.96 &  0.95 &  0.94 &  0.92 &  0.88 &  0.65  \\
   &            &       &       &       &       &       &       &       &       &       &       &       &       &  1.00 &  0.97 &  0.96 &  0.94 &  0.92 &  0.89 &  0.65  \\
   &$K^-$       &       &       &       &       &       &       &       &       &       &       &       &       &       &  1.00 &  0.96 &  0.95 &  0.93 &  0.90 &  0.65  \\
   &            &       &       &       &       &       &       &       &       &       &       &       &       &       &       &  1.00 &  0.94 &  0.92 &  0.89 &  0.64  \\
   &            &       &       &       &       &       &       &       &       &       &       &       &       &       &       &       &  1.00 &  0.91 &  0.88 &  0.63  \\
   &            &       &       &       &       &       &       &       &       &       &       &       &       &       &       &       &       &  1.00 &  0.87 &  0.63  \\
   &            &       &       &       &       &       &       &       &       &       &       &       &       &       &       &       &       &       &  1.00 &  0.60  \\
   &            &       &       &       &       &       &       &       &       &       &       &       &       &       &       &       &       &       &       &  1.00  \\
\hline\hline

\end{tabular}
\end{center}
\label{tab:ff_rho_syst_tt1}
\end{table*}
\endgroup

\begingroup
\squeezetable
\begin{table*}[htbp]
\caption{The tagged block of the systematic correlation matrix. The lines indicate mode boundaries. The modes are labeled by their final state hadrons. Within each submode, the $q^2$ intervals are ordered from lowest to highest (part II).}
\begin{center}
\begin{tabular}{cc|rrrrrrr|rrrrr rrrrr} \hline\hline
& & \multicolumn{17}{c}{Tagged} \\
& & \multicolumn{7}{c|}{$\pi^0$} & \multicolumn{10}{c}{$\bar{K}^0$}\\\hline

   &            & 0.19 &  0.25 &  0.30 &  0.34 &  0.40 &  0.46 &  0.56 &  0.46 &  0.54 &  0.55 &  0.57 &  0.58 &  0.59 &  0.58 &  0.59 &  0.55 &  0.46 \\
   &            & 0.18 &  0.25 &  0.30 &  0.35 &  0.40 &  0.48 &  0.58 &  0.43 &  0.53 &  0.55 &  0.57 &  0.59 &  0.60 &  0.60 &  0.62 &  0.57 &  0.48 \\
   &            & 0.18 &  0.24 &  0.30 &  0.34 &  0.40 &  0.47 &  0.58 &  0.41 &  0.51 &  0.54 &  0.56 &  0.58 &  0.60 &  0.59 &  0.62 &  0.56 &  0.47 \\
   &            & 0.17 &  0.24 &  0.30 &  0.34 &  0.40 &  0.47 &  0.58 &  0.40 &  0.51 &  0.54 &  0.56 &  0.58 &  0.60 &  0.59 &  0.62 &  0.56 &  0.47 \\
   &$\pi^-$     & 0.17 &  0.24 &  0.29 &  0.34 &  0.40 &  0.47 &  0.57 &  0.40 &  0.50 &  0.53 &  0.55 &  0.57 &  0.59 &  0.58 &  0.61 &  0.56 &  0.47 \\
   &            & 0.17 &  0.23 &  0.29 &  0.34 &  0.39 &  0.47 &  0.57 &  0.38 &  0.49 &  0.53 &  0.55 &  0.56 &  0.58 &  0.58 &  0.61 &  0.55 &  0.47 \\
   &            & 0.17 &  0.23 &  0.29 &  0.34 &  0.39 &  0.47 &  0.57 &  0.38 &  0.49 &  0.53 &  0.55 &  0.56 &  0.58 &  0.58 &  0.61 &  0.55 &  0.47 \\
   &            & 0.16 &  0.23 &  0.28 &  0.32 &  0.38 &  0.45 &  0.55 &  0.38 &  0.48 &  0.51 &  0.53 &  0.54 &  0.56 &  0.56 &  0.58 &  0.53 &  0.45 \\
   &            & 0.15 &  0.21 &  0.26 &  0.31 &  0.36 &  0.43 &  0.52 &  0.33 &  0.44 &  0.47 &  0.50 &  0.51 &  0.53 &  0.53 &  0.56 &  0.51 &  0.43 \\\cline{2-19}
   &            & 0.19 &  0.23 &  0.27 &  0.31 &  0.35 &  0.40 &  0.48 &  0.53 &  0.56 &  0.55 &  0.55 &  0.55 &  0.55 &  0.55 &  0.52 &  0.50 &  0.42 \\
   &            & 0.19 &  0.24 &  0.29 &  0.33 &  0.38 &  0.44 &  0.53 &  0.52 &  0.57 &  0.58 &  0.59 &  0.59 &  0.60 &  0.59 &  0.58 &  0.55 &  0.46 \\
   &            & 0.19 &  0.25 &  0.30 &  0.34 &  0.39 &  0.46 &  0.56 &  0.49 &  0.56 &  0.58 &  0.59 &  0.60 &  0.61 &  0.61 &  0.61 &  0.57 &  0.48 \\
   &            & 0.19 &  0.25 &  0.30 &  0.34 &  0.40 &  0.46 &  0.57 &  0.48 &  0.56 &  0.58 &  0.59 &  0.60 &  0.62 &  0.61 &  0.62 &  0.58 &  0.48 \\
   &$K^-$       & 0.18 &  0.24 &  0.30 &  0.34 &  0.40 &  0.47 &  0.57 &  0.47 &  0.55 &  0.58 &  0.59 &  0.60 &  0.62 &  0.61 &  0.63 &  0.58 &  0.49 \\
   &            & 0.18 &  0.24 &  0.29 &  0.34 &  0.40 &  0.47 &  0.57 &  0.45 &  0.54 &  0.57 &  0.59 &  0.60 &  0.61 &  0.61 &  0.63 &  0.58 &  0.48 \\
   &            & 0.18 &  0.24 &  0.29 &  0.34 &  0.39 &  0.46 &  0.56 &  0.44 &  0.54 &  0.56 &  0.58 &  0.59 &  0.61 &  0.60 &  0.62 &  0.57 &  0.48 \\
   &            & 0.17 &  0.23 &  0.28 &  0.33 &  0.38 &  0.45 &  0.55 &  0.42 &  0.52 &  0.54 &  0.56 &  0.57 &  0.59 &  0.58 &  0.60 &  0.55 &  0.47 \\
   &            & 0.16 &  0.22 &  0.27 &  0.32 &  0.37 &  0.44 &  0.54 &  0.39 &  0.49 &  0.52 &  0.54 &  0.55 &  0.57 &  0.57 &  0.59 &  0.54 &  0.46 \\
Tagged&            & 0.12 &  0.16 &  0.19 &  0.22 &  0.26 &  0.30 &  0.37 &  0.32 &  0.37 &  0.38 &  0.39 &  0.39 &  0.40 &  0.40 &  0.40 &  0.38 &  0.32 \\\cline{2-19}
   &            & 1.00 &  0.98 &  0.97 &  0.95 &  0.91 &  0.84 &  0.67 &  0.15 &  0.15 &  0.15 &  0.15 &  0.15 &  0.15 &  0.15 &  0.14 &  0.14 &  0.12 \\
   &            &      &  1.00 &  0.98 &  0.96 &  0.93 &  0.87 &  0.72 &  0.17 &  0.19 &  0.19 &  0.19 &  0.19 &  0.19 &  0.19 &  0.19 &  0.18 &  0.15 \\
   &            &      &       &  1.00 &  0.97 &  0.95 &  0.90 &  0.77 &  0.19 &  0.22 &  0.22 &  0.23 &  0.23 &  0.23 &  0.23 &  0.23 &  0.22 &  0.18 \\
   &$\pi^0$     &      &       &       &  1.00 &  0.95 &  0.91 &  0.80 &  0.22 &  0.25 &  0.26 &  0.26 &  0.26 &  0.27 &  0.27 &  0.27 &  0.25 &  0.21 \\
   &            &      &       &       &       &  1.00 &  0.93 &  0.85 &  0.24 &  0.28 &  0.29 &  0.30 &  0.30 &  0.31 &  0.31 &  0.31 &  0.29 &  0.25 \\
   &            &      &       &       &       &       &  1.00 &  0.89 &  0.28 &  0.33 &  0.34 &  0.35 &  0.35 &  0.36 &  0.36 &  0.37 &  0.34 &  0.29 \\
   &            &      &       &       &       &       &       &  1.00 &  0.33 &  0.39 &  0.41 &  0.42 &  0.43 &  0.44 &  0.44 &  0.45 &  0.42 &  0.35 \\\cline{2-19}
   &            &      &       &       &       &       &       &       &  1.00 &  0.96 &  0.94 &  0.93 &  0.91 &  0.90 &  0.89 &  0.84 &  0.82 &  0.69 \\
   &            &      &       &       &       &       &       &       &       &  1.00 &  0.96 &  0.96 &  0.95 &  0.94 &  0.93 &  0.89 &  0.87 &  0.73 \\
   &            &      &       &       &       &       &       &       &       &       &  1.00 &  0.96 &  0.95 &  0.95 &  0.94 &  0.91 &  0.88 &  0.74 \\
   &            &      &       &       &       &       &       &       &       &       &       &  1.00 &  0.95 &  0.95 &  0.94 &  0.92 &  0.88 &  0.74 \\
   &$\bar{K}^0$ &      &       &       &       &       &       &       &       &       &       &       &  1.00 &  0.95 &  0.94 &  0.91 &  0.88 &  0.74 \\
   &            &      &       &       &       &       &       &       &       &       &       &       &       &  1.00 &  0.94 &  0.92 &  0.88 &  0.74 \\
   &            &      &       &       &       &       &       &       &       &       &       &       &       &       &  1.00 &  0.91 &  0.88 &  0.74 \\
   &            &      &       &       &       &       &       &       &       &       &       &       &       &       &       &  1.00 &  0.86 &  0.72 \\
   &            &      &       &       &       &       &       &       &       &       &       &       &       &       &       &       &  1.00 &  0.69 \\
   &            &      &       &       &       &       &       &       &       &       &       &       &       &       &       &       &       &  1.00 \\
\hline\hline

\end{tabular}
\end{center}
\label{tab:ff_rho_syst_tt2}
\end{table*}
\endgroup


\clearpage



\begin{thebibliography}{99}


\bibitem{ckm} M.~Kobayashi and T.~Maskawa, Prog. Theor. Phys. {\bf 49}, 652 (1973).


\bibitem{VubVcb} M.~Artuso and E.~Barberio,   Phys. Lett. B {\bf 592}, 786 (2004);
                 M.~Battaglia and L.~Gibbons, Phys. Lett. B {\bf 592}, 793 (2004).


\bibitem{Nadia}
D. Cronin-Hennessey~{\it et al.}, [CLEO Collaboration],
     Phys. Rev. Lett. {\bf 100}, 251802 (2008), and
S.~Dobbs~{\it et al.}, [CLEO Collaboration], Phys. Rev. D {\bf 77}, 112005 (2008).

\bibitem{BSW} J.D. Richman and P. Burchat, Rev. Mod. Phys.
{\bf 67}, 893 (1995) and references therein.

\bibitem{Boyd} C. G. Boyd, B.  Grinstein and R. F. Lebed, Nucl. Phys. B {\bf 461}, 493 (1996).

\bibitem{Boyd-2} C. G. Boyd and M. J. Savage, Phys. Rev.
D {\bf 56}, 303 (1997) and references therein.

\bibitem{Arnesen} M.~C. Arnesen, B.~Grinstein, I.~Z. Rothstein,
and I.~Z. Stewart Phys. Rev. Lett. {\bf 95}, 071802 (2005).

\bibitem{rhill} T. Becher and R.J. Hill, Phys. Lett. B {\bf 633}, 61 (2006).

\bibitem{HQET} N. Isgur and M. B. Wise, Phys. Lett. B {\bf 232}, 113 (1989); {\bf 237}, 527 (1990); E. Eichten and B. Hill, Phys.
Lett. B {\bf 234}, 511 (1990); H. Georgi, Phys. Lett. B {\bf 240}, 447
(1990).

\bibitem{BABAR-06}
  B.~Aubert {\it et al.},  [BABAR Collaboration],
Phys. Rev. D {\bf 76}, 052005 (2007).

\bibitem{mere-q2-expansion} See, e.g.,  S.~Descotes-Genon and A.~Le Yaouanc,
  J.\ Phys.\ G {\bf 35}, 115005 (2008)
  .

\bibitem{CLEOIII}
 G.S. Huang {\em et al.}, [CLEO Collaboration], Phys. Rev. Lett. {\bf
94}, 011802 (2005).

\bibitem{FOCUS-05} J.M. Link {\em et al.}, [FOCUS Collaboration],
Phys. Lett. B {\bf 607}, 233 (2005).

\bibitem{BELLE-06} L. Widhalm {\em et al.}, [Belle Collaboration],
Phys. Rev. Lett. {\bf 97}, 061804 (2006).

\bibitem{modpole} D. Becirevic and A.B. Kaidelov. Phys. Lett,
B {\bf 478}, 417 (2000).

\bibitem{Hill05}
  R.~J.~Hill,
  Phys.\ Rev.\  D {\bf 73}, 014012 (2006).

\bibitem{Hill06FPCP}
  R.~J.~Hill,
{\it In the Proceedings of 4th Flavor Physics and CP Violation Conference (FPCP 2006), Vancouver, British Columbia, Canada, 9-12 Apr 2006, pp
027}
  [arXiv:hep-ph/0606023].

\bibitem{survey-of-models}
M. Wirbel, B. Stech and M. Bauer, Z. Phys. C {\bf 29}, 637 (1985); J. G.
Korner and G. A. Schuler, Z. Phys. C {\bf 38}, 511 (1988), Erratum-ibid,
C {\bf 41}, 690 (1988); M. Bauer and M. Wirbel, Z. Phys. C {\bf 42}, 671 (1989);
J.~G.~Korner, K.~Schilcher, M.~Wirbel and Y.~L.~Wu,
  Z.\ Phys.\  C {\bf 48}, 663 (1990)
; W. Jaus, Phys. Rev. D {\bf 41}, 3394 (1990);
W.~Jaus,
  Phys.\ Rev.\  D {\bf 53}, 1349 (1996)
  [Erratum-ibid.\  D {\bf 54}, 5904 (1996)]
; R. Aleksan, A. Le Yaouanc, L. Oliver, O. P`ene, and J.-C.
Raynal, Phys. Rev. D {\bf 51}, 6235 (1995); I. L. Grach, I. M.
Narodetskii and S. Simula, Phys. Lett. B{ \bf 385}, 317 (1996); H. M.
Choi and C. R. Ji, Phys. Lett. B {\bf 460}, 461 (1999); D. Melikhov and
B. Stech, Phys. Rev. D {\bf 62}, 014006 (2000); G. Amoros, S. Noguera, and
J. Portoles, Eur. Phys. J. C {\bf 27}, 243 (2003); S. Fajfer and J.
Kamenik, Phys. Rev. D {\bf 71}, 014020 (2005).

\bibitem{ISGW} N. Isgur, D. Scora, B. Grinstein, and M. B. Wise, Phys.
Rev. D {\bf 39}, 799 (1989).

\bibitem{ISGW2} D. Scora and N. Isgur, Phys. Rev. D{ \bf 52}, 2783 (1995).

\bibitem{QCD-sum-rules-1} T. M. Aliev, V. L. Eletsky, and Y. I. Kogan, Sov. J. Nucl. Phys.
{\bf 40}, 527 (1984); P. Ball, V. M. Braun, and H. G. Dosch, Phys. Rev.
D {\bf 44}, 3567 (1991).

\bibitem{QCD-sum-rules-2} A. Khodjamirian, R. Ruckl, S. Weinzierl, C. W. Winhart, and O.
Yakovlev, Phys. Rev. D{ \bf 62}, 114002 (2000).

\bibitem{Flynn} J. M. Flynn and C. T. Sachrajda, Heavy Flavours (2nd ed.), ed.
by A. J. Buras and M. Linder (World Scientific, Singapore).
Published in Adv. Ser. Direct. High Energy Phys. {\bf 15}, 402 (1998).

\bibitem{Abada}
A. Abada, D. Becirevic, P. Boucaud, J. P. Leroy, V. Lubicz, and F.
Mescia, Nucl. Phys. B {\bf 619}, 565 (2001).

\bibitem{Davies}
C.T.H. Davies {\it et al.} Phys. Rev. Lett. {\bf 92},
 022001 (2004);
Follana, E. {\it et al.} [HPQCD and UKQCD Collaborations] Phys. Rev. Lett. {\bf 100}, 062002 (2008).

\bibitem{unquenched_LQCD} C.~Aubin {\it et al.}, Phys. Rev. Lett.
{\bf 94}, 011601 (2005).

\bibitem{cleo_detector}
Y.~Kubota {\it et~al.}, {Nucl. Instrum. Meth. Phys. Res., Sect. A}
\textbf{320}, {66} ({1992}); D.~Peterson {\it et~al.}, {Nucl.
Instrum. Meth. Phys. Res., Sect. A} \textbf{478}, {142} ({2002});
M.~Artuso {\it et~al.}, {Nucl. Instrum. Meth. Phys. Res., Sect. A}
\textbf{554}, {147} ({2005}).

\bibitem{GEANT} R.~Brun {\it et al.}, {\tt Geant 3.21}, CERN Program
Library Long Writeup W5013, unpublished.

\bibitem{EvtGen} D.J. Lange, Nucl. Instrum. Methods Phys. Res.
Sect. A {\bf 462}, 152 (2001).

\bibitem{MkIII}
J.~Adler {\it et al.}, [Mark III Collaboration], Phys. Rev. Lett.
{\bf 62}, 1821 (1989).

\bibitem{DSemilBFs-2005}
     G.S.~Huang~{\it et al.}, [CLEO Collaboration],
     Phys. Rev. Lett. {\bf 95}, 181801~(2005);
     T.E.~Coan {\it et al.}, [CLEO Collaboration],
     Phys. Rev. Lett. {\bf 95}, 181802~(2005).

\bibitem{PDG2004}
S.~Eidelman {\it et al.}, [Particle Data Group],
Phys. Lett. B {\bf 592}, 1 (2004).

\bibitem{cleoc-Dtagging}
Q.~He {\it et al.}  [CLEO Collaboration],
  Phys.\ Rev.\ Lett.\  {\bf 95}, 121801 (2005)
  [Erratum-ibid.\  {\bf 96}, 199903 (2006)].

\bibitem{ARGUS} H.~Albrecht {\it et al.}, [ARGUS Collaboration],
  Phys. Lett. B {\bf 241}, 278 (1990).

\bibitem{extra_track_cut}
Events with extra tracks are generally in the tails of the U distributions. We choose
to veto these events rather than to rely on the MC to model them.
The estimation of the systematic uncertainty associated with this veto is
described in Sec.~\ref{BFSystErrs}.

\bibitem{subsidiarybf} We use $\epsilon(\ksenu)
= \epsilon(D^+ \to K^0_S e^+ \nu_e) \cdot 0.5 \cdot {\cal B}(K^0_S \to \pi^+ \pi^-)$,
where ${\cal B}(K^0_S \to \pi^+ \pi^-) = (69.20 \pm 0.05)\%$~\cite{PDG2006}.

\bibitem{PHOTOS}
E.~Barberio and Z.~Was, Comput. Phys. Commun.  {\bf 79}, 291
(1994).

\bibitem{PDG2006}
  W.~M.~Yao {\it et al.}  [Particle Data Group],
  J.\ Phys.\ G {\bf 33}, 1 (2006).


\bibitem{PDG2008}
C.~Amsler {\it et al.}, [Particle Data Group],
Phys. Lett. B {\bf 667}, 1 (2008).

\bibitem{BESII}  W. Ablikim {\em et al.,} [BES Collaboration],
Phys. Lett. B {\bf 597}, 39 (2004).

\bibitem{WWZ}
W.~Y. Wang, Y.~L. Wu, and M. Zhong, Phys. Rev. D {\bf 67} 014024 (2003).

\bibitem{BESII_ratio}
M.~Ablikim {\it et al.},[BES Collaboration],  Phys. Lett. B {\bf 608}, 24 (2005).

\bibitem{FOCUS_ratio}
J.M.~Link {\it et al.}, [FOCUS Collaboration],  Phys. Lett. B {\bf 598}, 33
(2004).



\bibitem{pikenu_focus}
J.M.~Link {\it et al.}, [FOCUS Collaboration],  Phys. Lett. B {\bf 607},
51 (2005).

\bibitem{errors} M.~Lefebvre, R.K.~Keeler, R.~Sobie, and J.~White,
 Nucl. Instrum. Methods A {\bf 451}, 520 (2000).

\bibitem{eadie}
    W.T.~Eadie {\em et al.}, Statistical Methods in Experimental Physics,
Elsevier, New York, 1988.

\bibitem{MARKIII-91} A. Bai {\em et al.}, [MARK III Collaboration],
Phys. Rev. Lett. {\bf 66}, 1011 (1991).

\bibitem{E691-89} J.~C. Anjos {\em et al.}, [E691 Collaboration], Phys. Rev. Lett. {\bf 62}, 1587 (1989).

\bibitem{CLEO-91} G.~Crawford {\em et al.}, [CLEO Collaboration],
Phys. Rev. D {\bf 44}, 3394 (1991).

\bibitem{CLEOII-93} A. Bean {\em et al.}, [CLEO Collaboration],  Phys.
Lett. B {\bf 317}, 647 (1993).

\bibitem{E687-95} P.~L. Frabetti {\em et al.}, [E687 Collaboration],
Phys. Lett. B {\bf317}, 647 (1995).

\bibitem{CQM-alpha} D. Melikhov and B. Stech, Phys. Rev. D {\bf62},
014006 (2000).


\bibitem{note:newVcx}
The $\cleoc$ untagged analysis~\cite{Nadia} does not present
$f_+(0)$ results. It uses
values of $|V_{cs}|$ and $|V_{cd}|$ from an earlier PDG summary~\cite{PDG2006}
to extract the parameters $a_i$ in the series parametrization (see Eq.~(\ref{eq:series}))
for each semileptonic mode separately, but does not provide isospin averages.


\bibitem{PDG2000} D.E. Groom {\em et al.}, [Particle Data Group], European Physical Journal C{\bf 15}, 1 (2000).

\bibitem{BESIIVcs} To obtain $\vcs$ we have
used the measured $\bdzke$ from~\cite{BESII} and $\Gamma / \vcs^2$ from~\cite{unquenched_LQCD}.

\bibitem{Patricia}
P. Ball, Phys. Lett. B {\bf 641}, 50 (2006).

\bibitem{bootstrap} B.~Efron, Ann. Stat. {\bf 7}, 1 (1979).

\bibitem{nompoleave} As we do not present averages
with Ref.~\cite{Nadia} for the simple pole and modified pole parametrizations,
the results presented in this work for $M_{\rm pole}^\pi$,
$M_{\rm pole}^K$,  $\alpha^\pi$, and  $\alpha^K$,
should not be considered to supersede those of
Ref~\cite{Nadia}.

\bibitem{cleoc} CLEO-c/CESR-c Taskforces and CLEO Collaboration,
    Cornell LEPP Preprint CLNS 01/1742 (2001).

\end{thebibliography}
\end{document}